\documentclass[aps,prd,onecolumn,nofootinbib,superscriptaddress,bibnotes,preprintnumbers]{revtex4-2}

\usepackage{style}

\begin{document}

\vspace*{-0.5cm}
\noindent\hfill\raisebox{0pt}[0pt][0pt]{\small FERMILAB-PUB-25-0811-T} \\ \\
\vspace*{-0.5cm}
\title{A Thermal Relic Encyclopedia: Dark Matter Candidates Coupled to Quarks}

\author{Dan Hooper\,\orcidlink{0009-0004-2456-1221}}
\email{dwhooper@wisc.edu}
\affiliation{Theoretical Physics Division, Fermi National Accelerator Laboratory, Batavia, IL 60510, USA}
\affiliation{Department of Astronomy \& Astrophysics and KICP, University of Chicago, Chicago, IL 60637}
\affiliation{Wisconsin IceCube Particle Astrophysics Center (WIPAC) and Department of Physics, University of Wisconsin, Madison 53707, USA}

\author{Gordan Krnjaic\,\orcidlink{0000-0001-7420-9577}}
\email{krnjaicg@fnal.gov}
\affiliation{Theoretical Physics Division, Fermi National Accelerator Laboratory, Batavia, IL 60510, USA}
\affiliation{Department of Astronomy \& Astrophysics and KICP, University of Chicago, Chicago, IL 60637}

\author{Tanner Trickle\,\orcidlink{0000-0003-1371-4988}}
\email{ttrickle@illinois.edu}
\affiliation{Theoretical Physics Division, Fermi National Accelerator Laboratory, Batavia, IL 60510, USA}
\affiliation{Department of Physics, Grainger College of Engineering, University of Illinois Urbana-Champaign, Urbana, IL 61801, USA}

\author{Isaac R. Wang\,\orcidlink{0000-0003-0789-218X}}
\email{isaacw@fnal.gov}
\affiliation{Theoretical Physics Division, Fermi National Accelerator Laboratory, Batavia, IL 60510, USA}

\date{\today}

\begin{abstract}
    Thermal freeze-out is a compelling framework for naturally generating the dark matter abundance. 
    We systematically study a broad range of dark matter and mediator particle combinations that can viably realize thermal freeze-out, focusing on models in which the mediator couples to Standard Model quarks. In each case, we calculate the relic density and consider existing constraints from accelerators, cosmology, direct detection, and indirect detection over the full range of dark matter and mediator masses. We present an encyclopedic catalog of matrix elements, cross sections, and decay rates which can be used as a reference for future studies of dark matter phenomenology.
\end{abstract}

\maketitle

\vspace{-0.7cm}
{\small
\begingroup
  \setlength{\parskip}{1.3pt}%
  \renewcommand{\baselinestretch}{1.0}\normalsize
  \tableofcontents
\endgroup
}

\clearpage
\thispagestyle{empty}
\vspace*{\fill}
{
    \centering
    \renewcommand{\arraystretch}{2}
    \vspace*{\fill}
    \vspace*{-10cm}
    \begin{tabularx}{\textwidth}{>{\centering\arraybackslash}p{4cm}C C C C}
        \toprule\multirow{2}{4cm}[-1ex]{\centering \textbf{Quark Flavor}} & \multicolumn{4}{c}{\textbf{Dark Matter and Mediator Spin Combinations}} \\
        \cmidrule{2-5}
        & \makecell{$\bf (0,0)$\\Sec.~\ref{sec:Dm0Med0}} & \makecell{$\bf (\frac{1}{2},0)$\\ Sec.~\ref{sec:Dm1/2Med0}} & \makecell{$\bf (0,1)$\\ Sec.~\ref{sec:Dm0Med1}} & \makecell{$\bf (\frac{1}{2},1)$ \\
        Sec.~\ref{sec:Dm1/2Med1}} \\
        \hline
        $u$ & Fig.~\ref{fig:u_6_panel} & Fig.~\ref{fig:u_12_panel} & Fig.~\ref{fig:u_scalar_vector_6_panel} & Fig.~\ref{fig:u_12_panel_2} \\
        \hline
        $d$ & Fig.~\ref{fig:d_6_panel} & Fig.~\ref{fig:d_12_panel} & Fig.~\ref{fig:d_scalar_vector_6_panel} & Fig.~\ref{fig:d_12_panel_2} \\
        \hline
        $c$ & Fig.~\ref{fig:c_6_panel} & Fig.~\ref{fig:c_12_panel} & Fig.~\ref{fig:c_scalar_vector_6_panel} & Fig.~\ref{fig:c_12_panel_2} \\
        \hline
        $s$ & Fig.~\ref{fig:s_6_panel} & Fig.~\ref{fig:s_12_panel} & Fig.~\ref{fig:s_scalar_vector_6_panel} & Fig.~\ref{fig:s_12_panel_2} \\
        \hline
        $b$ & Fig.~\ref{fig:b_6_panel} & Fig.~\ref{fig:b_12_panel} & Fig.~\ref{fig:b_scalar_vector_6_panel} & Fig.~\ref{fig:b_12_panel_2} \\
        \hline
        $t$ & Fig.~\ref{fig:t_6_panel} & Fig.~\ref{fig:t_12_panel} & Fig.~\ref{fig:t_scalar_vector_6_panel} & Fig.~\ref{fig:t_12_panel_2} \\
        \bottomrule
    \end{tabularx}
    \captionof{table}{A summary of where the reader can find the formulas and plots corresponding to each of the dark matter models considered. In the first row, we list the the sections that contain the formulas for each combination of dark matter and mediator spins, labeled as ``$(\text{dark matter spin}, \text{mediator spin})$". The following rows list where the corresponding results are presented, for couplings to each quark flavor.
    } 
    \label{tab:summary}
}
\vspace*{\fill}
\clearpage

\section{Introduction}
\label{sec:intro}

For decades, many of the most popular dark matter candidates have been stable particles that froze out of equilibrium in the early universe~\cite{Bertone:2016nfn,Cirelli:2024ssz}. It was widely appreciated that if such particles had roughly weak-scale masses and couplings, this process would naturally result in an abundance similar to the measured cosmological dark matter density. These considerations -- sometimes referred to as the weakly interacting massive particle (WIMP) miracle -- elevated WIMPs to the most studied class of dark matter candidates.

The view of the scientific community on this question has shifted significantly in recent years. Due to a series of null results from underground dark matter direct detection experiments~\cite{DarkSide:2018bpj,PICO:2019vsc,CRESST:2019jnq,LZ:2022lsv,XENON:2023cxc}, searches for dark matter annihilation products~\cite{DiMauro:2022hue,McDaniel:2023bju,Bergstrom:2013jra,Planck:2018vyg,Cuoco:2016eej}, and searches for dark matter particles at the Large Hadron Collider (LHC)~\cite{CMS:2023dof,CMS:2020ulv,CMS:2019ykj,CMS:2018zjv,CMS:2018mgb,CMS:2018ucw,CMS:2017jdm,ATLAS:2024kpy,ATLAS:2023ild,ATLAS:2021shl,ATLAS:2021jbf,ATLAS:2020uiq,ATLAS:2018nda,ATLAS:2017bfj}, many WIMP models have been ruled out. In response, less attention has been directed toward WIMPs in recent years, while interest in other dark matter candidates (such as axions and primordial black holes) has increased.

In this study, we revisit the paradigm of WIMP-like dark matter in an effort to systematically evaluate the range of models that remain experimentally viable at this time.
To this end, we consider a broad range of scalar and fermionic dark matter candidates coupled to a spin-0 or spin-1 mediator, focusing on thermal relics that freeze out to yield the observed abundance of dark matter. We further limit this study to dark matter candidates that annihilate through mediators that couple to Standard Model (SM) quarks. We do not limit our discussion to models with specific ultraviolet (UV) completions, but rather approach this problem from the context of a simplified models framework, in analogy with the philosophy in Ref.~\cite{Alwall:2008ag}.
More discussion of such models can be found in Refs.~\cite{Kahn:2016vjr,Batell:2017kty,Batell:2018fqo,Batell:2021xsi}.

In many models, we identify substantial regions of parameter space that predict a WIMP consistent with all existing constraints from accelerators, cosmology, direct detection, and indirect detection. We present for each case the formulas necessary to compute the dark matter relic abundance, as well as those relevant to experimental probes, including collider phenomenology, direct detection, indirect detection, rare meson decays, and constraints on the effective number of neutrino species, $\Delta N_{\rm eff}$. For certain interactions (spin-1 mediators with axial-vector couplings, for example), we note that additional constraints can arise from anomaly cancellation and other details that are sensitive to the model's UV-completion. In our treatment, we limit our analysis to the bounds that follow directly from the simplified interactions in our effective Lagrangians, noting that concrete realizations may face additional constraints. 

This paper is organized as follows. 
In Sec.~\ref{sec:formalism} we introduce all the mediator-SM interactions we will consider here.
In Sec.~\ref{sec:calcs}, we present the essential formalisms required to calculate the dark matter relic abundance and experimental observables. 
In Secs.~\ref{sec:Dm0Med0} to~\ref{sec:Dm1/2Med1}, we derive formulas for each of the dark matter models considered in this paper. 
In Sec.~\ref{sec:result}, we evaluate the parameter space of each of these models, identifying those regions that remain viable in light of current constraints.
In Table~\ref{tab:summary}, we summarize where the reader can find the formulas and results that correspond to each of dark matter models considered here.

\section{Mediator-Standard Model Interactions}
\label{sec:formalism}

In this study, we consider simplified models of spin-0 and spin-$1/2$ dark matter candidates coupled to either spin-0 or spin-1 mediators. We further assume that these bosonic mediators couple to the SM quarks at tree level, which generates effective interactions with other SM species at low energies. 
In Sec.~\ref{subsec:mediator_SM_interaction} we briefly summarize the mediator-SM interaction Lagrangians at the energy scale of interest relevant for annihilation processes, discuss their derivation using low-energy theorems and chiral perturbation theory (ChPT) in Sec.~\ref{sec:lowE}. In each of these cases, we assume that the couplings are flavor-diagonal in order to avoid flavor-violating currents. The interactions of the mediator with the dark matter, $\chi$, will be described later in Sec.~\ref{sec:dm_models}. 

\subsection{Mediator-Standard Model Annihilation Interactions}
\label{subsec:mediator_SM_interaction}

For light quark flavors, $u$ and $d$ (see App.~\ref{sec:chpt app} for more details), the Lagrangians describing the mediator-SM interactions at the relevant energy scales are given by
\begin{align}
    \mathcal{L}_\text{SM-med.}^\phi & = \phi \begin{cases}
         g_{f s} \bar{f} f + i g_{f p} \bar{f} \gamma^5  f, & E > \Lambda_\text{QCD} \\[2ex] 
         g_{fs} \, c_{\phi\gamma} \, F_{\mu \nu}F^{\mu \nu} + g_{fp} \, c'_{\phi\gamma} \, F_{\mu \nu} \tilde{F}^{\mu \nu}, & E < \Lambda_\text{QCD}
    \end{cases}
    \label{eq:L_phi_SM_med}
\end{align}

\begin{align}
    \mathcal{L}_\text{SM-med.}^V & = V_\mu \begin{cases}
         g_{f v} \, \bar{f} \gamma^\mu f + g_{f a} \, \bar{f} \gamma^\mu \gamma^5  f, & E > \Lambda_\text{QCD} \\[3ex]
         \displaystyle \frac{Q_f e g_{fv}}{16 \pi^2} \, V_{\mu \nu} F^{\mu \nu}, & E < \Lambda_\text{QCD}
     \end{cases}
     \label{eq:L_V_SM_med}
\end{align}
where $g_{fs}, g_{fp}, g_{fv}$, and $g_{fa}$ are the real-valued mediator-SM quark couplings, $f$ is a SM quark with electric charge $Q_f$, $\phi$ is the spin-0 mediator, $V_\mu$ is the spin-1 mediator, $F_{\mu \nu}$ is the electromagnetic field strength tensor, $\tilde{F}^{\mu \nu} \equiv \epsilon^{\alpha \beta \mu \nu} F_{\alpha \beta} / 2$, $V_{\mu \nu} = \partial_\mu V_\nu - \partial_\nu V_\mu$, and $\Lambda_\text{QCD}$ is the QCD scale. We adopt $\Lambda_\text{QCD} = 146 \, \text{MeV}$ and use ChPT to derive the effective mediator-photon coupling coefficients, $c_{\phi\gamma}$ and $c_{\phi\gamma}'$ in Sec.~\ref{sec:lowE}.
% Note that for $2m_e < m_V < \Lambda_{\rm QCD}$, the vector mediator cannot decay to pairs of gluons or photons and must decay through the kinetic mixing to two electrons, and for $m_V < 2m_e$, the only decay channel is $V\to 3 \gamma$. 
% which is generically long-lived and spoils the successful predictions of standard BBN~\cite{McDermott:2017qcg}. 

For heavy quark flavors, $c, s, t$, and $b$, their mass is another relevant scale. For energies below the quark mass, $m_f$, but above $\Lambda_\text{QCD}$, the mediator can also couple to gluons. Therefore the relevant interaction Lagrangians are
\begin{align}
    \mathcal{L}_\text{SM-med.}^\phi & = \phi \begin{cases}
         g_{f s} \bar{f} f + i g_{f p} \bar{f} \gamma^5  f, & E > m_f \\[2ex] 
         g_{fs} \, c_{\phi\gamma} \, F_{\mu \nu}F^{\mu \nu} + g_{fp} \, c'_{\phi\gamma} \, F_{\mu \nu} \tilde{F}^{\mu \nu} + g_{fs} \, c_{\phi g} \, G_{\mu \nu}^a G^{\mu \nu}_a + g_{fp} \, c'_{\phi g} \, G_{\mu \nu}^a \tilde{G}^{\mu \nu}_a, & m_f > E > \Lambda_\text{QCD} \\[2ex]
         g_{fs} \, c_{\phi\gamma} \, F_{\mu \nu}F^{\mu \nu} + g_{fp} \, c'_{\phi\gamma} \, F_{\mu \nu} \tilde{F}^{\mu \nu}, & E < \Lambda_\text{QCD} \, ,
    \end{cases}
\end{align}
where $G^{\mu \nu}_a$ is the gluon field strength tensor, $\tilde{G}^{\mu \nu}_a \equiv \epsilon^{\alpha \beta \mu \nu} G_{\alpha \beta}^a / 2$, and the spin-1 interaction Lagrangian is identical to Eq.~\eqref{eq:L_V_SM_med} with $\Lambda_{\rm QCD}$ replaced by $2 m_f$.\footnote{For simplicity, we ignore the logarithmic factor in the $E < \Lambda_\text{QCD}$ Lagrangian in Eq.~\eqref{eq:L_V_SM_med}.}

\subsection{Low-Energy Theorems and Chiral Perturbation Theory}
\label{sec:lowE}

We begin by discussing the calculation of the spin-0 mediator's coupling to photons via interactions with light quarks, as found in Eq.~\eqref{eq:L_phi_SM_med} for $E < \Lambda_\text{QCD}$. At these energies, we can use ChPT to derive the coupling of the mediator to the pions, and then use the charged pion-photon coupling to derive the interaction Lagrangian. We refer the readers to Refs.~\cite{Pich:1993uq,Scherer:2002tk,Scherer:2012xha} for a comprehensive review of ChPT.

ChPT informs us that the effective Lagrangian coupling the spin-0 mediator to pions is given by
\begin{align}
    \mathcal{L}_\text{ChPT} \supset \left( g_{us} + g_{ds} \right) B_0 \, \phi \, \pi ^- \pi ^+ + ( g_{up} - g_{dp} ) F_\pi B_0 \, \phi \, \pi^0 \,,
    \label{eq:L_ChPT}
\end{align}
where the first two terms are generated from scalar interactions, the last two are generated from pseudoscalar interactions, $F_\pi = 93.2~\rm MeV$, $B_0 \simeq m_\pi^2/(m_u + m_d) \simeq 2.6~\rm GeV$~\cite{Batell:2018fqo}, and $m_\pi \approx 135 \, \text{MeV}$ is the pion mass. 
The scalar mediator's interactions with a pair of photons is due to a charged pion loop, and the effective operator is given by,
\begin{align}
\label{eq:chpt-med0}
    \phi \, \pi^- \pi^+ \to  \frac{\alpha}{48 \pi m_\pi^2} \phi F_{\mu \nu}F^{\mu \nu}\,,
\end{align} 
where $\alpha$ is the electromagnetic fine structure constant. Pseudoscalar mediators can mix with the neutral pion, allowing them to interact with a pair of photons. Given the SM pion-photon effective interaction, $\mathcal{L}_\text{ChPT} \supset ( \alpha / 4\pi F_\pi) \, \pi^0 F_{\mu \nu}\tilde{F}^{\mu \nu}$~\cite{Schwartz:2014sze},
and the mixing term in Eq.~\eqref{eq:L_ChPT}, an effective $\phi \gamma \gamma$ interaction can be obtained by multiplying this by $g_{fp} F_\pi B_0$ and dividing by the propagating pion mass, resulting in the following substitution in Eq.~\eqref{eq:L_ChPT}:
\begin{align}
    F_\pi B_0  \, \phi \pi^0 \to \frac{B_0}{m_\pi^2} \frac{\alpha}{4\pi} \phi F_{\mu \nu} \tilde{F}^{\mu \nu}\,.
\end{align}

The heavy flavor effective operators are easier to compute, since these follow simply from low-energy theorems~\cite{Kniehl:1995tn}:
\begin{align}
    \phi \, \bar{f} f  \rightarrow  \, \frac{\alpha_s}{12 \pi m_f} \, \phi \, G^a_{\mu \nu} G_a^{\mu \nu} + \frac{N_c Q_f^2 \alpha}{6 \pi m_f} \, \phi \, F_{\mu \nu} F^{\mu \nu}~,~~~~~\phi \, i \bar{f} \gamma^5 f  \rightarrow - \,\frac{\alpha_s}{4 \pi m_f} \, \phi \, G^a_{\mu \nu} \widetilde{G}_a^{\mu \nu} - \frac{N_c Q_f^2 \alpha}{4 \pi m_f} \,\phi \, F_{\mu \nu} \widetilde{F}^{\mu \nu}\,, \label{eq:ps-quark-loop}
\end{align}
where $N_c = 3$ is the number of colors, and $\alpha_s$ is the strong coupling constant.

To summarize, the $\phi \gamma \gamma$ and $\phi g g$ interaction coefficients in Eq.~\eqref{eq:L_phi_SM_med} for light quark ($u$, $d$) couplings are given by
\begin{align}
    c_{\phi \gamma} = \frac{\alpha B_0}{48 \pi m_\pi^2} ~,~~~~~c_{\phi\gamma}' = \frac{\alpha B_0}{4 \pi m_\pi^2}~,
\end{align}
and for heavy quark ($c$, $s$, $t$, $b$) couplings,
\begin{align}
    c_{\phi \gamma} & = \frac{N_c Q_f^2 \alpha}{6 \pi m_f}~,~~~~~c_{\phi\gamma}' = - \frac{N_c Q_f^2 \alpha}{4 \pi m_f}~,~~~~~c_{\phi g} = \frac{\alpha_s}{12 \pi m_f} ~,~~~~~c_{\phi g}' = -\frac{\alpha_s}{4 \pi m_f} ~.%~~~~~~~~({\rm heavy}\, {\rm  quark} \, {\rm flavors}, \,c, s, t, b) 
\end{align}

\subsection{Mediator Decay Rates to Standard Model Particles}
\label{subsec:decay_rate_to_SM}

With the mediator-SM Lagrangians defined in Sec.~\ref{subsec:mediator_SM_interaction}, we can now compute the decay rate of the mediators into the relevant SM particles at each energy. For the light quark ($u$, $d$) couplings, the relevant decay rates are
\begin{eqnarray}
        \Gamma_{\phi}  &=& \begin{cases}
        \displaystyle \frac{N_c m_\phi }{8 \pi } \sqrt{1 - \frac{4 m_f^2}{m_\phi^2}} \left[ \, g_{f s}^2 \left( 1 - \frac{4 m_f^2}{m_\phi^2} \right)  + g_{f p}^2  \right]\ & \quad m_\phi > \Lambda_\text{QCD} \\[4ex] 
       \displaystyle \frac{m_\phi}{16 \pi} \left(\frac{\alpha B_0}{24 \pi} \frac{m_\phi}{m_\pi^2}\right)^2 \left[g_{fs}^2 + 144\, g_{fp}^2\right] & \quad m_\phi < \Lambda_\text{QCD} \, 
    \end{cases}
\end{eqnarray}
\begin{eqnarray}
    \Gamma_V  &=& 
    \begin{cases}
        \displaystyle \frac{N_c m_V}{12 \pi} \sqrt{1 - \frac{4 m_f^2}{m_V^2}} \left[ g_{f v}^2 \left( 1 + \frac{2 m_f^2}{m_V^2} \right) + g_{f a}^2 \, \left( 1 - \frac{4 m_f^2}{m_V^2} \right) \right]\, & m_V > \Lambda_\text{QCD} \, \\[4ex]
        \displaystyle
        \frac{m_V}{12 \pi} \sqrt{1 - \frac{4 m_e^2}{m_V^2}} \left(\frac{ Q_f e^2g_{fv}}{16\pi^2}\right)^2 \left( 1 + \frac{2 m_e^2}{m_V^2} \right)\, & m_V < \Lambda_{\rm QCD}\,.
    \end{cases}
    \label{eq:gamma_V_light}
\end{eqnarray}

For heavy quark ($c$, $s$, $t$, $b$) couplings, the decay rates are given by
\begin{align}
    \Gamma_{\phi} & = \begin{cases}
        \displaystyle \frac{N_c m_\phi }{8 \pi } \sqrt{1 - \frac{4 m_f^2}{m_\phi^2}} \left[ \, g_{f s}^2 \left( 1 - \frac{4 m_f^2}{m_\phi^2} \right)  + g_{f p}^2  \right]\,, & m_\phi > 2 m_f \\[4ex] 
       \displaystyle \frac{(N_c^2 - 1) m_\phi}{16 \pi} \left(\frac{\alpha_s}{6 \pi} \frac{m_\phi}{m_f}\right)^2\left[g_{fs}^2 + 9\, g_{fp}^2 \right] + \frac{m_\phi}{16 \pi} \left(\frac{N_c \alpha Q_f^2}{3 \pi} \frac{m_\phi}{m_f}\right)^2 \left[g_{fs}^2 + \frac{9}{4}\, g_{fp}^2\right]\,, & 2 m_f > m_\phi > \Lambda_\text{QCD} \\[4ex]
       \displaystyle \frac{m_\phi}{16 \pi} \left(\frac{N_c \alpha Q_f^2}{3 \pi} \frac{m_\phi}{m_f}\right)^2 \left[g_{fs}^2 + \frac{9}{4} \, g_{fp}^2 \right]\,, & m_\phi < \Lambda_\text{QCD} \, 
    \end{cases}
\end{align}
with an identical expression for the vector decay rate at $m_V > 2 m_f$, as in Eq.~\eqref{eq:gamma_V_light}. 

Note that a spin-1 mediator cannot decay to a pair of photons or gluons due to the Landau-Yang theorem~\cite{Landau:1948kw,Yang:1950rg}.
Thus, for $m_V < 2m_f$, decays to three photons or gluons may be the leading processes. For completeness, we comment on other possible decay channels.
If the mediator has a purely vector ($g_{fv} \neq 0, g_{av} = 0$) to fermions, it can also decay through an induced kinetic mixing interaction arising from fermion loops, so generically all vector mediators with $m_V > 2m_e$ can decay promptly to $V \to e^+e^-$ or to pairs of other kinematically accessible charged fermions.
For purely axial-vector couplings ($g_{fv} = 0, g_{av} \ne 0$), there is no kinetic mixing with $F_{\mu \nu}$ induced through $f$ loops due to the $\gamma^5$ vertex, and thus decaying into lighter charged fermion pairs through a kinetic mixing insertion is forbidden.
However, since purely axial-vector couplings are anomalous under the SM gauge group, theoretical consistency requires additional heavy ``anomalon" fields with SM charge and couplings to $V$ in order to cancel anomalies in the UV theory \cite{Kahn:2016vjr}. Generically, the presence of such states yields additional decay channels that enable the $V$ to decay promptly through higher-order operators. Furthermore, at energies below the scale of anomalon masses, the non-decoupling Wess-Zumino $V\gamma \gamma$ interaction allows decays of the form $V \to e^+e^- \gamma$ (and other charged fermion channels in association with a photon), even though $V \to \gamma\gamma$ decays are forbidden by the Landau-Yang theorem. Thus, despite some mild model dependence in the anomaly-induced interactions of the axial-vector mediator, its decays should be prompt for all masses in the $2m_\chi > m_V > 2m_e$ window, assuming couplings of the size required for dark matter freeze-out.

\section{Calculations and Constraints}
\label{sec:calcs}

We present the theoretical framework required to calculate a variety of relevant dark matter observables, including the thermal relic abundance (Sec.~\ref{subsec:relic_abundance}), cosmological constraints (Sec.~\ref{sec:Neff}), direct detection signals (Sec.~\ref{sec:direct}), indirect detection signals (Sec.~\ref{sec:indirect}), collider signals (Sec.~\ref{sec:collider}), and rare meson decays (Sec.~\ref{sec:meson}). In this section, we will use $\chi$ to denote both spin-0 and spin-1/2 dark matter candidates. 

\subsection{Relic Abundance}
\label{subsec:relic_abundance}

The number density of dark matter particles, $n_\chi$, evolves according to the standard Boltzmann equation~\cite{Hooper:2024avz},
\begin{align}
    \frac{dn_{\chi}}{dt} + 3Hn_{\chi} =  - \langle \sigma v \rangle [n_\chi^2-(n_\chi^{\rm eq})^2]\,,
    \label{eq:boltzmann}
\end{align}
where $H = 1.66 \sqrt{g_*} \, T^2/M_{\rm Pl}$ is the rate of Hubble expansion in radiation domination, $g_*$ is the effective number of relativistic degrees of freedom, $T$ is the temperature of the thermal bath, and $M_{\rm Pl} \simeq 1.22 \times 10^{19} \, \text{GeV}$ is the Planck mass.
In equilibrium (and in the absence of any chemical potential), the number density of $\chi$ particles is given by
\begin{align}
    n_{\chi}^{\rm eq} = \frac{g_{\rm spin}}{2\pi^2}
    \int_{m_{\chi}}^\infty \dd E \,  \frac{E(E^2-m_\chi^2)^{1/2}}{e^{E/T}\pm 1} \,,
\end{align}
where $g_{\rm spin}$ and $m_{\chi}$ are the number of internal degrees-of-freedom and the mass of the dark matter particle, respectively.\footnote{We adopt the notation $g_{\rm spin}$ to avoid later confusion with the couplings appearing in the interaction Lagrangians in Sec.~\ref{sec:dm_models}.}
The plus (minus) sign in the denominator applies to the case of fermions (bosons).
If $\chi$ is not its own antiparticle, the total dark matter relic density is $n_{\rm DM} = n_\chi + n_{\bar{\chi}} = 2 n_\chi$, where $\bar{\chi}$ is the antiparticle of $\chi$.

The thermally-averaged cross section for dark matter annihilation to the final state $f\bar{f}$, with a mass $m_f$, is given by~\cite{Gondolo:1990dk}
\begin{align}
    \langle \sigma v \rangle = \frac{x}{128 \pi g_{\rm spin}^2  m_\chi^5 K_2^2(x) } \int_{s_0}^\infty \frac{\dd s}{\sqrt{s}}   \, \sqrt{s  - 4 m_f^2 } \sqrt{s - 4 m_\chi^2}\, |\mathscr{M}|^2(s) \, K_1\left( \frac{ \sqrt{s}}{ T} \right) \,,
    \label{eq:sigma_v_general}
\end{align}
where $\sqrt{s}$ is the total energy of the collision in the center-of-momentum frame, $s_0 = 4 \, \text{max} \{ m_f^2, m_\chi^2 \}$, $x \equiv m_\chi/T$ is a widely-used time variable for dark matter freeze-out, and $K_1$ and $K_2$ are modified Bessel functions of the second kind.

The squared amplitude in this expression, $|\mathscr{M}|^2$, is summed over spin and polarization states and averaged over the direction of the outgoing particles in the center-of-mass frame. This can be written 
\begin{align}
    \label{eq:Msq}
    |\mathscr{M}|^2(s)  \equiv  \sum_{\substack{ \text{spins}                                                     \\ \text{pols} } } \frac{1}{4 \pi} \int |\mathcal{M}|^2 \, \dd \Omega_{\text{cm}} 
                      =  \frac{1}{s F_{f\chi}(s)} \sum_{\substack{ \text{spins} \\ \text{pols} } } \int_{t_-}^{t_+}  |\mathcal{M}|^2(s, t) \,  \dd t \, ,
\end{align}
where $F_{f\chi}(s) \equiv s^{-1} \sqrt{s - 4 m_f^2} \sqrt{s - 4 m_\chi^2} $, $|\mathcal{M}|^2(s, t)$ is the squared amplitude as a function of Mandelstam variables, and $t_\pm = m_f^2 + m_\chi^2 + s \left( -1 \pm F_{f\chi}(s) \right) / 2$.
Note that for $n$ identical particles in the final state, $|\mathscr{M}|^2$ should be divided by a symmetry factor, $n!$.

As we are interested in cold dark matter candidates, we can be confident that freeze-out will occur when the dark matter particles are non-relativistic, such that $x \equiv m_\chi/T \gg 1$.
In this limit, $\langle \sigma v\rangle$ in Eq.~\eqref{eq:sigma_v_general} can be approximated as $\langle \sigma v\rangle \approx \langle \sigma v \rangle_\text{NR}$, where,
\begin{align}
    \label{eq:sigmav NR}
    \langle \sigma v \rangle_\text{NR} & \equiv \int_{v_\text{min}}^\infty \dd v \, ( \sigma v)_\text{NR}  \, f_\text{MB}(v) ,             
\end{align}
and $(\sigma v)_\text{NR}$ can be written as
\begin{equation}
     (\sigma v)_\text{NR}  \equiv \frac{1}{32\sqrt{2} \pi v} \frac{1}{g_{\rm spin}^2 m_\chi^2} \left[ \left( \frac{s}{m_\chi^2} \right)^{3/4} F_{f\chi}(s) \, |\mathscr{M}|^2(s) \right]_{s \rightarrow 4 m_\chi^2 (1 + v^2 / 8)^2}\, . \label{eq:sigmav NR non-avg}     
\end{equation}
The quantity $v_{\rm min}$ can be determined by solving  $4m_\chi^2 (1+v_{\rm min}^2/8)^2 = s_0$ and $f_\text{MB}$ is the unit normalized ($\int \dd v f_\text{MB}(v) = 1$) Maxwell-Boltzmann velocity distribution,
\begin{align}
    f_\text{MB}(v) = \frac{x^{3/2} \, v^2}{2\sqrt{\pi}} \,  e^{- x v^2 / 4} \, .
\end{align}
Eq.~\eqref{eq:sigmav NR} can be derived by taking the $x \gg 1$ limit of Eq.~\eqref{eq:sigma_v_general} and making the variable redefinition, $s = 4 m_\chi^2 (1 + v^2 / 8)^2$. Written in the form of Eq.~\eqref{eq:sigmav NR} makes the meaning of the different cross sections clear: $(\sigma v)_\text{NR}$ is the non-relativistic cross section multiplied by the relative velocity, and $\langle \sigma v \rangle_\text{NR}$ is it's velocity-averaged counterpart. 

To find the cold dark matter relic abundance we solve Eq.~\eqref{eq:boltzmann} assuming $\langle \sigma v\rangle \approx \langle \sigma v \rangle_\text{NR}$, using the procedure outlined in Refs.~\cite{Steigman:2012nb,Steigman:2013yua}. The couplings which produce the observed relic abundance are found by requiring that the dark matter energy density today is $m_\chi n_\text{DM} = \Omega_\text{DM} \rho_c$, where $\Omega_\text{DM} h^2 \approx 0.12$ and $\rho_c \approx 2.78 \times 10^{11} h^2 M_\odot \, \text{Mpc}^{-3}$~\cite{ParticleDataGroup:2024cfk}.

\subsection{Cosmological Constraints}
\label{sec:Neff}

If the dark matter and mediator transfer entropy to SM species after the epoch of neutrino decoupling at $T_{\rm dec} \approx$ 2.3 MeV \cite{Enqvist:1991gx}, the neutrino-photon temperature ratio can differ markedly from the SM prediction, $T_\nu/T_\gamma = (4/11)^{1/3}$ \cite{Boehm:2012gr,Ho:2012ug,Steigman:2013yua}. Deviations from this value can affect the light element yields from Big Bang Nucleosynthesis (BBN) and the temperature anisotropies of the Cosmic Microwave Background (CMB), which both agree well with SM predictions and can be used to constrain new species that reached thermal equilibrium in the early universe. 

For dark matter candidates with masses between a few MeV and a few tens of MeV, the gradual departure from chemical equilibrium does not fully complete by the time of neutrino decoupling. Thus, for electromagnetically coupled dark matter, the residual entropy that is transferred to SM species after this time increases the photon temperature, but does not affect the neutrino temperature, thereby predicting $T_\nu/T_\gamma < (4/11)^{1/3}$.
The resulting impact on BBN and CMB observables can be parameterized in terms of the effective number of neutrino species, $N_{\rm eff}$, defined through the relation 
\begin{equation}
\rho_R = \rho_\gamma \left[  1 + \frac{7}{8} \left( \frac{4}{11} \right)^{4/3} N_{\rm eff}   \right],
\end{equation}
where $\rho_R$ is the total radiation energy density,  $\rho_\gamma = \pi^2 T^4/15$ is the photon energy density, $N_{\rm eff} = N_\nu + \Delta N_{\rm eff}$ is the effective number of neutrino species, $N_\nu = 3.043$
is the prediction of the SM with three neutrino species~\cite{Cielo:2023bqp}, and $\Delta N_{\rm eff}$ parametrizes the effects of new physics. Including the effect of dark sector species thermalized with photons and electrons, we can write~\cite{Boehm:2012gr}
\begin{equation}
     N_{\rm eff} = N_\nu \left[ 1 + \frac{7}{22}\sum_{i} \frac{ g_{\mathrm{spin}, i}}{2} \, F\left( \frac{m_i}{T_{\rm dec}}\right) \right]^{-4/3},
\end{equation}
where the sum is over all dark sector states of mass $m_i$, $g_{\mathrm{spin}, i}$ is the corresponding number of degrees of freedom, and 
we have defined 
\begin{equation}
    F(x) = \frac{30}{7\pi^4} \int_x^\infty \dd y \, \frac{ (4y^2 - x^2 )  \sqrt{y^2 - x^2}   }{ e^y \pm 1}, 
\end{equation}
where the $\pm$ correspond to fermions and bosons, respectively. Note that for electromagnetically coupled species (as opposed to neutrino-coupled species), the entropy transfer to photons results in a negative contribution to $\Delta N_{\rm eff}$.
In our analysis, we use the Planck measurement of $N_{\rm eff} = 2.99 \pm 0.17$ \cite{Planck:2018vyg} to place 2$\sigma$ constraints on the dark matter and mediator contributions to this quantity.

\subsection{Direct Detection}
\label{sec:direct}

The effective field theory of non-relativistic dark matter-nuclei scattering was developed in Refs.~\cite{Fitzpatrick:2012ix,Cirelli:2013ufw} (see also, Ref.~\cite{Kumar:2013iva}). These references provide the algorithm needed to match any dark matter-quark interaction to the dark matter-nucleon interaction that governs the dark matter-nuclei scattering rate. Furthermore, experimental collaborations have recently been publishing constraints on the dark matter-nucleon couplings as described in the effective field theory~\cite{XENON:2017fdd,LZ:2023lvz}. Using these results, we are able to place limits on the coupling parameters for each of the dark matter models discussed in Sec.~\ref{sec:dm_models}. 

To find the dark matter-nucleon couplings, we follow the procedure outlined in Refs.~\cite{Fitzpatrick:2012ix,Cirelli:2013ufw}. An example of this matching procedure for a vector mediator coupled to the $u$-quark and dark matter currents, $J^\mu_f \equiv \bar{f} \gamma^\mu f$, is done in three steps, as follows:
\begin{align}
    V_\mu ( g_{uv} J^\mu_u + g_{\chi v} J^\mu_\chi)~~\rightarrow~~\frac{g_{uv} g_{\chi v}}{m_V^2} J^\mu_u J_{\chi \, \mu}~~\rightarrow~~\frac{g_{uv} g_{\chi v}}{m_V^2} (2 J^\mu_p + J^\mu_n) J_{\chi \, \mu}~~\rightarrow~~\frac{ 4 m_\chi g_{uv} g_{\chi v}}{m_V^2} 1_\chi \left( 2 m_p \, 1_p + m_n 1_n \right) \, .
    \label{eq:example_EFT_match}
\end{align}
In the first step we integrate out the heavy mediator, which is appropriate for nearly all dark matter and mediator masses we consider here, since the typical momentum transfer is $\lesssim 100 \, \text{MeV}$. In the second step, we match from the quark level description to the nucleon level description using the dictionary in Eq. 45 of Ref.~\cite{Cirelli:2013ufw}. In the third step, we evaluate the relevant spinors in the non-relativistic limit of the dark matter-nucleon interaction to arrive at the effective non-relativistic dark matter-nucleon interaction. From the last term we can identify the (dimensionless) coefficients of the $\mathcal{O}^N_1 = 1_\chi 1_N$ operator as,
\begin{align}
    c^p_1 = \frac{ 8 g_{uv} g_{\chi v} m_p m_\chi}{m_V^2}~,~~~~~c^n_1 = \frac{ 4 g_{uv} g_{\chi v} m_n m_\chi}{m_V^2} \, .
    \label{eq:nr_coefficients}
\end{align}

The same steps of the matching procedure can be applied to all of the models discussed in Sec.~\ref{sec:dm_models}, and the different interaction structures lead to many more operators being generated, e.g., $\mathcal{O}_4 = \mathbf{S}_\chi \cdot \mathbf{S}_N~,~\mathcal{O}_{11} = i \mathbf{S}_\chi \cdot \mathbf{q} / m_N$, where $\mathbf{S}$ is the spin operator. To set limits on the Lagrangian couplings appearing in the non-relativistic coefficients, we use the LZ collaboration's recent results as presented in Ref.~\cite{LZ:2023lvz}.\footnote{The coefficients in Ref.~\cite{LZ:2023lvz} are dimensionful due to different conventions. To convert between them we note that their coefficients, $c_\text{LZ}$, are related to the ones here, $c$ (Eq.~\eqref{eq:nr_coefficients}), by $c = 4 m_\chi m_N \, c_\text{LZ}$.}

To develop intuition for the size of couplings needed to generate detectable elastic scattering cross sections, it is useful to directly calculate the elastic scattering rate of dark matter particles with a target. This can be expressed as
\begin{align}
\frac{dR}{dE_{R}} = N_T \, \frac{\rho_\chi}{m_\chi} \int 
\dd^3\mathbf{v} \, v \,
\frac{d\sigma_N}{dE_R}  f(\mathbf{v}) ,
\end{align}
where $N_T$ is the number of nuclei in the target, $d\sigma_N/dE_R$ is the dark matter's differential scattering cross section with a target nucleus (per unit recoil energy, $E_R = q^2 / 2 m_T$, where $q$ is the momentum transfer and $m_T$ is the target mass), $\mathbf{v}$ is the velocity of a dark matter particle relative to the target, and $f(\mathbf{v})$ is the velocity distribution of dark matter particles, normalized such that $\int f(\mathbf{v}) \, \dd^3\mathbf{v}=1$. Direct detection experiments traditionally present their results adopting a boosted Maxwell-Boltzmann velocity distribution with $v_0=238 \, {\rm km/s}$, truncated at an escape velocity of 544 km/s, and normalized to a local density of $\rho_{\chi} = 0.3 \, {\rm GeV/cm}^3$~\cite{Baxter:2021pqo}. 

Given the non-relativistic nature of dark matter particles in the halo of the Milky Way, we will first focus on those elastic scattering cross sections that are finite in the $q \rightarrow 0, v \rightarrow 0$ limit. 
There are three classes of operators that yield finite scattering cross sections in the $q \rightarrow 0, v \rightarrow 0$ limit. The first of these is the case of vector interactions, which yield a spin-independent cross section of the following form:
\begin{align}
\bar{\sigma}_N = \frac{\mu_{\chi N}^2 \lambda^2_{\chi}}{\pi m_V^4} \bigg[Z (2g_{uv}+g_{dv}) +(A-Z) (g_{uv}+2g_{dv}) \bigg]^2,
\end{align}
where $\bar{\sigma}_N \equiv \int \dd E_R \, d \sigma_N / d E_R$ is the integrated scattering cross section, $\mu_{\chi N} \equiv m_{\chi} \,m_N/(m_{\chi}+m_N)$ is the reduced mass of the dark matter-nucleus system, and $A$ and $Z$ are the atomic mass and charge of the nucleus, respectively. The coupling of the dark matter to the mediator is $\lambda_{\chi}=g_{X}/2$ for spin-0 dark matter and $\lambda_{\chi}=g_{\chi v}$ for the spin-$1/2$ dark matter. $g_{qv}$ is the vector coupling of a given quark species to the mediator. These couplings will be defined in terms of the Lagrangian parameters in Eqs.~\eqref{eq:L DM0 med1} and~\eqref{eq:L DM1/2 med1}.

Next, we will consider scalar interactions, which yield a spin-independent cross section of the following form:
\begin{align}
\bar{\sigma}_N = \frac{\mu_{\chi N}^2 \lambda^2_{\chi}}{\pi m_\phi^4} \bigg[Z \tilde{f}_{p} +(A-Z) \tilde{f}_n \bigg]^2,
\end{align}
where the couplings to protons and neutrons are given by
\begin{align}
\frac{\tilde{f}_{p,n}}{m_{p,n}} = \sum_{q=u,d,s}  f^{p,n}_{T_q} g_{qs} +\frac{2}{27} \, f_{TG} \sum_{q=c,b,t} g_{qs}.
\end{align}
The coupling of the dark matter to the mediator is $\lambda_{\chi}=\lambda_X/2$ for spin-0 dark matter and $\lambda_{\chi}=g_{\chi s}$ for the spin-$1/2$ dark matter. $g_{qs}$ is the scalar coupling of a given quark species to the mediator. These couplings will be defined in terms of the Lagrangian parameters in Eqs.~\eqref{eq:L DM0 med0} and~\eqref{eq:L DM1/2 med0}. The quantities $f^{p,n}_{T_q}$ are the matrix elements which characterize the light quark content of the proton and neutron (see Table~\ref{Table:direct}). The gluon content of the nucleon is related to these matrix elements by $f_{TG}=1-f^p_{T_u}-f^p_{T_d}-f^p_{T_s} = 1-f^n_{T_u}-f^n_{T_d}-f^n_{T_s}$. The first term in this expression accounts for scattering with light quarks in the target, while the second term arises from scattering with gluons through a loop of heavy quarks. 

\begin{table}[t]
    \centering
    \begin{tabularx}{\textwidth}{CCCCCCCC}
        \toprule
         & $f_{T_u}^{p,n}$ & $f^{p,n}_{T_d}$ & $f^{p,n}_{T_s}$ & $f^{p,n}_{TG}$ &  $\Delta^{p,n}_u$ & $\Delta^{p,n}_d$ & $\Delta^{p,n}_s$ \\\cmidrule{2-8}
         % \cmidrule{}\\
         Protons & 0.023 & 0.033& 0.26& 0.68 &0.84 & $-0.43$ & $-0.08$ \\
         Neutrons & 0.018 & 0.042& 0.26 & 0.68 & 0.84&$-0.43$ & $-0.08$ \\
         \bottomrule
    \end{tabularx}
\caption{Nuclear coefficients governing the dark matter-nucleon interaction strength~\cite{Belanger:2008sj,Fitzpatrick:2012ix,Cirelli:2013ufw}.}
\label{Table:direct}
\end{table}

Thirdly is the case of axial interactions, which yield the following spin-dependent cross section:
\begin{align}
\bar{\sigma}_N = \frac{4 \mu_{\chi N}^2 g_{\chi a}^2}{\pi m_V^4} J_N (J_N+1) \bigg(\frac{\langle S_p\rangle}{J_N} \, \tilde{a}_p  + \frac{\langle S_n\rangle}{J_N} \, \tilde{a}_n \bigg)^2,
\end{align}
where $\langle S_{p, n}\rangle$ is the expectation value of the spin content of the proton or neutron group in the nucleus and $J_N$ is the spin of the nucleus. As an example, for the case of $^{131}$Xe, $J_N=1/2$,  $\langle S_{p}\rangle =-0.005$, and $\langle S_{n}\rangle = -0.20$~\cite{Fitzpatrick:2012ix}. The effective proton and neutron couplings  are given by
\begin{align}
\tilde{a}_{p,n} = \sum_{q=u,d,s} g_{qa} \, \Delta^{p,n}_q,
\end{align}
where the axial couplings of the dark matter and quarks are defined in terms of the Lagrangian parameters in Eq.~\eqref{eq:L DM1/2 med1}. The values of the spin content of the nucleon, $\Delta^{p,n}_q$ are provided in Table~\ref{Table:direct}.

In many cases, the interactions of the dark matter with quarks lead to an elastic scattering cross section with nuclei that do not vanish in the $q \rightarrow 0, v \rightarrow 0$ limit. In such cases, we calculate the elastic scattering cross section by following the procedure outlined in Refs.~\cite{Fitzpatrick:2012ix,Cirelli:2013ufw,Kumar:2013iva}, as described earlier in this section. 

\subsection{Indirect Detection}
\label{sec:indirect}

If dark matter is a thermal relic, it could annihilate at an appreciable rate in the present-day universe. These annihilation reactions can produce potentially observable fluxes of gamma rays, neutrinos, or charged cosmic rays (including electrons, protons, nuclei, and their antiparticles). Efforts to observe the products of dark matter annihilation or decay are collectively known as indirect detection. Here, we focus on annihilation constraints based on energy injection into the cosmic microwave background, gamma-ray observations of dwarf spheroidal galaxies, and the positron content of the cosmic-ray spectrum. 

Any annihilation of dark matter particles during the era of recombination would have transferred energy to the baryonic matter of our universe, impacting the ionization history and consequently the observed characteristics of the cosmic microwave background. Measurements from the Planck collaboration allow us to place the following constraint~\cite{Planck:2018vyg}:
\begin{align}
\frac{\langle \sigma v \rangle \, f_{\rm eff}}{m_{\chi}} < 3.2 \times 10^{-28} \, {\rm cm}^3 \, {\rm s^{-1}} \, {\rm GeV^{-1}},
\end{align}
where $\langle \sigma v \rangle$ is the thermally-averaged dark matter annihilation cross section (Eq.~\eqref{eq:sigma_v_general}) evaluated at the time and temperature of recombination, and $f_{\rm eff}$ is the fraction of the energy injected in this process that is transferred to the intergalactic medium~\cite{Slatyer:2015jla}. Although $f_{\rm eff}$ can be as large as $\sim 0.4-0.6$ for electromagnetic annihilation channels, this quantity is typically somewhat smaller for annihilations to quarks, gauge bosons, or Higgs bosons, $f_{\rm eff} \sim 0.2$. For a thermal relic with a velocity-independent annihilation cross section, $\langle \sigma v \rangle \approx 2 \times 10^{-26} \, {\rm cm}^3/{\rm s}$, this measurement typically allows us to restrict $m_{\chi} > (10-15) \, {\rm GeV}$.

The dark matter halo of the Milky Way contains a large a number of subhalos, the largest of which contain stars. These so-called dwarf galaxies are highly dark matter dominated systems, making them attractive targets for indirect dark matter searches. 
The flux of gamma rays, $\phi_\gamma$, generated through the process of dark matter annihilation can be expressed as 
\begin{align}
    \label{eq:indirect flux}
    \frac{d \phi_{\gamma}}{dE_{\gamma}} =  \frac{\langle \sigma v \rangle}{8 \pi m_\chi^2} \frac{dN_{\gamma}}{dE_{\gamma}} \int_{\Delta \Omega} \dd\Omega\ \int_{\rm los} \dd l \, \rho_\chi^2(l,\Omega) \,  ,
\end{align}
where $dN_{\gamma}/dE_{\gamma}$ is the spectrum of gamma rays produced in a single dark matter annihilation reaction, and is often calculated using \verb|PYTHIA|~\cite{Sjostrand:2006za} or other event generating software. The integral of the square of the dark matter density over the observed solid angle and line-of-sight (los) is known as the $J$-factor. The $J$-factors of many Milky Way dwarf galaxies have been estimated from the radial profiles of their stellar velocity dispersions~\cite{Martinez:2013els,Bonnivard:2014kza,Geringer-Sameth:2014yza}. This information, combined with observations of these systems using gamma-ray telescopes, allows us to place upper limits on the dark matter annihilation cross section.

Observations of Milky Way dwarf galaxies with the Fermi Gamma-Ray Space Telescope have been used to place constraints on the dark matter annihilation cross section. For a thermal relic with a velocity-independent annihilation cross section, $\langle \sigma v \rangle \approx 2 \times 10^{-26} \, {\rm cm}^3/{\rm s}$, the authors of Ref.~\cite{McDaniel:2023bju} found that this data could restrict $m_{\chi} > 60 \, {\rm GeV}$ for annihilation to $b\bar{b}$ (other hadronic final states yield similar constraints) and $m_{\chi} > 45 \, {\rm GeV}$ for annihilation to $\tau^+ \tau^-$. In that study, each dwarf galaxy was treated as a point source, and these constraints are weakened somewhat when more realistic extended profiles are adopted. For $\langle \sigma v \rangle \approx 2 \times 10^{-26} \, {\rm cm}^3/{\rm s}$, Ref.~\cite{DiMauro:2022hue} obtains constraints of $m_{\chi} > 40 \, {\rm GeV}$ for annihilation to $b\bar{b}$ and $m_{\chi} > 30 \, {\rm GeV}$ for annihilation to $\tau^+ \tau^-$. We also note that these studies each identified a small excess of gamma-ray emission from this sample of dwarf galaxies (at the $\sim 3\sigma$ level), favoring dark matter parameters that can consistently account for the long-standing Galactic Center Gamma-Ray Excess~\cite{Goodenough:2009gk,Hooper:2010mq,Hooper:2011ti,Daylan:2014rsa,Calore:2014xka,Fermi-LAT:2017opo}.

In most models, dark matter annihilations produce equal amounts of matter and antimatter. Dark matter particles annihilating in the local volume of the Galactic Halo can thus contribute to the  antimatter component of the cosmic ray spectrum, providing us with a signal to search for with space-based cosmic-ray detectors. Studies of the cosmic-ray positron spectrum, as measured by AMS onboard the International Space Station, have yielded constraints on the dark matter annihilation cross section to leptonic final states. In particular, if the dark matter is a thermal relic with a velocity-independent annihilation cross section, $\langle \sigma v \rangle \approx 2 \times 10^{-26} \, {\rm cm}^3/{\rm s}$, such considerations rule out masses below 300 GeV, 150 GeV, and 50 GeV for annihilations to $e^+e^-$, $\mu^+ \mu^-$, and $\tau^+ \tau^-$, respectively~\cite{Bergstrom:2013jra,John:2021ugy}. In the case of dark matter annihilating to quarks, constraints can also be derived from the cosmic-ray antiproton spectrum~\cite{Cholis:2019ejx,Calore:2022stf}.

\subsection{Accelerator Searches}
\label{sec:collider}

Mediators can be produced at accelerator experiments, including LHC, LEP, and fixed-target experiments.
There are multiple LHC searches for particles that mediate interactions between dark and visible matter \cite{Boveia:2018yeb}. Such particles can be produced on-shell at hadron colliders and then decay either to SM fermions, yielding a visible final state, or to dark matter particles, appearing as missing energy (MET). Typically, these results place an upper limit on the quantity $\sigma(pp \to \phi+X) \mathrm{BR}(\phi \to  F)$, where $\phi$ denotes the mediator,  $F$ denotes a given final state (dark or visible), and $X$ is an associated product tagged in the search. The rate of such events scales as
\begin{align}
    \sigma \times \mathrm{BR} \propto g_{\rm SM}^2 \times \frac{   \Gamma_{\rm F}}{  \Gamma_{\rm SM} + \Gamma_{\rm DM}}\,,
\end{align}
where $g_{\rm SM}$ is the mediator's coupling to the relevant SM state(s), and $\Gamma_{\rm F}$ is the mediator's decay rate to the final state, $F$. The mediator to SM and DM decay widths, $\Gamma_{\rm SM }$, $\Gamma_{\rm DM}$, are discussed in detail in Sec.~\ref{subsec:decay_rate_to_SM} and Sec.~\ref{sec:dm_models}, respectively. The mediators can also be produced off-shell, contributing to the visible final states.
The production cross section thus only depends on the coupling to the SM fermions, $\sigma \propto g_\text{SM}^2$, and is independent of the branching ratio.
Collider searches typically focus on mediators that couple to $b$-quarks, $t$-quarks, or light flavors including $u$, $d$, $c$, and $s$.
We refer the reader to Ref.~\cite{ATLAS:2024kpy} for a comprehensive review, while summarizing the typical search strategies and results here. Our results are summarized below according to the quark flavor to which the mediator predominantly couples.

{\bf Light Flavors ($u$, $d$, $c$, $s$):} Searches are typically performed for a vector boson assuming a democratic coupling to all light flavors~\cite{ATLAS:2018hbc,ATLAS:2024qqm,CMS:2025ybc,ATLAS:2019fgd,ATLAS:2021kxv}.
Here, we reinterpret these search results by rescaling the PDF factor for each quark flavor using \verb|ManeParse|~\cite{Clark:2016jgm}.
The dominant production channel is quark–antiquark annihilation. In the limit of negligible light-quark masses, the production rate is similar for scalar and vector mediators, so the qualitative behavior does not depend on the mediator spin.
Refs.~\cite{ATLAS:2018hbc,ATLAS:2024qqm,CMS:2025ybc,ATLAS:2019fgd} perform LHC searches for dijet resonances between 100 GeV to 5 TeV.
Ref.~\cite{ATLAS:2021kxv} uses LHC data to search for invisibly decaying particles with masses in the 10 GeV - 1 TeV range, which is translated into limits on the mass and coupling of spin-0 mediators in Ref.~\cite{ATLAS:2024kpy}.
In our analysis, we interpret these results and apply them to the models explored in this work.

{\bf Bottom quarks:}
The mediator is produced at tree level through its coupling to the bottom quark, with the spin-0 and spin-1 mediators sharing the same production diagrams.
Searches have been performed in the invisible decay channel for $10~\mathrm{GeV} \leq m_\phi \leq 500~\mathrm{GeV}$~\cite{ATLAS:2021yij},
and in the visible decay channel for $100~\gev < m_\phi < 1300~\gev$~\cite{CDF:2019jwx,CMS:2018hir}.
To interpret these results, we calculate the production cross section, $\sigma(pp\to b \phi +X)$, using \verb|Madgraph-v3.5.3| with the default PDFs~\cite{Alwall:2014hca}.

{\bf Top quarks:}
Scalar and pseudoscalar mediators coupled to top quarks can be produced at the LHC through a tree-level top quark coupling and through a loop-induced coupling to gluons.
These particles have been searched for in invisible decays, in off-shell $\bar{t}t$ production, and in a $\bar{t}t$ resonance~\cite{ATLAS:2019wdu,ATLAS:2024rcx,Esser:2023fdo,ATLAS:2024kpy,ATLAS:2020lks,ATLAS:2022abz,ATLAS:2024bjr,CMS:2024yhz,ATLAS:2021uiz}.
Diphoton resonances also constrain the mediator-top quark coupling.
We again run \verb|Madgraph-v3.5.3| to calculate required cross sections with the model file provided in Ref.~\cite{ALPUFO}.
We find that the production cross sections for scalar and pseudoscalar couplings are nearly identical and apply the constraint for both cases when only one of these two is available, calculating the branching ratio separately.

Together, we apply the following constraints on spin-0 mediators coupled to top quarks:
\begin{itemize}
    \item \textbf{$\bar{t}t$+MET:} The mediator is produced resonantly associated with a pair of top quarks, then decays invisibly. Ref.~\cite{Esser:2023fdo} provides a constraint on the pseudoscalar coupling to top quarks assuming invisible decay for 10 MeV to 2 TeV, while ATLAS data~\cite{ATLAS:2019wdu,ATLAS:2024rcx} provides a stronger constraint between 10 GeV and 500 GeV. We combine these results and apply the strongest limit for the covered mass range. The spin-1 mediator has qualitatively similar production diagrams, so the results for this case can be directly translated from the spin-0 analysis.
    \item \textbf{$\bar{t}t$ production:} Ref.~\cite{Esser:2023fdo} provides a constraint on the pseudoscalar coupling to the top quark for pseudoscalar masses from 10 MeV to 2 TeV based on ATLAS data.
    This includes both non-resonant production and resonant production, with the former dominating the low-mass range.
    We use $m_\phi \simeq 700~\rm GeV$ as the critical value to separate non-resonant and resonant production, where the shape of the constraint curve in Ref.~\cite{Esser:2023fdo} begins to change significantly.
    The constraint on this coupling can be directly matched onto our coupling to top quarks for non-resonant production, while the resonant production case must be translated according to the branching ratio. Ref.~\cite{ATLAS:2024kpy} interprets the $\bar{t}t$-resonance search to the case of an axial-vector mediator in Ref.~\cite{ATLAS:2020lks}, and
 we directly adopt the results of this analysis. As above, the limits on the vector mediator can be derived from the corresponding limits on the spin-0 scenario.
    \item \textbf{Diphoton resonances:} Ref.~\cite{ATLAS:2022abz} performed a search for diphoton resonances between 10 and 70 GeV, while Refs.~\cite{ATLAS:2024bjr,CMS:2024yhz} provide two independent searches for masses between 70 and 110 GeV. Ref.~\cite{ATLAS:2021uiz} searched for resonances between 200 GeV and 3 TeV. For our analysis, we use \verb|MadGraph| to calculate the LHC production cross section and \verb|Delphes|~\cite{deFavereau:2013fsa} to estimate the selection efficiency of the analysis to extract an upper limit on the fiducial cross section. Since a spin-1 mediator cannot decay into a pair of photons, diphoton resonance limits do not apply to this scenario.
\end{itemize}

In addition to LHC search, light mediators with $m_\phi < m_Z$ can also be produced on-shell in $Z \to \bar q q \phi$ decays and contribute to the hadronic width of the $Z$ boson. From LEP data at the $Z$-pole, the inclusive hadronic width is measured to be $\Gamma(Z \to \rm hadrons) = 1744.4 \pm 2.0\;\text{MeV}$~\cite{ParticleDataGroup:2024cfk}, so in our analysis we demand that corrections from radiative mediator emission satisfy the requirement, $\Gamma(Z \to \bar q q \phi) <  4.0$ MeV, independently of whether the mediator decays to quarks or to dark matter in the final state. Note that the hadronic $Z$-width is an inclusive observable, so adding missing energy to the final state will also contribute as long as there are hadrons from the quarks, which are always produced in these 
decays. Here we calculate the $Z \to \bar q q \phi$ branching fraction following the procedure outlined in Ref.~\cite{Cesarotti:2024rbh} adapted to quark final states. 

Data from fixed-target accelerator experiments have also been used to place constraints on sub-GeV dark matter and mediator particles~\cite{Krnjaic:2022ozp}. For these probes, a proton ~\cite{Batell:2009di,deNiverville:2018dbu,Jordan:2018gcd}, electron~\cite{Batell:2014mga,Izaguirre:2013uxa,Izaguirre:2014bca,BDX:2016akw,LDMX:2025pkp}, or muon beam~\cite{NA64:2024klw,Kahn:2018cqs,Krnjaic:2024ols} impinges onto a fixed target and produces new particles which are detected through their impact on either the recoiling particle beam, or on their energy injection in a downstream detector through scattering or decay. Most of the existing bounds on such scenarios assume specific mediator particles (e.g. kinetically mixed dark photons or leptophilic gauge bosons), which do not easily generalize to the scenarios that we consider here. Thus, although these constraints could potentially apply to some of the low-mass parameter space that we will consider in this study, 
extracting the relevant bounds would typically require a dedicated analysis, which we leave for future work.
However, the vector boson mediator always acquires a mixing with the SM photon through a loop of SM particles.
All dark photon searches can thus apply to the vector boson mediator scenario, and those bounds are adopted in this work.

\subsection{Rare Meson Decays}
\label{sec:meson}

Sub-GeV mediators and dark matter particles can be produced in rare meson decays, whose branching fractions are strongly constrained by experimental searches. 
Here we consider the constraints on spin-0 mediators, while the vector boson mediator is more strongly constrained by the loop-induced mixing with the photon, as discussed in Sec.~\ref{sec:collider}.
The axial-vector mediator, on the other hand, is UV sensitive and is currently less investigated.
We thus omit this constraint in the case of the axial-vector coupling.

We focus here on the limits derived from the most commonly studied rare $B$ and $K^+$ decays.
If the mediator decays invisibly, the branching ratio of $K$ and $B$ mesons are constrained through the following~\cite{BaBar:2013npw,NA62:2021zjw}:
\begin{align}
    \mathrm{BR}(B \to K + \mathrm{inv})& \lesssim 3.2 \times 10^{-5},~\,\,\,\,\,\,\,\,\,\, m_\phi \lesssim 5~\rm GeV, \\
    \mathrm{BR}(K^+ \to \pi^+ + \mathrm{inv})& \lesssim 10^{-11}, \,\,\,\,\,\,\,\,\,\, m_\phi \lesssim 110~\mathrm{MeV} \text{ and }m_\phi \in [160,260]~\rm MeV.
\end{align}
If the mediator instead decays into two photons, it contributes to $K^+ \to \pi^+ \gamma\gamma$.
This channel is constrained both by the measured $K^+ \to \pi^+  \gamma\gamma$ rate~\cite{NA62:2023olg} and by the inclusive search for $K^+ \to \pi^+  \phi$ decays, which is sensitive to all possible $\phi$ decay modes~\cite{Yamazaki:1984vg}:
\begin{align}
    \mathrm{BR}(K^+ \to \pi^+ + \phi \to \pi^+ \gamma \gamma)& \lesssim 10^{-9}, \,\,\,\,\,\,\,\,\,\, m_\phi \in [200,350]~\rm MeV,\\
    \mathrm{BR}(K^+ \to \pi^+ + X)& \lesssim 10^{-6}, \,\,\,\,\,\,\,\,\,\, m_\phi \lesssim 110~\rm MeV.
\end{align}
Decays where the mediator subsequently produces a pair of quarks or contributes to $B \to K  \gamma\gamma$ are not currently constrained.
Following Ref.~\cite{Dolan:2014ska}, we additionally impose an inclusive bound, $\mathrm{BR}(B \to K  \phi) \lesssim 10^{-2}$, to avoid large corrections to the total $B$-meson width.

To apply these limits, we compute the mediator branching ratios into DM or $\gamma\gamma$ and select the appropriate experimental constraint.
For couplings that reproduce the correct DM relic abundance, the mediator decays promptly in all cases except when it couples only to the top quark.
In that scenario, the mediator may decay outside of the BaBar detector, where the limit of $B \to K + \rm inv$ is measured.
We therefore treat $\phi \to \gamma \gamma$ as invisible if the decay length exceeds 1 meter, which is roughly the dimension of the BaBar detector.

For the convenience of the reader, we define some useful quantities here before introducing the decay rates.
The masses of $B$ meson, $K$ meson, and pion are $m_B = 5.28~\rm GeV$, $m_K = 494~\rm MeV$, and $m_\pi \simeq 135~\rm MeV$, respectively.
The Kaon decay constant is $F_K \simeq 113~\rm MeV$.
The Higgs vacuum expectation value is $v = 246.22~\rm GeV$.
We use $V_{ij}$ to label CKM matrix elements.
We define the kinematic factor $\lambda(a,b,c) \equiv a^2 + b^2 + c^2 - 2ab - 2bc - 2ac$.
Specifically, the nuclear form factor for $B$-meson is given by~\cite{Ball:2004ye}
\begin{align}
    f_0(q^2) = \frac{0.33}{1-  q^2/(38~\rm GeV^2)}\,.
\end{align}
Calculations involving light quarks typically rely on ChPT, as described in Sec.~\ref{sec:lowE} and App.~\ref{sec:chpt app}.

\begin{itemize}

\item {\bf \boldmath Up quarks:} For the scalar coupling, the dominant constraints come from the tree-level contribution to $K^+$ decay.
The decay rate is~\cite{Batell:2018fqo}
\begin{align}
    \Gamma_{K\to \pi\phi} = \frac{g_{fs}^2 F_\pi^2 F_K^2 B_0^2}{8\pi m_{K} v^4}|V_{ud}^* V_{us}|^2 \lambda^{1/2}\left(1, \frac{m_\pi^2}{m_{K}^2}, \frac{m_\phi^2}{m_{K}^2}\right)\,.
\end{align}

For the pseudoscalar coupling, the process is the same as in the scalar case.
The decay rate is~\cite{Guerrera:2021yss}
\begin{align}
    \Gamma_{K \to \pi \phi} = \frac{g_{fp}^2 F_\pi^2 F_K^2 m_K^3}{64 \pi m_u^2 v^4} |V_{us}^* V_{ud}|^2 \lambda^{1/2}\left(1, \frac{m_\pi^2}{m_{K}^2}, \frac{m_\phi^2}{m_{K}^2}\right)\,.
\end{align}

\item{\bf \boldmath Down quarks:} For the scalar coupling, the dominant constraints come from the tree-level contribution to $K^+$ decay.
The decay rate is
\begin{align}
    \Gamma_{K \to \pi \phi} = \frac{g_{fs}^2 F_\pi^2 F_K^2 B_0^2}{8\pi m_{K} v^4}|V_{ud}^* V_{us}|^2 \left(\frac{m_K^2}{m_K^2 - m_\pi^2}\right)^2 \lambda^{1/2}\left(1, \frac{m_{\pi}^2}{m_{K}^2}, \frac{m_\phi^2}{m_{K}^2}\right) \,.
\end{align}

For the pseudoscalar coupling, the process is the same as in the scalar case.
The decay rate is~\cite{Guerrera:2021yss}
\begin{align}
    \Gamma_{K \to \pi \phi} = \frac{g_{fp}^2 F_\pi^2 F_K^2 m_\pi^5}{\pi m_d^2 m_K^2 v^4}  |V_{us}^* V_{ud}|^2 \lambda^{1/2}\left(1, \frac{m_\pi^2}{m_{K}^2}, \frac{m_\phi^2}{m_{K}^2}\right)\,.
\end{align}

\item{\bf \boldmath Strange quarks:} For the scalar coupling, the dominant constraints come from the tree-level contribution to $K^+$ decay.
The decay rate is
\begin{align}
    \Gamma_{K \to \pi \phi} = \frac{g_{fs}^2 F_\pi^2 F_K^2 B_0^2}{8\pi m_{K} v^4}|V_{ud}^* V_{us}|^2 \left(\frac{m_\pi^2}{m_K^2 - m_\pi^2}\right)^2 \lambda^{1/2}\left(1, \frac{m_{\pi}^2}{m_{K}^2}, \frac{m_\phi^2}{m_{K}^2}\right)\,.
\end{align}

For the pseudoscalar coupling, the process is the same as in the scalar case.
The decay rate is
\begin{align}
    \Gamma_{K \to \pi \phi} = \frac{g_{fp}^2 F_\pi^2 F_K^2 m_K^3}{64 \pi m_s^2 v^4} |V_{us}^* V_{ud}|^2 \lambda^{1/2}\left(1, \frac{m_\pi^2}{m_{K}^2}, \frac{m_\phi^2}{m_{K}^2}\right)\,.
\end{align}

\item {\bf \boldmath Charm quarks:} A charm-coupled mediator can be emitted from tree-level diagrams involving charm mesons in $B_c^+$ or $D^+$ decays.
The corresponding decay rates have not been thoroughly investigated in the literature, however, so we leave this case for future work, while presenting the $B$ meson decay results involving the loop-level process.

For the scalar coupling, the induced $B$ decay arises from a loop diagram; a bottom quark is converted into a strange quark with a $W$-boson and an up-type quark running in the loop.
A charm-coupled mediator can then be emitted from the charm quark inside the loop.
The corresponding decay rate is calculated as~\cite{Knapen:2017xzo}
\begin{align}
    \Gamma_{B\to K \phi} &= \frac{9 g_{fs}^2 m_b^2 m_c^2 |V_{cs}^*V_{cb}|^2}{2048 \pi^5 m_B v^4} f_0(m_\phi^2)^2 \left(\frac{m_B^2 - m_K^2}{m_b - m_s}\right)^2 \lambda^{1/2}\left(1, \frac{m_K^2}{m_B^2}, \frac{m_\phi^2}{m_B^2}\right)\,.
\end{align}

For the pseudoscalar coupling, the decay is driven by the same diagram~\cite{Dolan:2014ska},
\begin{align}
    \Gamma_{B \to K \phi}   &= \frac{g_{fp}^2 m_c^2 (m_b - m_s)^2  |V_{cs}^*V_{cb}|^2}{4096 \pi^5 m_B v^4}f_0(m_\phi^2)^2 \left(\frac{m_B^2 - m_K^2}{m_b - m_s}\right)^2 \lambda^{1/2}\left(1, \frac{m_K^2}{m_B^2}, \frac{m_\phi^2}{m_B^2}\right)\,.
\end{align}

In both cases, one can derive the $K$ decay rate by replacing $B \to K$, $K \to \pi$, $b \to s$, and $s \to d$. This yields constraints on $m_\phi \lesssim 260~\rm MeV$.

\item {\bf Bottom quarks:} The dominant constraint comes from $B$ meson decay.
For the scalar coupling, the decay process arises at one loop, with the scalar emitted from the external bottom quark line that extends outside the loop. 
The decay rate for this process is~\cite{Dolan:2014ska}%\TT{What is $f_0$? If it's the form factor $f_0 \rightarrow f$}
\begin{align}
    \Gamma_{B\to K \phi} &= \sum_{q=u,c,t}\frac{g_{fs}^2 m_q^4 |V_{qs}^* V_{qb}|^2}{8192 \pi^5 m_B v^4}f_0(m_\phi^2)^2 \left(\frac{m_B^2 - m_K^2}{m_b - m_s}\right)^2 \lambda^{1/2}\left(1, \frac{m_K^2}{m_B^2}, \frac{m_\phi^2}{m_B^2}\right)\,.
\end{align}

For the pseudoscalar coupling, the decay arises from the same diagram topology and the following width~\cite{Dolan:2014ska},
\begin{align}
    \Gamma_{B\to K \phi} &= \sum_{q=u,c,t}\frac{9g_{fp}^2m_q^4 (m_b^2 + m_s^2)^2 |V_{qs}^* V_{qb}|^2}{8192 \pi^5 m_B v^4 (m_b^2 - m_s^2)^2}f_0(m_\phi^2)^2 \left(\frac{m_B^2 - m_K^2}{m_b - m_s}\right)^2 \lambda^{1/2}\left(1, \frac{m_K^2}{m_B^2}, \frac{m_\phi^2}{m_B^2}\right)\,.
\end{align}

Here, the summation runs over all the up-type quarks inside the loop.
In both cases, the top-loop contribution dominates.

\item {\bf Top quarks:} For the scalar coupling, $B$ decay is induced through a loop diagram with the mediator emitted from the top quark propagator inside the loop, similar to the case of couplings to the charm quark.
The decay rate satisfies~\cite{Knapen:2017xzo}
\begin{align}
    \Gamma_{B\to K \phi} &= \frac{9 g_{fs}^2 m_b^2 m_t^2 |V_{ts}^*V_{tb}|^2}{2048 \pi^5 m_B v^4} f_0(m_\phi^2)^2 \left(\frac{m_B^2 - m_K^2}{m_b - m_s}\right)^2 \lambda^{1/2}\left(1, \frac{m_K^2}{m_B^2}, \frac{m_\phi^2}{m_B^2}\right)\,.
\end{align}

For the pseudoscalar coupling, the decay arises from a similar diagram topology~\cite{Dolan:2014ska},
\begin{align}
    \Gamma_{B \to K \phi}   &= \frac{g_{fp}^2 m_t^2 (m_b - m_s)^2  |V_{ts}^*V_{tb}|^2}{4096 \pi^5 m_B v^4}f_0(m_\phi^2)^2 \left(\frac{m_B^2 - m_K^2}{m_b - m_s}\right)^2 \lambda^{1/2}\left(1, \frac{m_K^2}{m_B^2}, \frac{m_\phi^2}{m_B^2}\right)\,.
\end{align}

In both cases, one can derive the $K$ decay rate by replacing $B \to K$, $K \to \pi$, $b \to s$, and $s \to d$. This yields constraints on $m_\phi \lesssim 260~\rm MeV$.

\end{itemize}

\section{Dark Matter Models}
\label{sec:dm_models}
\subsection{Spin-0 Dark Matter, Spin-0 Mediator}
\label{sec:Dm0Med0}

A spin-0 mediator, $\phi$, couples with a spin-0 dark matter candidate, $X$, and SM fermions with the following Lagrangian:
\begin{align}
    \label{eq:L DM0 med0}
    \mathcal{L} \supset a \lambda_X X^\star X \, \phi + \phi \, \left( g_{fs} \bar{f} f + i g_{f p} \bar{f} \gamma^5 f \right)\,,
\end{align}
where $\lambda_X \equiv m_X g_X$ is a dimension-one coupling and $a$ is a symmetry factor which equals 1 for complex scalar dark matter and 1/2 for a real scalar.
This symmetry factor ensures the same Feynman rule for real and complex scalars.
The number of internal degrees-of-freedom in this case is $g_{\rm spin} = 1$.

If $m_\phi > 2 m_X$, the mediator can decay into a pair of dark matter particles with the following width,
\begin{align}
    \Gamma_{\phi \rightarrow X^* X} &= a \frac{\lambda_X^2}{16 \pi m_\phi} \sqrt{1 - \frac{4 m_X^2}{m_\phi^2}}\,.
\end{align}
The corresponding decay rates to SM particles are given in Sec.~\ref{subsec:decay_rate_to_SM}.

The squared amplitudes for dark matter annihilation to fermions, photons, gluons, and mediators are given by
\begin{align}
    |\mathscr{M}|^2_{f\bar{f}} &= \frac{N_c \lambda_X^2}{(s - m_\phi^2)^2 + m_\phi^2 \Gamma_\phi^2}\bigg( \, g_{f s}^2\left[ \, 2 \left( s - 4 m_f^2 \right) \, \right] + g_{f p}^2 \left[\, 2s \,\right] \, \bigg)\, , \\
    |\mathscr{M}|_{\gamma\gamma}^2 & = \frac{ \lambda_X^2}{(s - m_\phi^2)^2 + m_\phi^2 \Gamma_\phi^2} \left( \frac{N_c Q_f^2 \alpha}{ 3 \pi} \frac{s}{m_f} \right)^2 \Bigg( \, g_{f s}^2 +  \frac{9}{4} g_{f p}^2 \, \Bigg)\,,   \\
    |\mathscr{M}|_{gg}^2           & = \frac{\lambda_X^2 (N_c^2 - 1)}{(s - m_\phi^2)^2 + m_\phi^2 \Gamma_\phi^2} \left( \frac{\alpha_s}{6 \pi} \frac{s}{m_f} \right)^2  \Bigg( \, g_{f s}^2 + 9 \, g_{f p}^2  \, \Bigg)\,, \\
    |\mathscr{M}|_{\gamma\gamma,\rm ChPT}^2 & = \frac{\lambda_X^2}{(s - m_\phi^2)^2 + m_\phi^2 \Gamma_\phi^2} \left(\frac{\alpha B_0}{24 \pi} \frac{s}{m_\pi^2}\right)^2  \Bigg( \, g_{f s}^2+ 144 g_{f p}^2\Bigg)\,,
    \end{align}
    \begin{align}
    |\mathscr{M}|^2_{\phi \phi} &= \lambda _X^4 \left(\frac{1}{m_X^2 \left(s-4 m_\phi ^2\right)+m_\phi^4}+\frac{4 \tanh^{-1} \left(\sqrt{\left(s-4 m_X^2\right) \left(s-4 m_\phi^2\right)}/(s-2 m_\phi^2)\right)}{\left(s-2 m_\phi^2\right) \sqrt{\left(s-4 m_X^2\right) \left(s-4 m_\phi^2\right)}}\right)\,, 
        \end{align}
 where $\Gamma_\phi$ is  mediator's total width. The subscript ``ChPT'' denotes the squared amplitude resulting from a charged pion loop, appropriate for energy scales below $\Lambda_{\rm QCD}$, as discussed in Sec.~\ref{sec:lowE}.

We can calculate the dark matter annihilation cross section using Eq.~\eqref{eq:sigmav NR non-avg}.
We take the non-relativistic limit and expand $(\sigma v)_\text{NR}$ in powers of relative velocity.
For simplicity, we also take the limit of $m_f \to 0$.
The amplitudes given above yield the following results:
\begin{align}
    (\sigma v)_{f\bar{f}} &\approx \frac{N_c \lambda _X^2}{128 \pi (m_\phi^2 - 4m_X^2)^2} \left(g_{fs}^2+g_{fp}^2\right)\left[32+ \left(\frac{20m_X^2 + 11m_\phi^2}{m_\phi^2 - 4m_X^2}\right) v^2\right]\,, \\
    (\sigma v)_{\gamma \gamma} &\approx \frac{\lambda _X^2}{64 \pi} \frac{m_X^2}{(m_\phi^2 - 4m_X^2)^2}\left(\frac{N_c Q_f^2 \alpha}{3 \pi m_f}\right)^2 \left(g_{fs}^2  + \frac{9}{4} g_{fp}^2\right) \left(32+ \left(\frac{19m_\phi^2 - 12 m_X^2}{m_\phi^2 - 4m_X^2}\right) v^2\right)\,, \\
    (\sigma v)_{gg} &\approx \frac{\lambda _X^2 (N_c^2 - 1)}{64 \pi} \frac{m_X^2}{(m_\phi^2 - 4m_X^2)^2} \left(\frac{\alpha_s}{6 \pi m_f}\right)^2 \left(g_{fs}^2 + 9 g_{fp}^2\right) \left(32+ \left(\frac{19m_\phi^2 - 12 m_X^2}{m_\phi^2 - 4m_X^2}\right) v^2\right)\,, \\
    (\sigma v)_{\gamma \gamma, \rm ChPT} &\approx \frac{\lambda _X^2}{64 \pi} \frac{m_X^2}{(m_\phi^2 - 4m_X^2)^2} \left(\frac{\alpha B_0}{24 \pi m_\pi^2}\right)^2\left(g_{fs}^2  + 144 g_{fp}^2\right) \left(32+ \left(\frac{19m_\phi^2 - 12 m_X^2}{m_\phi^2 - 4m_X^2}\right) v^2\right)\,, \\
    (\sigma v)_{\phi \phi} &\approx \frac{\lambda_X^4 \sqrt{1-\frac{m_\phi^2}{m_X^2}}}{16 \pi m_X^2 \left( m_\phi^2-2 m_X^2\right)^2} \left(1 - \frac{92 m_X^6 - 136 m_X^4 m_\phi^2 + 35 m_X^2 m_\phi^4 - 3 m_\phi^6}{96 \left(m_X^2 - m_\phi^2\right) \left(m_\phi^2 - 2m_X^2\right)^2} v^2\right) \,.
\end{align}

\subsection{Spin-1/2 Dark Matter, Spin-0 Mediator}
\label{sec:Dm1/2Med0}

A spin-0 mediator, $\phi$, couples with a spin-1/2 dark matter candidate, $\chi$, and SM fermions with the following Lagrangian:
\begin{align}
    \label{eq:L DM1/2 med0}
    \mathcal{L} \supset a \left( g_{\chi s} \bar{\chi} \chi + i g_{\chi p} \bar{\chi} \gamma^5 \chi \right) \phi + \phi \left( g_{f s} \bar{f} f + i g_{f p} \bar{f} \gamma^5  f\right),
\end{align}
where $a$ is a symmetry factor which is equal to 1 for dark matter in the form of a Dirac fermion and 1/2 for a Majorana fermion.
This symmetry factor ensures the same Feynman rule for Dirac and Majorana fermions.
The number of internal degrees-of-freedom in this case is $g_{\rm spin} = 2$.

If $m_\phi > 2 m_\chi$, the mediator $\phi$ can decay into a pair of dark matter particles with the following width,
\begin{align}
    \Gamma_{\phi \rightarrow \bar{\chi} \chi}(m_\phi) & = a\frac{m_\phi}{8 \pi} \sqrt{1 - \frac{4 m_\chi^2}{m_\phi^2}} \left( \, g_{\chi s}^2 \left( 1 - \frac{4 m_\chi^2}{m_\phi^2} \right) + g_{\chi p}^2  \, \right)\,.
\end{align}
The corresponding decay rates to SM particles are given in Sec.~\ref{subsec:decay_rate_to_SM}.

The squared amplitudes for dark matter annihilation to fermions, photons, gluons, and mediators are given by
\begin{align}
    |\mathscr{M}|^2_{f\bar{f}} = \frac{N_c}{(s - m_\phi^2)^2 + m_\phi^2 \Gamma_\phi^2}   \Bigg( \, & g_{\chi s}^2 \, g_{f s}^2 \left[ \, 4\, s^2 \left(1 - \frac{4 m_f^2}{s} \right) \left( 1 - \frac{4 m_\chi^2}{s} \right) \, \right] 
    +  g_{\chi p}^2 \, g_{f p}^2 \left[ \, 4\,s^2 \, \right] \nonumber\\
    + \,                                                                                         & g_{\chi s}^2 \, g_{f p}^2 \left[ \, 4 \,s^2 \left(1 - \frac{4 m_\chi^2}{s}\right) \, \right] 
    + g_{\chi p}^2 \, g_{f s}^2 \left[ \,  4 \, s^2 \left( 1 - \frac{4 m_f^2}{s} \right) \, \right] \, \Bigg)\,,
\end{align}
\begin{align}
    |\mathscr{M}|^2_{\gamma\gamma} = \frac{2N_c^2}{(s - m_\phi^2)^2 + m_\phi^2 \Gamma_\phi^2} \left( \frac{Q_f^2 \alpha}{ 3 \pi m_f} \right)^2   \Bigg( \, & \,  g_{\chi s}^2 \, \left( g_{f s}^2 + \frac{9}{4} g_{fp}^2 \right) s^3 \left(1 - \frac{4 m_\chi^2}{s} \right)  
    + g_{\chi p}^2 \, \left( g_{f s}^2 + \frac{9}{4} g_{f p}^2 \right) s^3 \Bigg)\,,
\end{align}
\begin{align}
    |\mathscr{M}|_{gg}^2 = \frac{2 (N_c^2 -1) }{(s - m_\phi^2)^2 + m_\phi^2 \Gamma_\phi^2} \left( \frac{\alpha_s}{ 6 \pi m_f} \right)^2 \Bigg( \, &\, g_{\chi s}^2 \, \left( g_{f s}^2 + 9 g_{fp}^2 \right) s^3 \left(1 - \frac{4 m_\chi^2}{s} \right) + \, g_{\chi p}^2 \, \left( g_{f s}^2 + 9 g_{fp}^2 \right)  s^3  \Bigg)\,,
\end{align}
\begin{align}
    |\mathscr{M}|^2_{\gamma\gamma,\rm ChPT} = \frac{2}{(s - m_\phi^2)^2 + m_\phi^2 \Gamma_\phi^2} \left( \frac{\alpha B_0}{24 \pi m_\pi^2} \right)^2 \Bigg( \, & g_{\chi s}^2 \left( g_{f s}^2 + 144 g_{fp}^2 \right)  s^3 \left(1 - \frac{4 m_\chi^2}{s} \right) + g_{\chi p}^2 \, \left( g_{f s}^2 + 144 g_{fp}^2 \right) \, s^3\, \Bigg)\,,
\end{align}
\begin{align}
    |\mathscr{M}|^2_{\phi \phi} = \, & g_{\chi s}^4 \bigg[-4-\frac{2 (4m_\chi^2 - m_\phi^2)^2}{m_\chi^2(s-4m_\phi^2) + m_\phi^4} + 4A \bigg(s^2 + 16 s m_{\chi }^2-4 m_{\phi }^2 \left(s+4 m_{\chi }^2\right)-32 m_{\chi }^4+6 m_{\phi }^4\bigg)\bigg] \nonumber\\
    + & g_{\chi s}^2 g_{\chi p}^2 \bigg[-8 + 4 \frac{(4m_\chi^2 - m_\phi^2) m_\phi^2}{m_\chi^2(s-4m_\phi^2) + m_\phi^4} +8A \bigg(s(s+8m_\chi^2) - 4 m_\phi^2 (s+ 2m_\chi^2) + 6 m_\phi^4\bigg)\bigg] \nonumber\\
    + & g_{\chi p}^4 \bigg[-4-\frac{2 m_\phi^4}{m_\chi^2(s-4m_\phi^2) + m_\phi^4} + 4 A \bigg(s^2 - 4sm_\phi^2 + 6 m_\phi^4\bigg)\bigg]
\end{align}
where we have defined,
\begin{align}
    A & \equiv \frac{\tanh^{-1}\left(\sqrt{\left(s-4 m_\chi^2\right) \left(s-4 m_\phi^2\right)}/(s-2 m_\phi^2)\right)}{\left(s-2 m_{\phi }^2\right) \sqrt{\left(s-4 m_{\chi }^2\right) \left(s-4 m_{\phi }^2\right)}} \, .
\end{align}
The subscript ``ChPT'' denotes the squared amplitude resulting from a charged pion loop, as appropriate at energy scales below $\Lambda_{\rm QCD}$.

We can calculate the dark matter annihilation cross section, following Eq.~\eqref{eq:sigmav NR non-avg}.
We take the non-relativistic limit and expand $\sigma v$ in powers of velocity.
For simplicity, we also take the limit of $m_f \to 0$.
The amplitudes given above yield the following results:
\begin{align}
    (\sigma v )_{f\bar{f}} \approx \frac{N_c}{256 \pi} \frac{m_\chi^2}{(m_\phi^2 - 4m_X^2)^2} \bigg( & g_{\chi s}^2 \, \left( g_{f s}^2 + g_{fp}^2 \right) \left[ \, 32 \, v^2 \, \right] 
    +  g_{\chi p}^2 \, \left( g_{fs}^2 + g_{f p}^2 \right) \left[ \, 128 + \left(\frac{76m_\phi^2 - 48m_\chi^2}{m_\phi^2 - 4m_\chi^2}\right) v^2 \, \right] 
\end{align}
\begin{align}
    (\sigma v)_{\gamma \gamma} \approx \frac{N_c^2}{256 \pi} \frac{m_\chi^2}{(m_\phi^2 - 4 m_\chi^2)^2} \left( \frac{Q_f^2 \alpha}{ 3 \pi} \frac{m_\chi}{m_f} \right)^2  \bigg( & g_{\chi s}^2 \, \left( g_{f s}^2 + \frac{9}{4} g_{fp}^2 \right) \, \left[ \, 64 \, v^2 \, \right] \nonumber\\
    +& g_{\chi p}^2 \, \left( g_{f s}^2 + \frac{9}{4} g_{fp}^2 \right)\,  \left[ \, 256 + \left(\frac{216m_\phi^2 - 352m_\chi^2}{m_\phi^2 - 4m_\chi^2}\right) \, v^2 \, \right] \bigg)
\end{align}

\begin{align}
    (\sigma v)_{gg} \approx \frac{N_c^2 - 1}{256 \pi} \frac{m_\chi^2}{(m_\phi^2 - 4 m_\chi^2)^2 } \left( \frac{\alpha_s}{ 6 \pi} \frac{m_\chi}{m_f} \right)^2 \bigg(& g_{\chi s}^2 \, g_{f s}^2 \, \left[ \, 64 \, v^2 \, \right] 
    + \frac{9}{4} g_{\chi p}^2 \, g_{f p}^2 \left[ \, 1024 + \left(\frac{864m_\phi^2 - 1408m_\chi^2}{m_\phi^2 - 4m_\chi^2}\right) \, v^2 \, \right] \nonumber \\
    + \, & \frac{9}{4} g_{\chi s}^2 \, g_{f p}^2 \left[ \, 256 \, v^2 \, \right] 
    + g_{\chi p}^2 \, g_{f s}^2 \left[ \, 256 + \left(\frac{216m_\phi^2 - 352m_\chi^2}{m_\phi^2 - 4m_\chi^2}\right)\, v^2 \, \right] \bigg)\,, \\
    (\sigma v)_{\gamma \gamma,\rm ChPT} \approx \frac{1}{256 \pi}\frac{m_\chi^2}{(m_\phi^2 - 4 m_\chi^2)^2 }  \left( \frac{\alpha B_0}{24 \pi m_\pi^2} \right)^2 \bigg(& g_{\chi s}^2 \, g_{f s}^2 \,\left[ \, 64 \, v^2 \, \right]  
    + 36 g_{\chi p}^2 \, g_{f p}^2 \, \left[ \, 1024 + \left(\frac{864m_\phi^2 - 1408m_\chi^2}{m_\phi^2 - 4m_\chi^2}\right) \, v^2  \, \right] \nonumber \\
    + \,                                                                                                                                                                   & 36 g_{\chi s}^2 \, g_{f p}^2 \, \left[ \, 256 \, v^2 \, \right]
    +  g_{\chi p}^2 \, g_{f s}^2\,  \left[ \, 256 + \left(\frac{216m_\phi^2 - 352m_\chi^2}{m_\phi^2 - 4m_\chi^2}\right) \, v^2 \, \right] \bigg)
\end{align}
\begin{align}
    (\sigma v)_{\phi \phi} \approx \frac{m_\chi^2 \sqrt{1-\frac{m_{\phi }^2}{m_{\chi }^2}}}{2 \pi  \left(m_\phi^2-2 m_\chi^2\right)^2}\Biggl(&g_{\chi p}^2 g_{\chi s}^2 \bigg(1 + \frac{3\left[4 m_\chi^6-23 m_\chi^2 m_{\phi }^4+8 m_\chi^4 m_\phi^2+7 m_\phi^6\right]}{96 \left(m_\chi^2 -m_\phi^2\right) \left(2m_\chi^2 - m_\phi^2\right)^2} v^2\bigg) \nonumber \\
    + & g_{\chi p}^4 \frac{8\left[m_\chi^2-m_\phi^2\right]^3}{96 \left(m_\chi^2 -m_\phi^2\right) \left(2m_\chi^2 - m_\phi^2\right)^2} v^2 + g_{\chi s}^4 \frac{8\left[9 m_\chi^6+10 m_\chi^2 m_{\phi }^4-17 m_\chi^4 m_\phi^2-2 m_{\phi }^6\right]}{96 \left(m_\chi^2 -m_\phi^2\right) \left(2m_\chi^2 - m_\phi^2\right)^2} v^2 \Biggr)\,.
\end{align}

\subsection{Spin-0 Dark Matter, Spin-1 Mediator}
\label{sec:Dm0Med1}

A spin-1 mediator, $V_\mu$, couples with a spin-0 dark matter candidate $X$, and SM fermions with the following Lagrangian
\begin{align}
    \label{eq:L DM0 med1}
    \mathcal{L} \supset i g_X \left( X^* \partial^\mu X - X \partial^\mu X^*  \right) V_\mu + g_X^2 V_\mu V^\mu X^* X + V_\mu \left( g_{f v} \bar{f} \gamma^\mu f + g_{f a} \bar{f} \gamma^\mu \gamma^5  f \right)\,.
\end{align}
Unlike the spin-0 mediator case, a spin-1 mediator cannot couple to a real scalar.
The number of internal degrees-of-freedom in this case is $g_{\rm spin}=1$.

Note that, in isolation,  axial-vector couplings to SM fermions are UV-sensitive as these interactions generically induce triangle anomalies in the absence of additional SM charged ``anomalons." For examples of UV sensitivity in axial-vector limits, see Refs.~\cite{Agrawal:2020lea,Kahn:2016vjr}. To remain agnostic to the details of the UV-completion, here we present only the direct laboratory bounds on the axial-vector interaction, noting that specific realizations generically face nontrivial model-dependent limits.

If $m_V > 2 m_X$, the mediator $V$ can decay into a pair of dark matter particles with the following width,
\begin{align}
    \Gamma_{V \rightarrow X^* X} &= \frac{g_X^2}{48 \pi} m_V \left( 1 - \frac{4 m_X^2}{m_V^2} \right)^{3/2}\,.
\end{align}
The corresponding decay rates into SM particles are given in Sec.~\ref{subsec:decay_rate_to_SM}.

The squared amplitudes for dark matter annihilation to quark fermions, electrons (via kinetic mixing), and mediators are given by
\begin{align}
    |\mathscr{M}|^2_{f\bar{f}} &= \frac{N_c}{(s - m_V^2)^2 + m_V^2 \Gamma_V^2} \frac{4}{3} (s - 4 m_X^2) \, g_X^2 \Bigg(g_{f v}^2 \left( s + 2 m_f^2 \right) + g_{f a}^2 \left( s - 4 m_f^2 \right) \Bigg)\,, \\
    |\mathscr{M}|^2_{ee} &= \frac{N_c}{(s - m_V^2)^2 + m_V^2 \Gamma_V^2} \frac{4}{3} (s - 4 m_X^2) \, g_X^2 \bigg(\frac{Q_f e^2 g_{f v}}{16\pi^2}  \bigg)^2\left( s + 2 m_e^2 \right) \,,
    \end{align}
    \begin{align}
    |\mathscr{M}|^2_{VV} &= 4 g_X^4 \bigg(1 + \frac{\left(m_V^2-4 m_X^2\right){}^2}{4 \left(s m_X^2-4 m_V^2 m_X^2+m_V^4\right)} +A\left(m_V^2 \left(4 s-8 m_X^2\right)-8 s m_X^2+m_V^4+16 m_X^4\right) \bigg)\,,
\end{align}
where we have defined,
\begin{align}
    A \equiv \frac{\tanh ^{-1}\left(\sqrt{\left(s-4 m_V^2\right) \left(s-4 m_X^2\right)}/(s-2 m_V^2)\right)}{\left(s-2 m_V^2\right) \sqrt{\left(s-4 m_V^2\right) \left(s-4 m_X^2\right)}} \, ,
\end{align}
and $\Gamma_V$ is the total decay width of the mediator.

We can calculate the dark matter annihilation cross section, following Eq.~\eqref{eq:sigmav NR non-avg}.
We take the non-relativistic limit and expand $\sigma v$ in powers of velocity.
For simplicity, we also take the limit of $m_f \to 0$.
The amplitudes given above yield the following results:
\begin{align}
    (\sigma v)_{f\bar{f}} &\approx \frac{ g_X^2}{6 \pi} \frac{m_X^2}{(m_V^2 - 4m_X^2)^2} (g_{fa}^2 + g_{fv}^2) v^2\,,\\
    (\sigma v)_{ee} &\approx \frac{g_X^2}{6 \pi} \frac{m_X^2}{(m_V^2 - 4m_X^2)^2} \left(\frac{Q_f e^2 g_{fv}}{16\pi^2}\right)^2 v^2\,,\\
    (\sigma v)_{VV} &\approx \frac{g_X^4 \sqrt{1-\frac{m_V^2}{m_X^2}}}{16 \pi m_X^2\left(2 m_X^2 - m_V^2\right)^2} \bigg(-8 m_V^2 m_X^2+3 m_V^4+8 m_X^4 \nonumber \\
    & - \frac{736 m_X^{10} - 2080 m_X^8 m_V^2 + 2284 m_X^6 m_V^4 - 1224 m_X^4 m_V^6 + 257 m_X^2 m_V^8 - 9 m_V^{10}}{96 \left(m_X^2 - m_V^2\right) \left(2m_X^2 - m_V^2\right)^2} v^2 \bigg)\,.
\end{align}

\subsection{Spin-1/2 Dark Matter, Spin-1 Mediator}
\label{sec:Dm1/2Med1}

A spin-1 mediator, $V_\mu$, couples with a spin-1/2 dark matter candidate $\chi$, and SM fermions with the following Lagrangian
\begin{align}
    \label{eq:L DM1/2 med1}
    \mathcal{L} \supset \left( g_{\chi v} \, \bar{\chi} \gamma^\mu \chi + a  g_{\chi a} \, \bar{\chi} \gamma^\mu \gamma^5 \chi \right) V_\mu + V_\mu \left( g_{f v} \bar{f} \gamma^\mu f + g_{f a} \bar{f} \gamma^\mu \gamma^5  f \right)\,,
\end{align}
where $a$ is a symmetry factor which equals 1 for a Dirac fermion and $1/2$ for a Majorana fermion.
Note that if $\chi$ is a Majorana fermion, the vector coupling $g_{\chi v}$ is forbidden.
The number of internal degrees-of-freedom in this case is $g_{\rm spin}=2$.

We note that some of the axial-vector bounds can be UV-sensitive.
The UV model for the axial-vector coupling to the SM fermion is somewhat complicated.
One example is shown in Ref.~\cite{Agrawal:2020lea}, which also claims $g_{fa} < m_V/m_f$.
We do not impose this bound through this work, but rather remind the reader that the available parameter space of axial-vector coupling may be further constrained by the UV model, and in some cases, the limit can be stringent.

If $m_V > 2 m_X$, the mediator $V_\mu$ can decay into a pair of dark matter particles with the following width,
\begin{align}
    \Gamma_{V \rightarrow \bar{\chi} \chi} &= \frac{m_V}{12 \pi} \sqrt{1 - \frac{4 m_\chi^2}{m_V^2}} \left( g_{\chi v}^2 \left[ 1 + \frac{2 m_\chi^2}{m_V^2} \right] + g_{\chi a}^2 \, \left[ 1 - \frac{4 m_\chi^2}{m_V^2} \right] \right)\,.
\end{align}
We refer the readers to Sec.~\ref{subsec:decay_rate_to_SM} for the decay rate to SM particles.

The squared amplitudes for dark matter annihilation to fermions, photons, gluons, and mediators are given by
\begin{align}
\label{eq:Mff-vector}
    |\mathscr{M}|^2_{f\bar{f}} = \frac{N_c s^2}{(s - m_V^2)^2 + m_V^2 \Gamma_V^2} \frac{16}{3} \Bigg(& g_{\chi v}^2 \, g_{f v}^2 \left[ \,  \, \left( 1 + \frac{2 m_\chi^2}{s} \right) \left( 1 + \frac{2 m_f^2}{s} \right) \, \right]                                                \nonumber  \\
    + \,                                                                                            & g_{\chi a}^2 \, g_{f a}^2 \left[ \, \left( 1 - \frac{4 m_\chi^2}{s} \right) \left( 1 - \frac{4 m_f^2}{s} \right) + 12 \frac{m_\chi^2 m_f^2}{m_V^4 s^2} (s - m_V^2)^2 \, \right] \nonumber\\
    + \,                                                                                            & g_{\chi v}^2 \, g_{f a}^2 \left[ \, \left( 1 + \frac{2 m_\chi^2}{s} \right) \left( 1 - \frac{4 m_f^2}{s} \right) \, \right]  + g_{\chi a}^2 \, g_{f v}^2 \left[ \, \left( 1 - \frac{4 m_\chi^2}{s} \right) \left( 1 + \frac{2 m_f^2}{s} \right)  \, \right]\Bigg)
\end{align}
\begin{align}
    |\mathscr{M}|^2_{ee} & = \frac{N_c s^2}{(s - m_V^2)^2 + m_V^2 \Gamma_V^2} \frac{16}{3} \left(
    \frac{Q_f e^2 g_{fv}}{16\pi^2}\right)^2\Bigg[g_{\chi v}^2 \left(1 + \frac{2 m_\chi^2}{s}\right)+ g_{\chi a}^2 \left(1 - \frac{4 m_\chi^2}{s}\right)\Bigg]\left(1 + \frac{2 m_f^2}{s}\right) 
\end{align}
\begin{align}
    |\mathscr{M}|^2_{VV} = 16\Bigg(& g_{\chi v}^4 \bigg[-1-\frac{(2m_\chi^2 + m_V^2)^2}{m_\chi^2(s-4m_V^2) + m_V^4} + 2 A \left(s^2 + 4 s m_{\chi }^2-8 m_\chi^4-8 m_\chi^2 m_V^2 +4 m_V^4\right)\bigg] \nonumber\\
    + & g_{\chi v}^2 g_{\chi a}^2 \bigg[-6 - \frac{2(4 s m_\chi^2 - 8 m_\chi^4 - 10 m_\chi^2 m_V^2 + 3 m_V^4)}{m_\chi^2(s-4m_V^2) + m_V^4} \nonumber\\
    & \hspace{3em}+ 4 \frac{A}{m_V^2} \bigg(m_V^2 \left(3 s^2 -20 s m_{\chi }^2+16 m_{\chi }^4\right) +2 s^2 m_{\chi }^2 - 20 m_V^4 m_{\chi }^2+12 m_V^6\bigg)\bigg] \nonumber\\
    + & g_{\chi a}^4 \bigg[ \frac{2s m_\chi^2 - 8 m_V^2 m_\chi^2 + m_V^4}{m_V^4} - \frac{(4m_\chi^2 - m_V^2)^2}{m_\chi^2 (s-4m_V^2) + m_V^4} \nonumber \\
    & \hspace{3em}+ 2 \frac{A}{m_V^4}  \bigg(-4 s^2 m_{\chi }^4+s m_V^4 \left(s-12 m_{\chi }^2\right) +4 s m_V^2 m_{\chi }^2 \left(4 m_{\chi }^2+s\right)-16 m_V^6 m_{\chi }^2+4 m_V^8\bigg)\bigg]
    \Bigg)\,,
\end{align}
where we have defined,
\begin{align}
    A \equiv \frac{\tanh^{-1}\left(\sqrt{\left(s-4 m_V^2\right) \left(s-4 m_{\chi }^2\right)}/(s-2 m_V^2)\right)}{\left(s-2 m_V^2\right) \sqrt{\left(s-4 m_V^2\right) \left(s-4 m_{\chi }^2\right)}} \, .
\end{align}

We can calculate the dark matter annihilation cross section, following Eq.~\eqref{eq:sigmav NR non-avg}.
We take the non-relativistic limit and expand $\sigma v$ in powers of velocity.
For simplicity, we also take the limit of $m_f \to 0$.
The amplitudes given above yield the following results:
The results are\footnote{Note that there is an extra $s$-wave contribution to the annihilation cross section proportional to $m_f^2$ in the case that the DM and SM quark couplings are axial; see Eq.~\eqref{eq:Mff-vector}. This term can dominate the top-coupled annihilation cross section if $m_\chi$ is slightly heavier than $m_t$, explaining the vertical non-secluded band appearing in Fig.~\ref{fig:t_12_panel_2}.}
\begin{align}
    (\sigma v)_{f\bar{f}} \approx \frac{N_c}{768 \pi} \frac{m_\chi^2}{(m_V^2 - 4m_\chi^2)^2}  \bigg( & g_{\chi v}^2 \, \left( g_{f v}^2 + g_{fa}^2 \right) \left[ \, 768 + \left(\frac{392 m_V^2 - 32m_\chi^2}{m_V^2 - 4m_\chi^2}\right) \,v^2  \, \right] +  g_{\chi a}^2 \, \left( g_{f a}^2 + g_{fv}^2 \right)\left[ \, 128\, v^2 \, \right] \bigg)\,,
\end{align}
\begin{align}
    (\sigma v)_{ee} \approx \frac{N_c}{768 \pi} \frac{m_\chi^2}{(m_V^2 - 4m_\chi^2)^2} &\left(
    \frac{Q_f e^2 g_{fv}}{16\pi^2}\right)^2 \bigg(  g_{\chi v}^2 \, \left[ \, 768 + \left(\frac{392 m_V^2 - 32m_\chi^2}{m_V^2 - 4m_\chi^2}\right) \,v^2  \, \right] + g_{\chi a}^2 \,\left[ \, 128 \, v^2 \, \right] \bigg)\,,
\end{align}
\begin{align}
    (\sigma v)_{VV} \approx \frac{1}{256\pi m_\chi^2} \bigg(&g_{\chi v}^4 (32+19 v^2) + \, g_{\chi a}^4 \left[32 + \frac{128m_\chi^4 -23 m_V^4}{3 m_V^2}v^2\right] \nonumber \\
    + &\, g_{\chi a}^2 g_{\chi v}^2 \left[256 \left(\frac{m_\chi^2}{m_V^2}\right)-320 + \left(\frac{136 m_\chi^2}{3 m_V^2}+ 2\right) v^2\right]
    \bigg)\,.
\end{align}

\section{Results}
\label{sec:result}

Each of the simplified models considered here is described by four parameters: the dark matter and mediator masses, and the dark matter-mediator and  SM quark-mediator couplings. 
To illustrate the allowed parameter space in a two-dimensional figure, two of these parameters must be fixed. For each model we choose to fix one of the couplings (or the ratio of them); specifically we will consider three scenarios: $g_{\rm SM} = \sqrt{4\pi}$, $g_{\rm SM} = g_{\rm DM}$, and $g_{\rm DM} = \sqrt{4\pi}$, where $g_{\rm SM}$ ($g_{\rm DM}$) is the mediator's coupling to SM quarks (DM).
Additionally, we fix the remaining coupling by finding the value needed to achieve the observed dark matter density today (when possible).
This allows us to illustrate the allowed parameter space for each model in the mediator mass, dark matter mass plane. See Table~\ref{tab:summary} for links to all figures.

The constraints on each model are applied sequentially: once a parameter point is found to be excluded, we do not check any further constraints on that point. 
Parameter points that are allowed remain unshaded in white (or in white with black dots, in the case of secluded dark matter).
The list of constraints we consider includes:
\begin{itemize}
\item \textbf{Resonant Annihilation:} When $m_\phi \simeq 2 m_\chi$, annihilation becomes resonant and requires a dedicated numerical treatment~\cite{Griest:1990kh,Hooper:2013qjx}; we leave this for future work.  
This region is shaded \textbf{\color{gray!50!black}dark gray}.
\item \textbf{Cosmology:} For $m_\phi$ or $m_\chi\lesssim 10~\rm MeV$, these particles remain in thermal equilibrium into the era of BBN, violating constraints on $\Delta N_{\rm eff}$ (Sec.~\ref{sec:Neff}). This region is shaded \textbf{\color{violet}purple}.
\item \textbf{Perturbativity:} When the mediator decay width exceeds its mass ($\Gamma_\phi > m_\phi$), perturbative unitarity is violated. This bound is maximally constraining for scenarios in which the SM coupling is set to $ g_{\rm SM} = \sqrt{4\pi}$ and the mediator is sufficiently heavy to have tree-level decays into SM quarks. 
Furthermore, the upper limit on the couplings we consider for the next constraints is found by satisfying perturbative unitarity.
This region is shaded \textbf{\color{gray}gray}.
\item \textbf{Overproduction:} There are regions of parameter space in which no combination of perturbative couplings can achieve the observed relic density and lead to overproduction of dark matter. This issue typically arises when either the dark matter or mediator mass becomes large. We also exclude any regions with $m_{\chi} > 110 \, {\rm TeV}$, following from the requirement of partial wave unitarity~\cite{Griest:1989wd,Smirnov:2019ngs,Blum:2014dca}. This region is shaded \textbf{\color{red}red}.
\item \textbf{Underproduction:} When $m_\chi > m_\phi$ and the coupling to dark matter is $\sqrt{4\pi}$, the secluded annihilation channel can be too efficient, driving the relic density below the observed value. This region is shaded \textbf{\color{blue}blue}.
\item \textbf{Direct Detection:} Direct detection constraints from the LZ collaboration in Ref.~\cite{LZ:2022lsv} (Sec.~\ref{sec:direct}). This region is shaded \textbf{\color{ForestGreen}light green}.
\item \textbf{Indirect Detection:} Constraints on the dark matter annihilation cross section based on measurements of the cosmic microwave background by the Planck collaboration~\cite{Planck:2018vyg}, of dwarf galaxies by the Fermi gamma-ray telescope~\cite{DiMauro:2022hue}, and of the cosmic-ray antiproton spectrum by AMS~\cite{Cholis:2019ejx} (Sec.~\ref{sec:indirect}). This region is shaded \textbf{\color{green!40!black} dark green}.
\item \textbf{Accelerators:} Constraints from the LHC, $Z$-pole measurements at LEP, and from fixed-target experiments in the case of spin-1 mediators with vector coupling to quarks (Sec.~\ref{sec:collider}). 
This region is shaded \textbf{\color{blue!40!black}dark blue}.
\item \textbf{Rare Meson Decays:} Constraints from kaon and $B$-meson decays (Sec.~\ref{sec:meson}) to spin-0 mediators (spin-1 mediators are typically more constrained by other probes). This region is shaded \textbf{\color{blue!50!white}light blue}.
\item \textbf{Secluded Annihilation:} Regions in which the dark matter annihilates predominantly (i.e., more than half of the time) to a pair of mediators, independently of the SM coupling, are marked with black dots. In these regions of parameter space, the thermal relic abundance {\it can} match the observed dark matter density, but this abundance does not depend appreciably on the coupling to SM particles. 
\end{itemize}

\section{Summary}
\label{sec:summary}

\begin{figure}[!htbp]
    \centering
    \includegraphics[width=\textwidth]{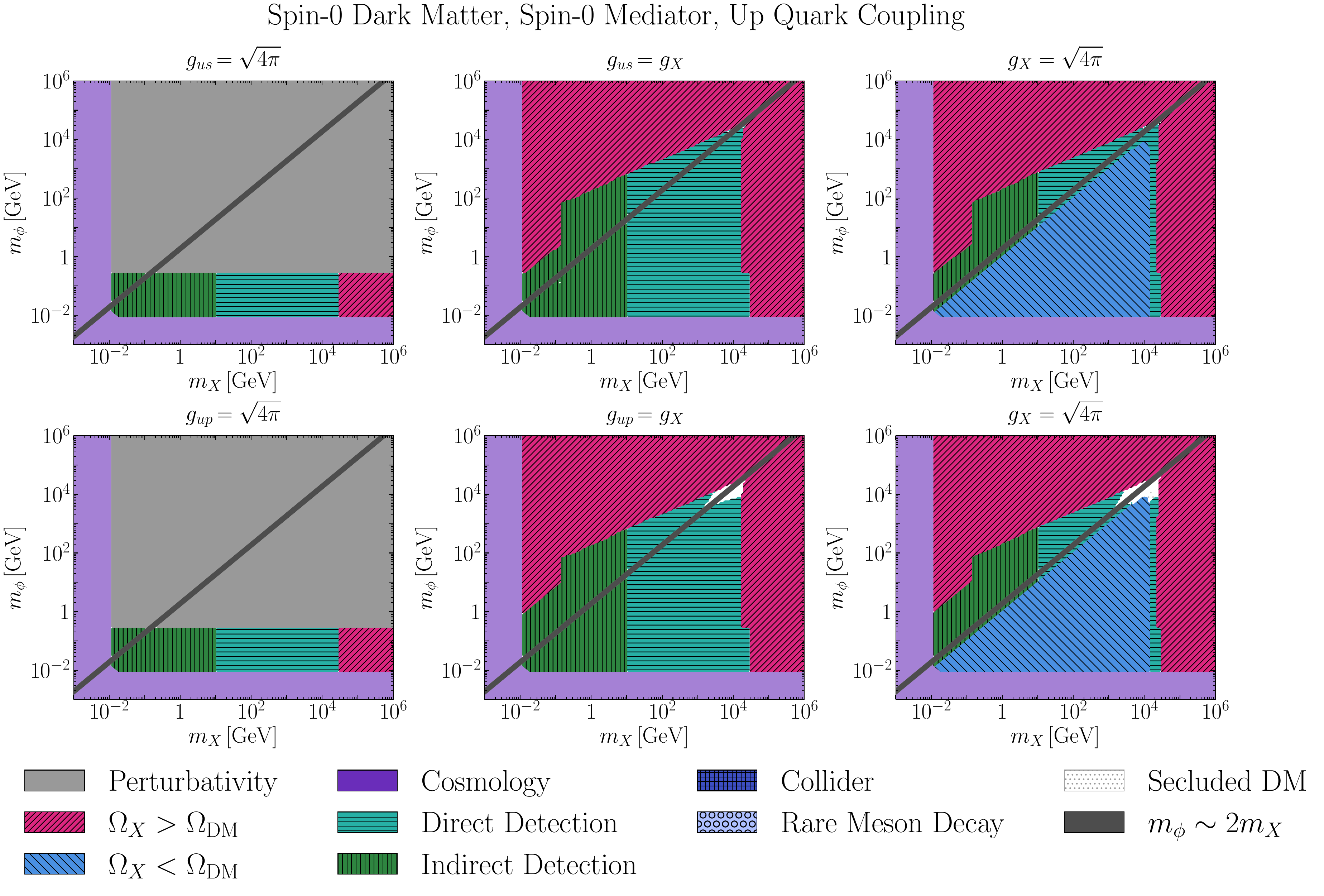}
    \caption{Results for the case of spin-0 dark matter with a spin-0 mediator that couples to up quarks. The top and bottom frames correspond to scalar and pseudoscalar couplings, respectively. The left, center, and right frames show results for mediator couplings that are maximal to quarks, equal to dark matter and quarks, and maximal to dark matter, respectively. In each case, we set the product of these two couplings to obtain a thermal relic abundance that is equal to the measured dark matter density. The meanings of the various colored regions are summarized in Sec.~\ref{sec:result} and the model is described in Sec.~\ref{sec:Dm0Med0}. The viable regions of parameter space are shown in white (and in white with black dots). See Table~\ref{tab:summary} for links to other figures.}
    \label{fig:u_6_panel}
\end{figure}

\begin{figure}[!htbp]
    \centering
    \includegraphics[width=\textwidth]{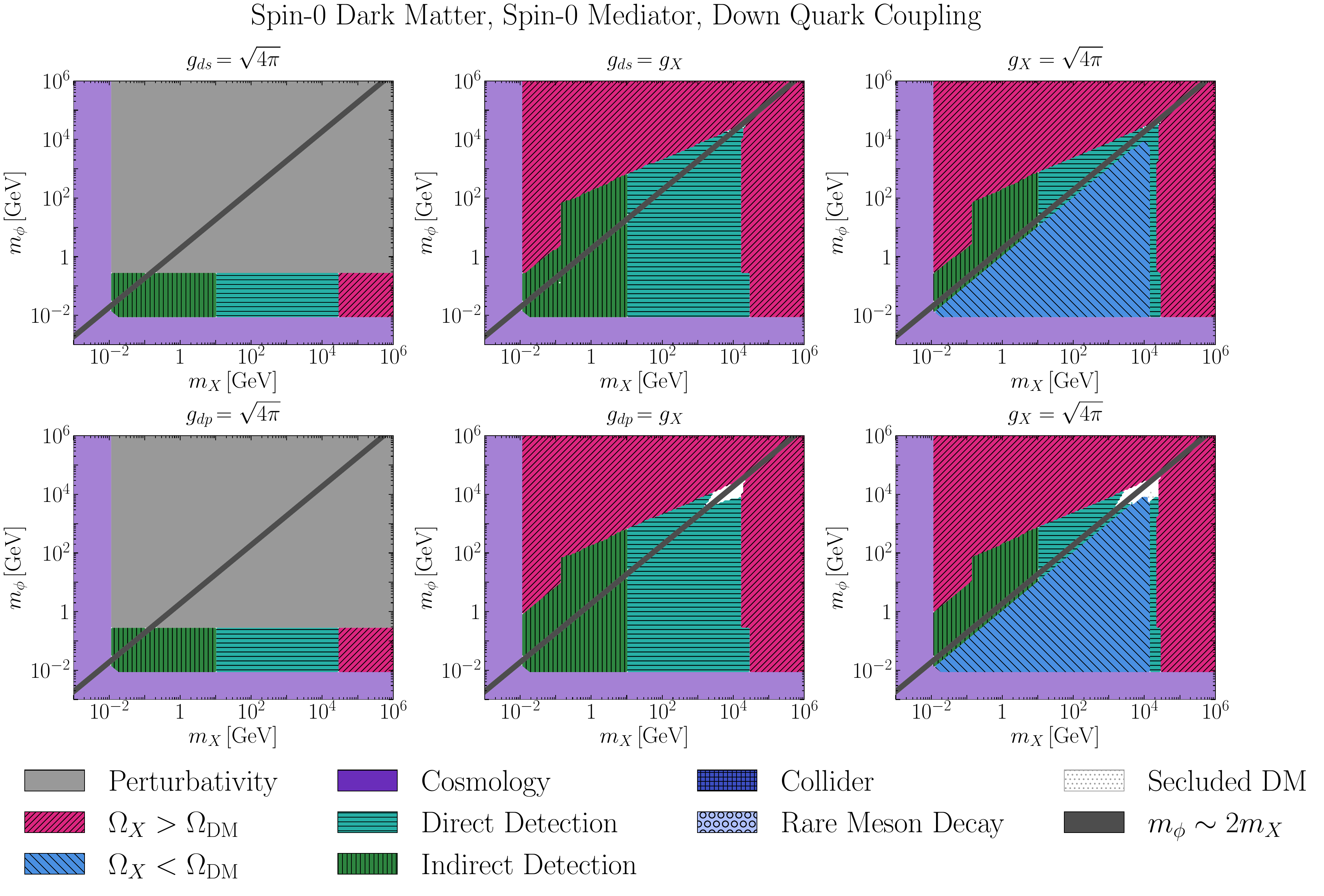}
    \caption{As in Fig.~\ref{fig:u_6_panel}, but for spin-0 dark matter with a spin-0 mediator that couples to down quarks, for scalar or pseudoscalar couplings. The meanings of the various colored regions are summarized in Sec.~\ref{sec:result} and the model is described in Sec.~\ref{sec:Dm0Med0}. The viable regions of parameter space are shown in white (and in white with black dots). See Table~\ref{tab:summary} for links to other figures.}
    \label{fig:d_6_panel}
\end{figure}

\begin{figure}[!htbp]
    \centering
    \includegraphics[width=\textwidth]{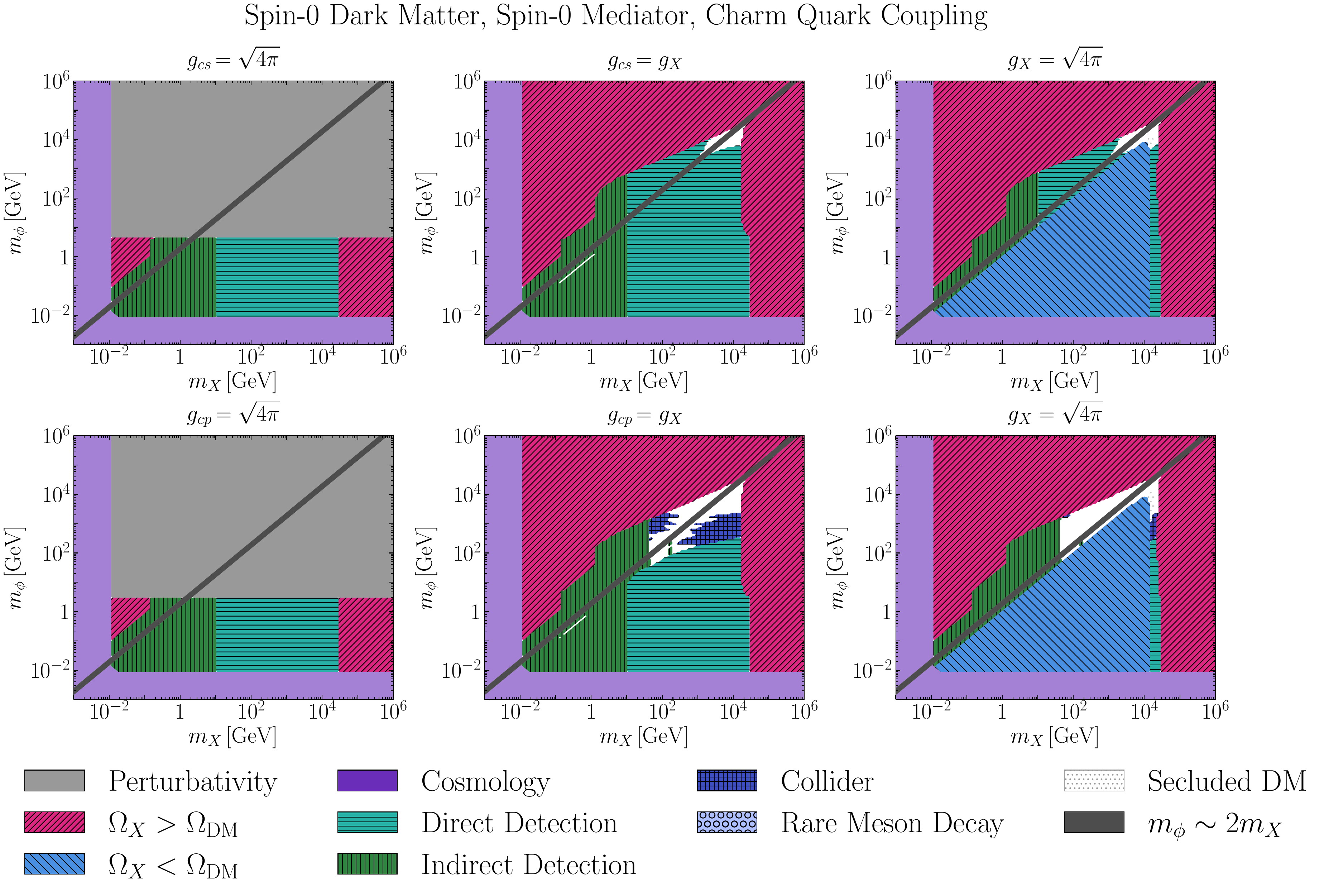}
    \caption{As in Fig.~\ref{fig:u_6_panel}, but for spin-0 dark matter with a spin-0 mediator that couples to charm quarks, for scalar or pseudoscalar couplings. The meanings of the various colored regions are summarized in Sec.~\ref{sec:result} and the model is described in Sec.~\ref{sec:Dm0Med0}. The viable regions of parameter space are shown in white (and in white with black dots). See Table~\ref{tab:summary} for links to other figures.}
    \label{fig:c_6_panel}
\end{figure}

\begin{figure}[!htbp]
    \centering
    \includegraphics[width=\textwidth]{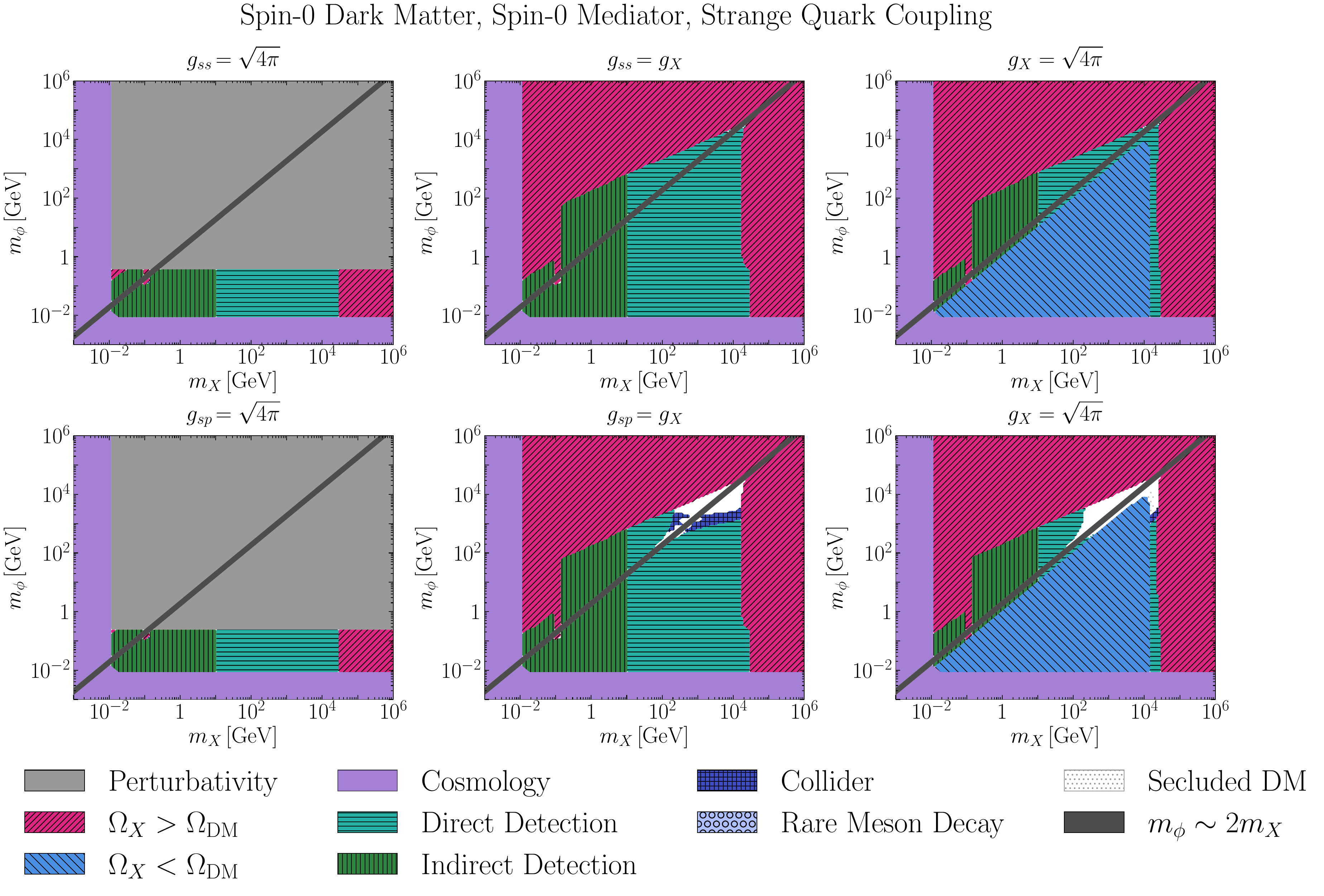}
    \caption{As in Fig.~\ref{fig:u_6_panel}, but for spin-0 dark matter with a spin-0 mediator that couples to strange quarks, for scalar or pseudoscalar couplings. The meanings of the various colored regions are summarized in Sec.~\ref{sec:result} and the model is described in Sec.~\ref{sec:Dm0Med0}. The viable regions of parameter space are shown in white (and in white with black dots). See Table~\ref{tab:summary} for links to other figures.}
    \label{fig:s_6_panel}
\end{figure}

\begin{figure}[!htbp]
    \centering
    \includegraphics[width=\textwidth]{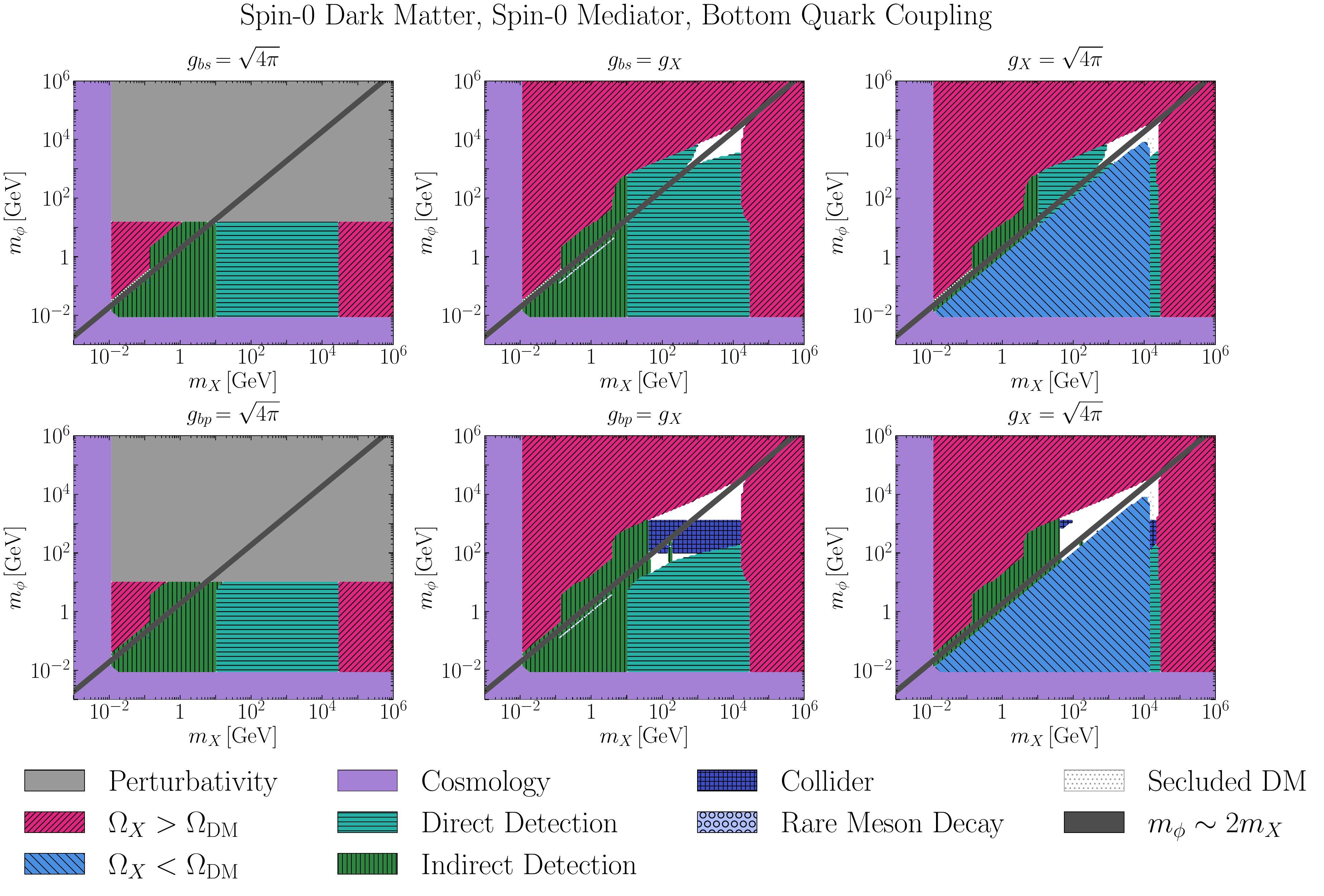}
    \caption{As in Fig.~\ref{fig:u_6_panel}, but for spin-0 dark matter with a spin-0 mediator that couples to bottom quarks, for scalar or pseudoscalar couplings. The meanings of the various colored regions are summarized in Sec.~\ref{sec:result} and the model is described in Sec.~\ref{sec:Dm0Med0}. The viable regions of parameter space are shown in white (and in white with black dots). See Table~\ref{tab:summary} for links to other figures.}
    \label{fig:b_6_panel}
\end{figure}

\begin{figure}[!htbp]
    \centering
    \includegraphics[width=\textwidth]{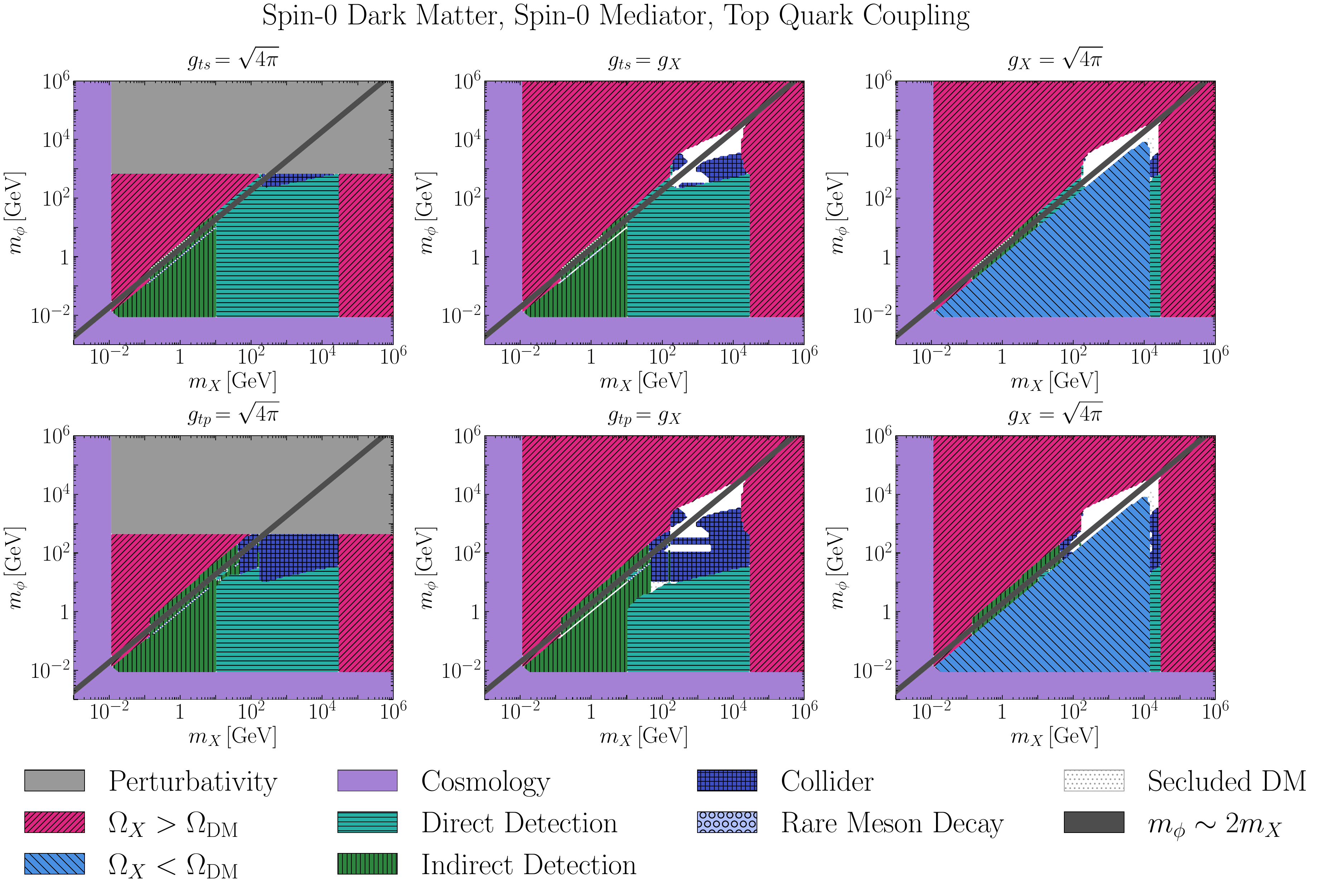}
    \caption{As in Fig.~\ref{fig:u_6_panel}, but for spin-0 dark matter with a spin-0 mediator that couples to top quarks, for scalar or pseudoscalar couplings. The meanings of the various colored regions are summarized in Sec.~\ref{sec:result} and the model is described in Sec.~\ref{sec:Dm0Med0}. The viable regions of parameter space are shown in white (and in white with black dots). See Table~\ref{tab:summary} for links to other figures.}
    \label{fig:t_6_panel}
\end{figure}

\begin{figure}[!htbp]
    \centering
    \includegraphics[width=\textwidth]{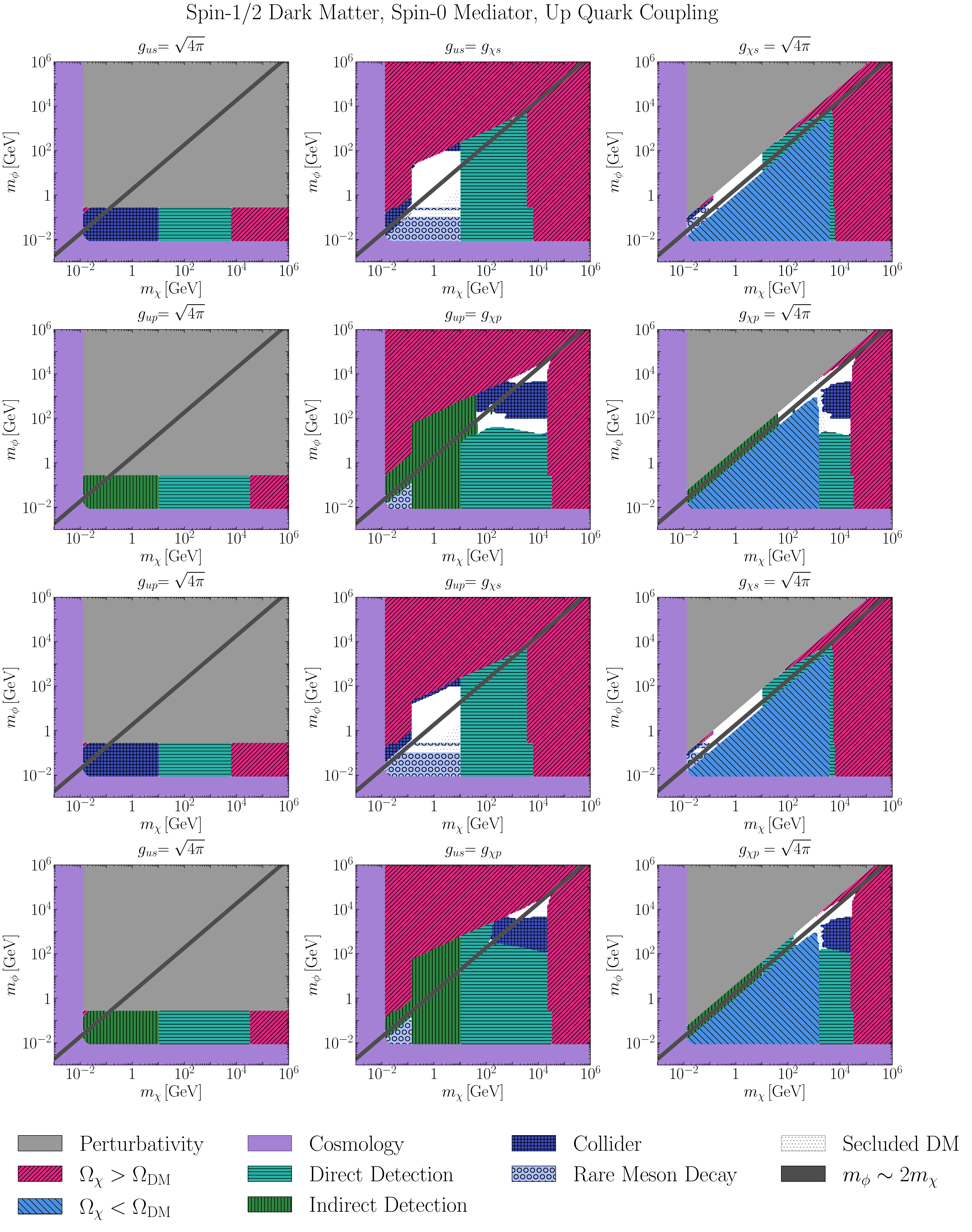}
    \caption{As in Fig.~\ref{fig:u_6_panel}, but for spin-1/2 dark matter with a spin-0 mediator that couples to up quarks, for various combinations of scalar and pseudoscalar couplings. The meanings of the various colored regions are summarized in Sec.~\ref{sec:result} and the model is described in Sec.~\ref{sec:Dm1/2Med0}. The viable regions of parameter space are shown in white (and in white with black dots). See Table~\ref{tab:summary} for links to other figures.}
    \label{fig:u_12_panel}
\end{figure}

\begin{figure}[!htbp]
    \centering
    \includegraphics[width=\textwidth]{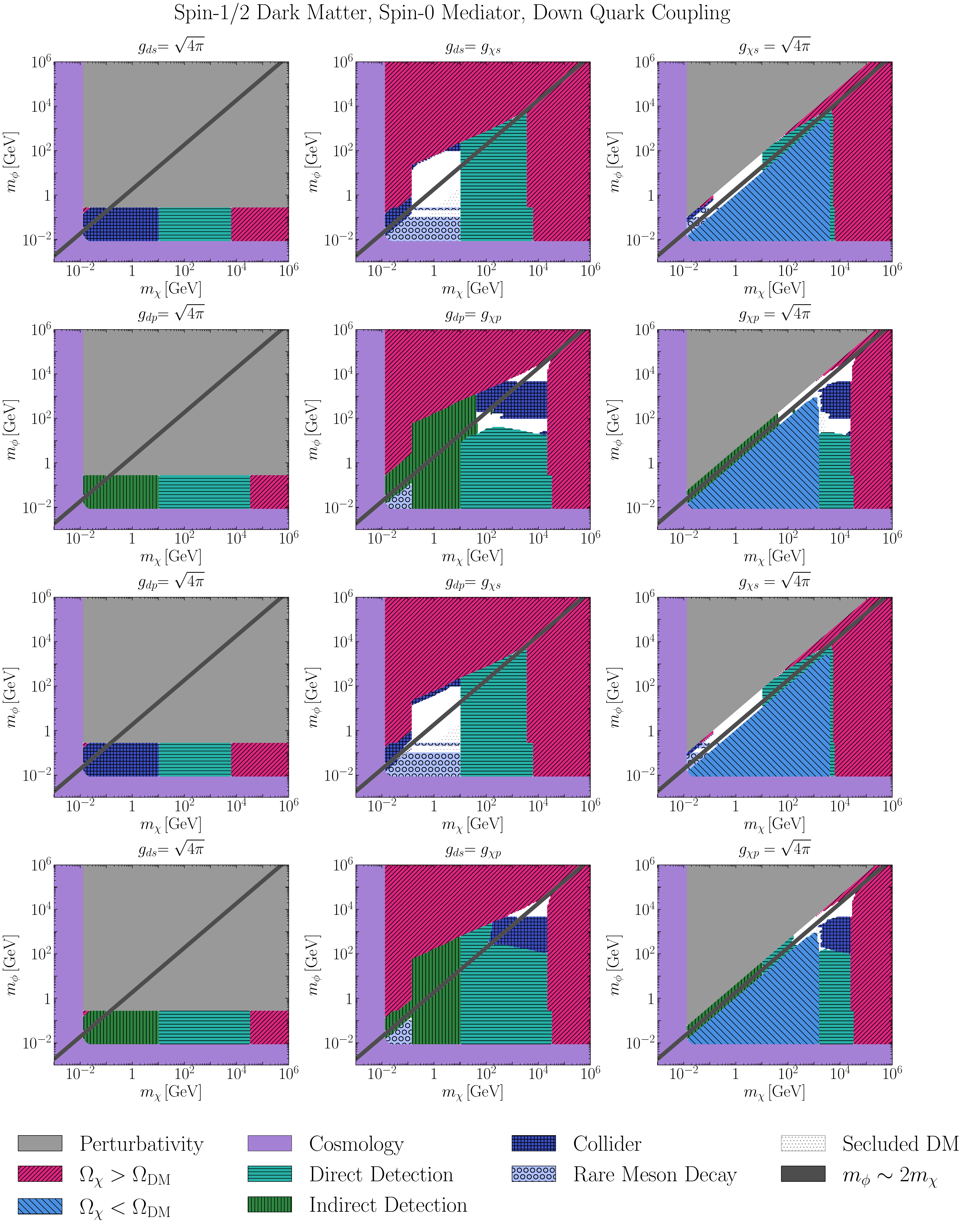}
    \caption{As in Fig.~\ref{fig:u_6_panel}, but for spin-1/2 dark matter with a spin-0 mediator that couples to down quarks, for various combinations of scalar and pseudoscalar couplings. The meanings of the various colored regions are summarized in Sec.~\ref{sec:result} and the model is described in Sec.~\ref{sec:Dm1/2Med0}. The viable regions of parameter space are shown in white (and in white with black dots). See Table~\ref{tab:summary} for links to other figures.}
    \label{fig:d_12_panel}
\end{figure}

\begin{figure}[!htbp]
    \centering
    \includegraphics[width=\textwidth]{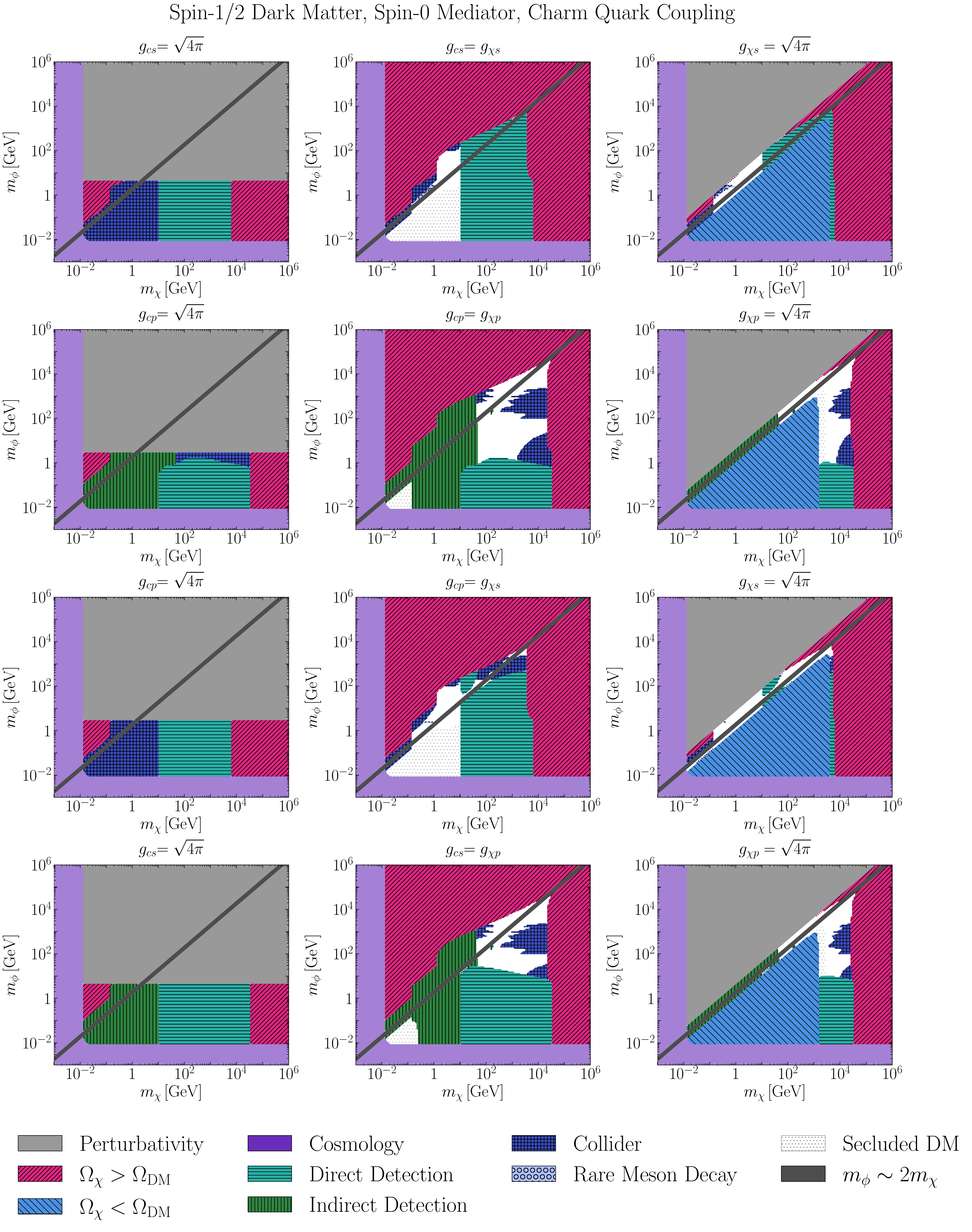}
    \caption{As in Fig.~\ref{fig:u_6_panel}, but for spin-1/2 dark matter with a spin-0 mediator that couples to charm quarks, for various combinations of scalar and pseudoscalar couplings. The meanings of the various colored regions are summarized in Sec.~\ref{sec:result} and the model is described in Sec.~\ref{sec:Dm1/2Med0}. The viable regions of parameter space are shown in white (and in white with black dots). See Table~\ref{tab:summary} for links to other figures.}
    \label{fig:c_12_panel}
\end{figure}

\begin{figure}[!htbp]
    \centering
    \includegraphics[width=\textwidth]{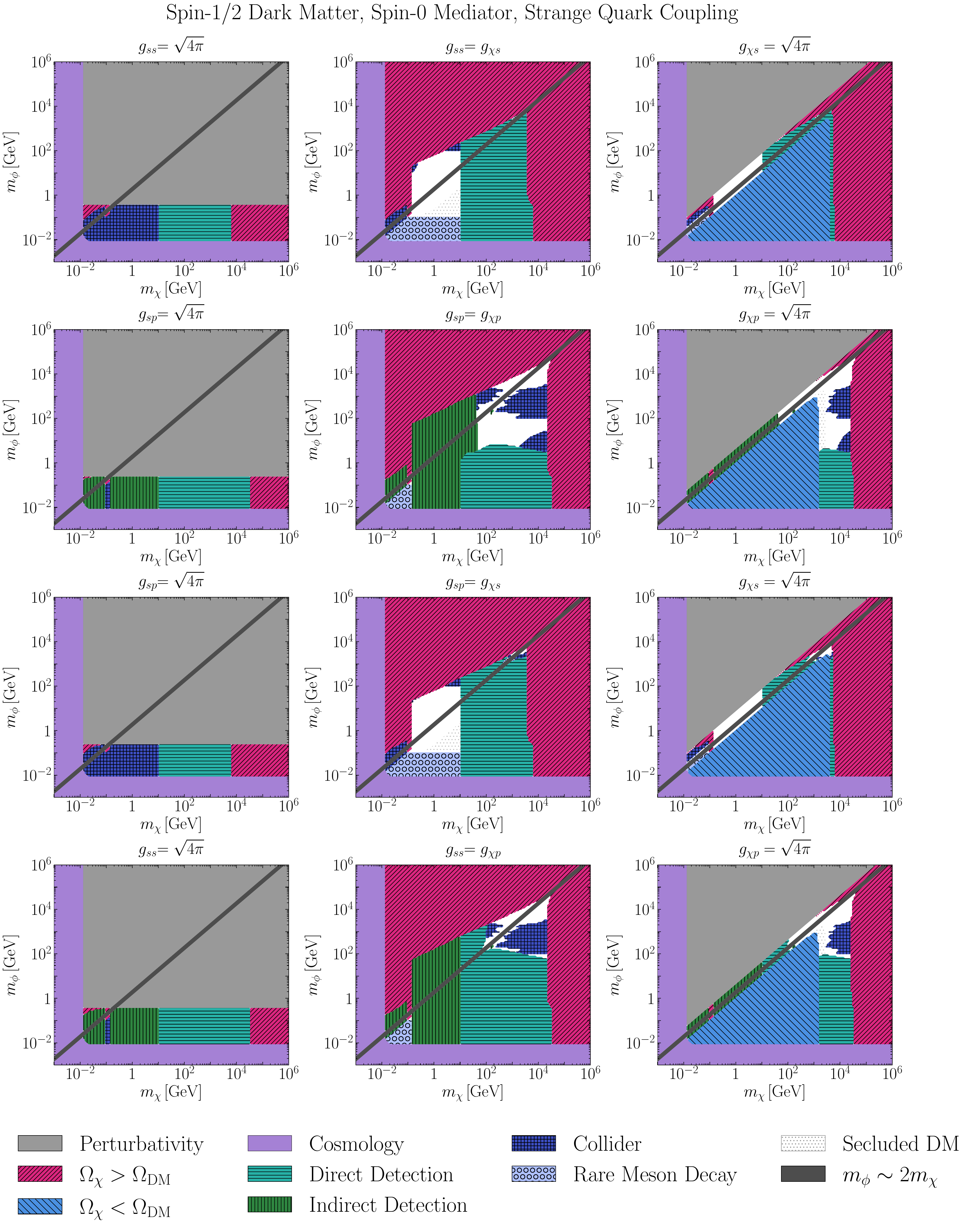}
    \caption{As in Fig.~\ref{fig:u_6_panel}, but for spin-1/2 dark matter with a spin-0 mediator that couples to strange quarks, for various combinations of scalar and pseudoscalar couplings. The meanings of the various colored regions are summarized in Sec.~\ref{sec:result} and the model is described in Sec.~\ref{sec:Dm1/2Med0}. The viable regions of parameter space are shown in white (and in white with black dots). See Table~\ref{tab:summary} for links to other figures.}
    \label{fig:s_12_panel}
\end{figure}

\begin{figure}[!htbp]
    \centering
    \includegraphics[width=\textwidth]{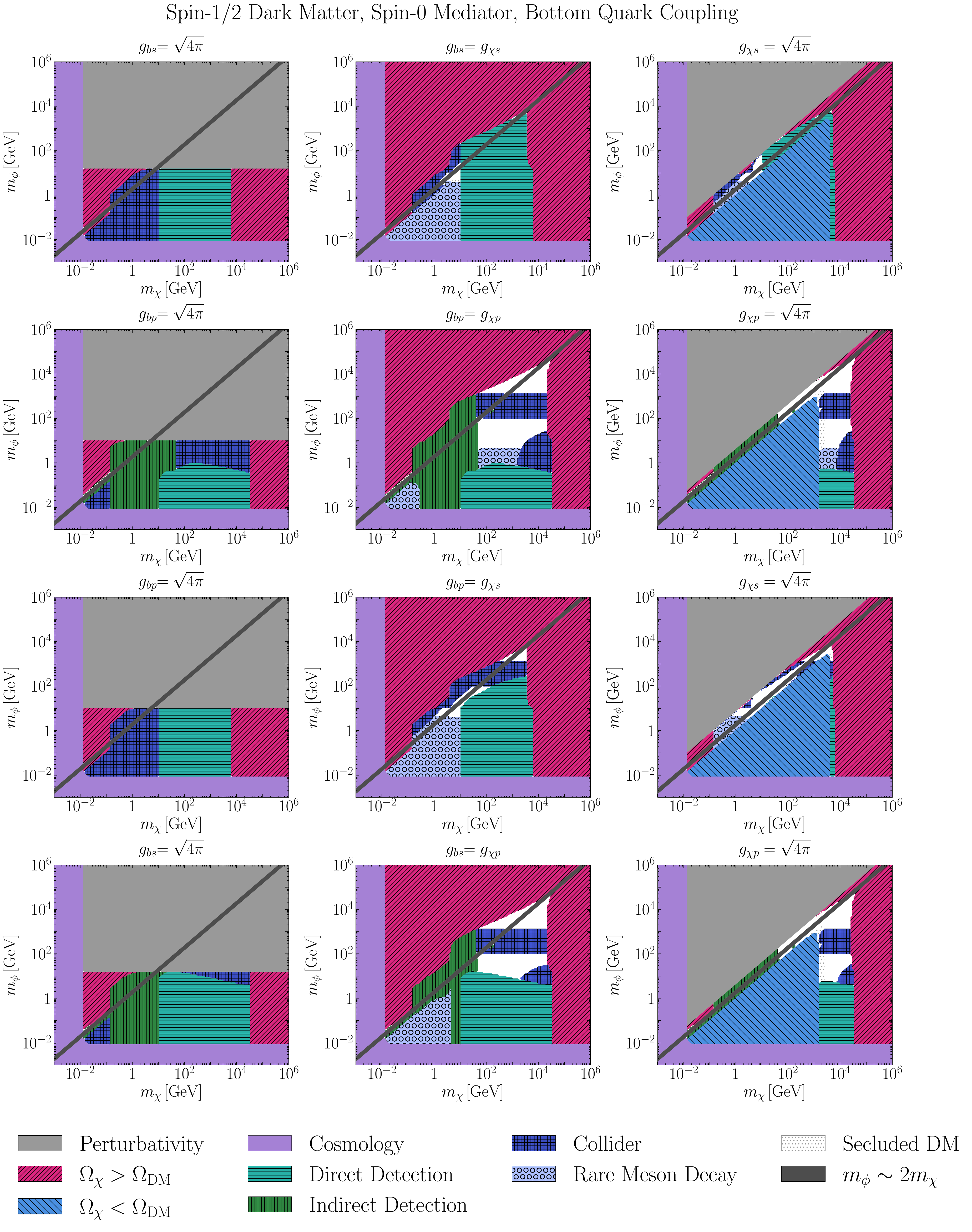}
    \caption{As in Fig.~\ref{fig:u_6_panel}, but for spin-1/2 dark matter with a spin-0 mediator that couples to bottom quarks, for various combinations of scalar and pseudoscalar couplings. The meanings of the various colored regions are summarized in Sec.~\ref{sec:result} and the model is described in Sec.~\ref{sec:Dm1/2Med0}. The viable regions of parameter space are shown in white (and in white with black dots). See Table~\ref{tab:summary} for links to other figures.}
    \label{fig:b_12_panel}
\end{figure}

\begin{figure}[!htbp]
    \centering
    \includegraphics[width=\textwidth]{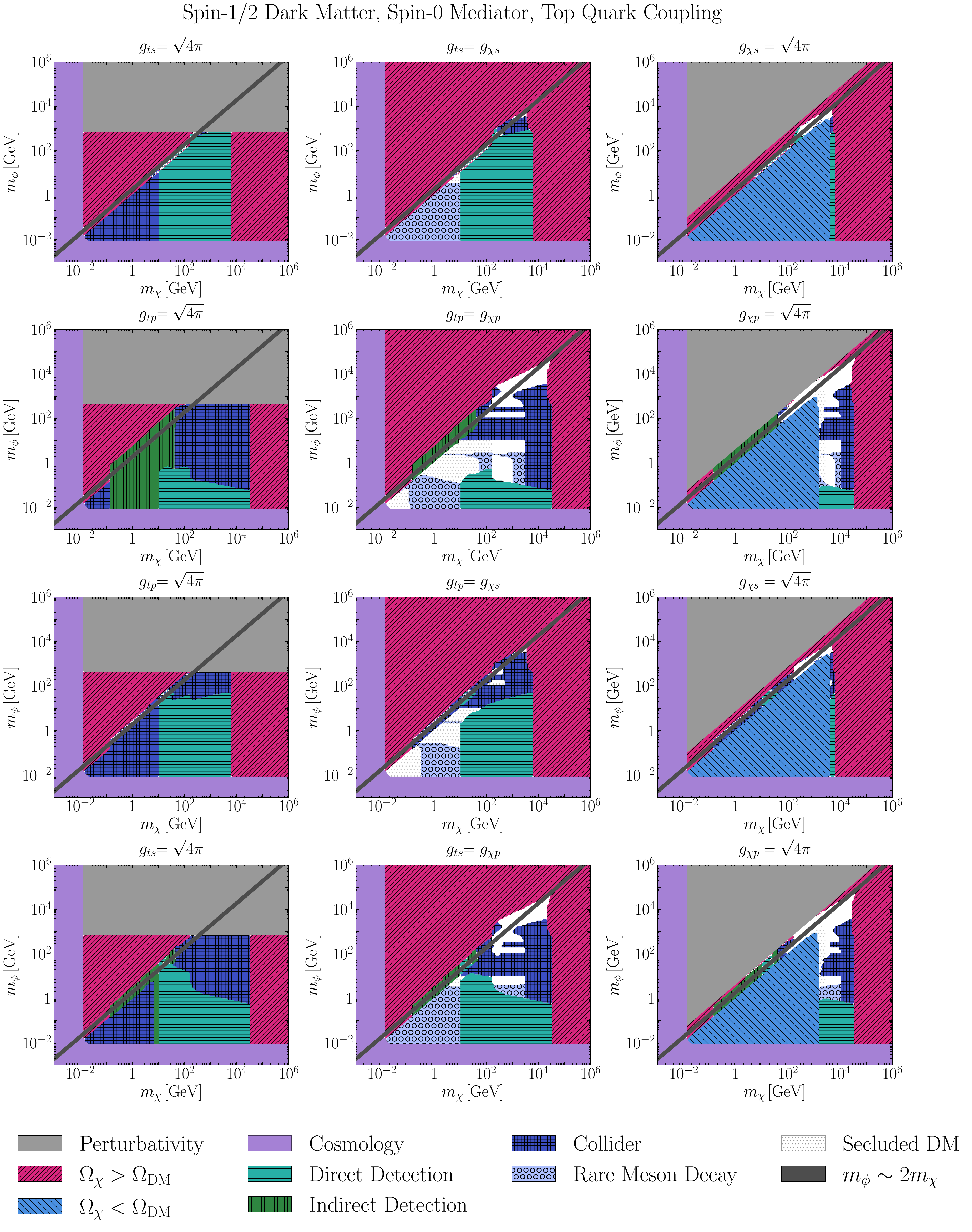}
    \caption{As in Fig.~\ref{fig:u_6_panel}, but for spin-1/2 dark matter with a spin-0 mediator that couples to top quarks, for various combinations of scalar and pseudoscalar couplings. The meanings of the various colored regions are summarized in Sec.~\ref{sec:result} and the model is described in Sec.~\ref{sec:Dm1/2Med0}. The viable regions of parameter space are shown in white (and in white with black dots). See Table~\ref{tab:summary} for links to other figures.}
    \label{fig:t_12_panel}
\end{figure}

\begin{figure}[!htbp]
    \centering
    \includegraphics[width=\textwidth]{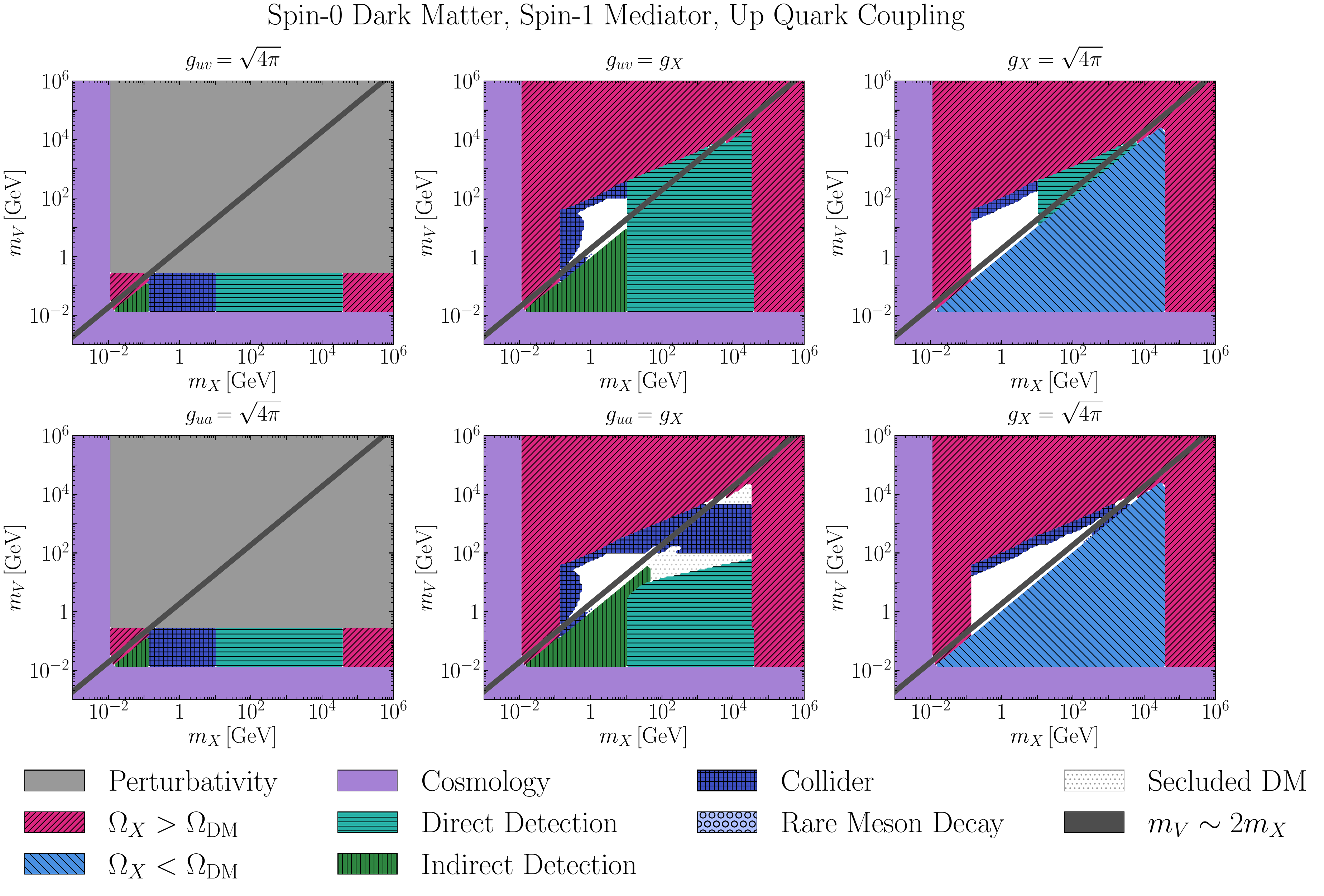}
    \caption{As in Fig.~\ref{fig:u_6_panel}, but for spin-0 dark matter with a spin-1 mediator that couples to up quarks, for vector or axial couplings. The meanings of the various colored regions are summarized in Sec.~\ref{sec:result} and the model is described in Sec.~\ref{sec:Dm0Med1}. The viable regions of parameter space are shown in white (and in white with black dots). See Table~\ref{tab:summary} for links to other figures.}
    \label{fig:u_scalar_vector_6_panel}
\end{figure}

\begin{figure}[!htbp]
    \centering
    \includegraphics[width=\textwidth]{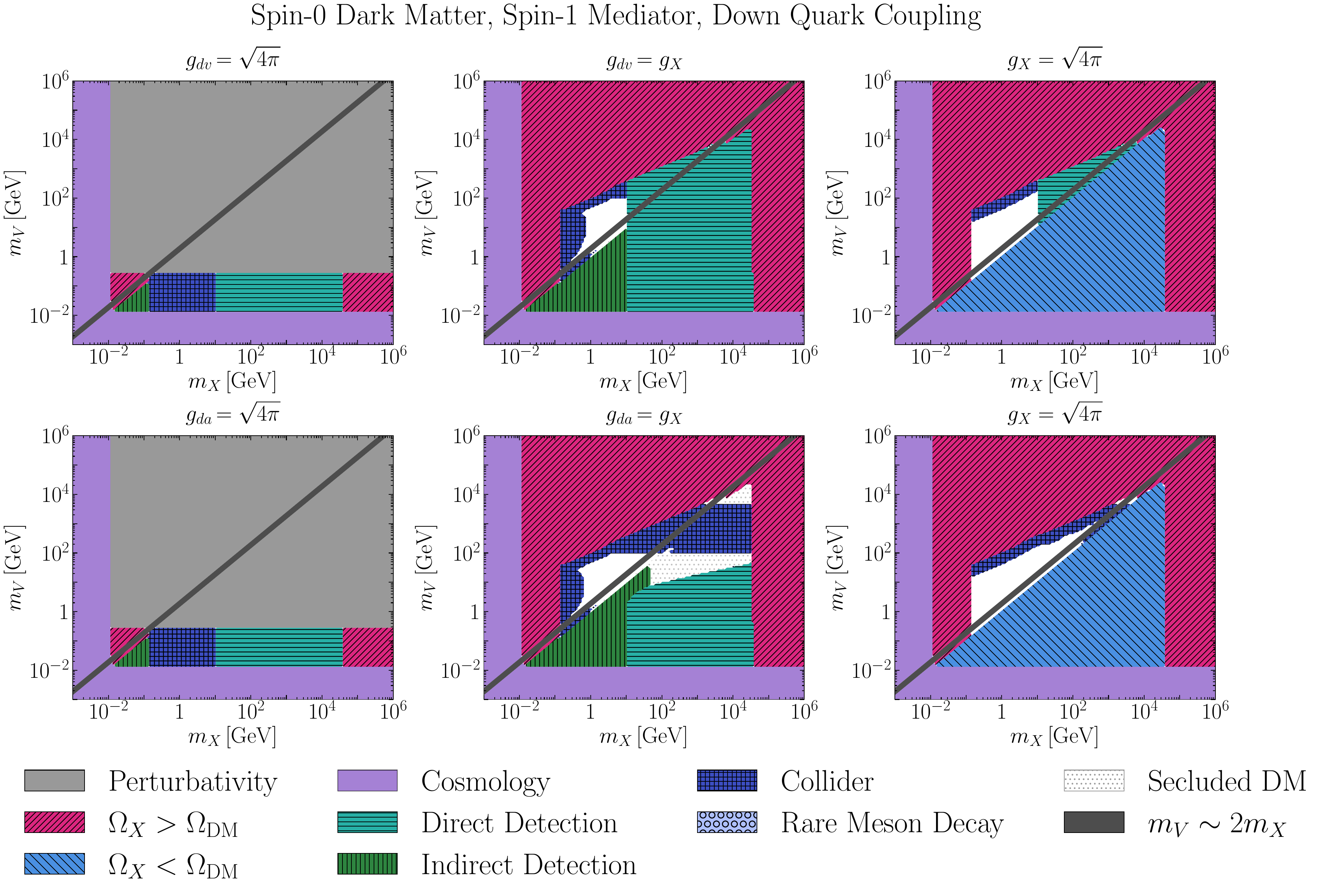}
    \caption{As in Fig.~\ref{fig:u_6_panel}, but for spin-0 dark matter with a spin-1 mediator that couples to down quarks, for vector or axial couplings. The meanings of the various colored regions are summarized in Sec.~\ref{sec:result} and the model is described in Sec.~\ref{sec:Dm0Med1}. The viable regions of parameter space are shown in white (and in white with black dots). See Table~\ref{tab:summary} for links to other figures.}
    \label{fig:d_scalar_vector_6_panel}
\end{figure}

\begin{figure}[!htbp]
    \centering
    \includegraphics[width=\textwidth]{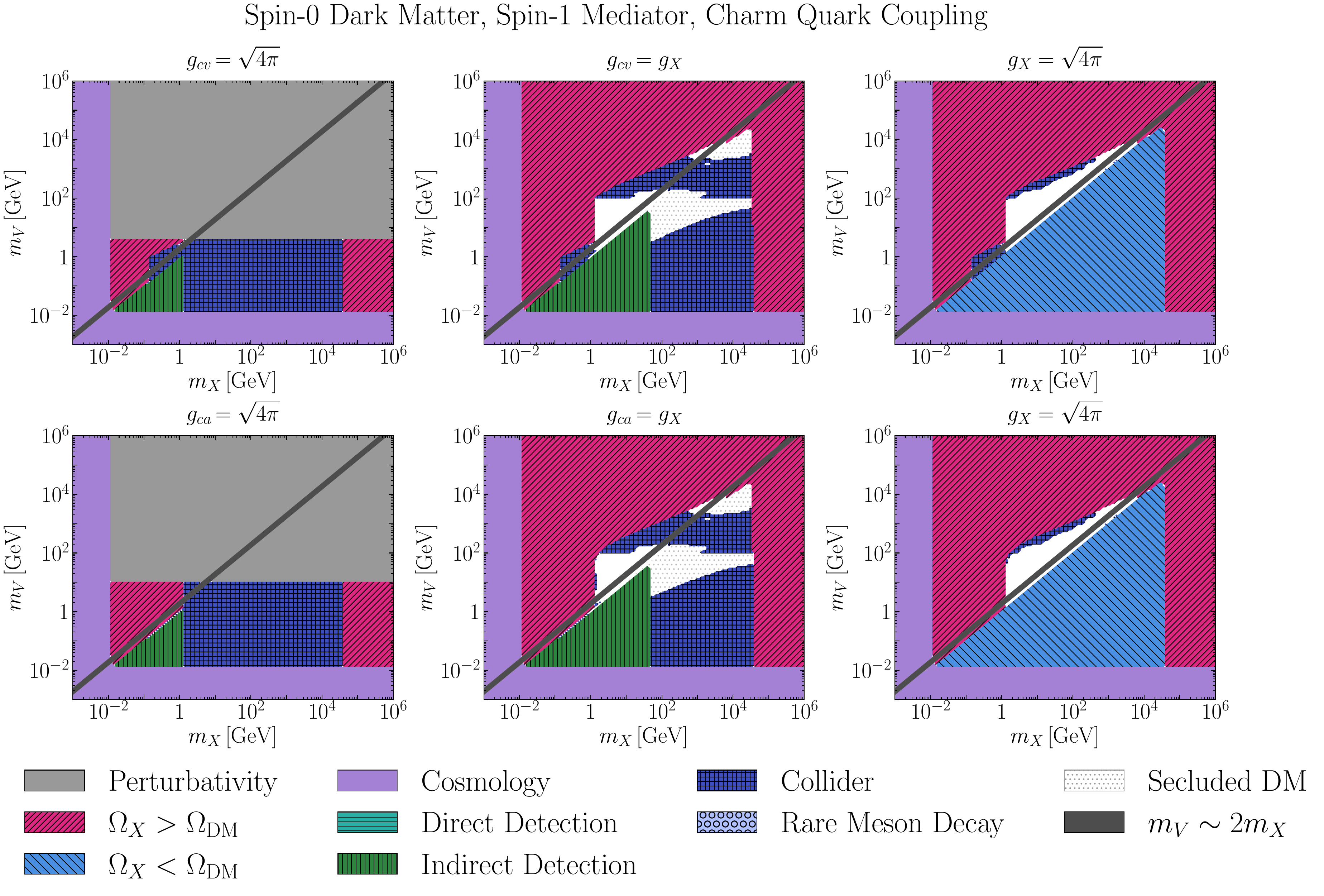}
    \caption{As in Fig.~\ref{fig:u_6_panel}, but for spin-0 dark matter with a spin-1 mediator that couples to charm quarks, for vector or axial couplings. The meanings of the various colored regions are summarized in Sec.~\ref{sec:result} and the model is described in Sec.~\ref{sec:Dm0Med1}. The viable regions of parameter space are shown in white (and in white with black dots). See Table~\ref{tab:summary} for links to other figures.}
    \label{fig:c_scalar_vector_6_panel}
\end{figure}

\begin{figure}[!htbp]
    \centering
    \includegraphics[width=\textwidth]{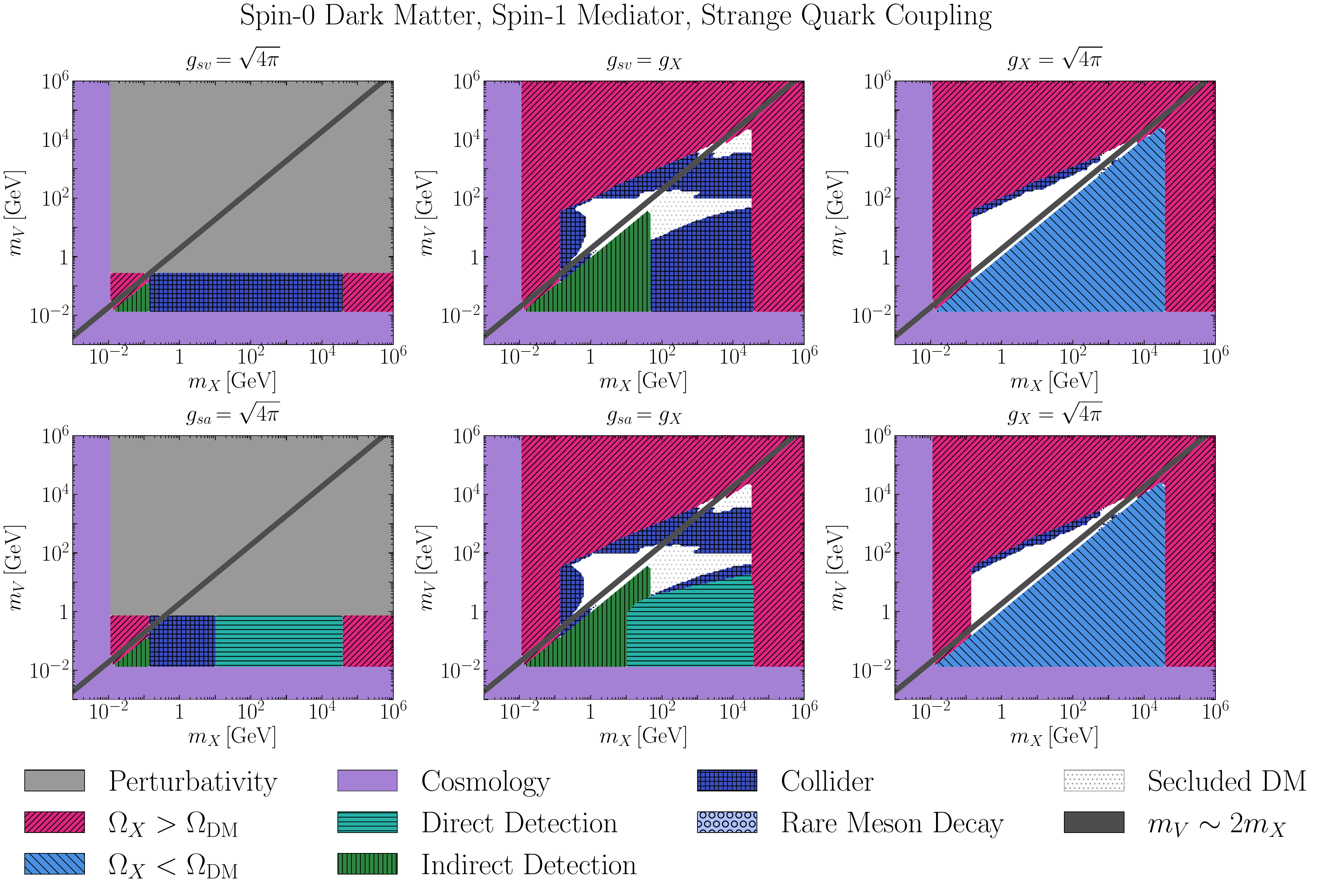}
    \caption{As in Fig.~\ref{fig:u_6_panel}, but for spin-0 dark matter with a spin-1 mediator that couples to strange quarks, for vector or axial couplings. The meanings of the various colored regions are summarized in Sec.~\ref{sec:result} and the model is described in Sec.~\ref{sec:Dm0Med1}. The viable regions of parameter space are shown in white (and in white with black dots). See Table~\ref{tab:summary} for links to other figures.}
    \label{fig:s_scalar_vector_6_panel}
\end{figure}

\begin{figure}[!htbp]
    \centering
    \includegraphics[width=\textwidth]{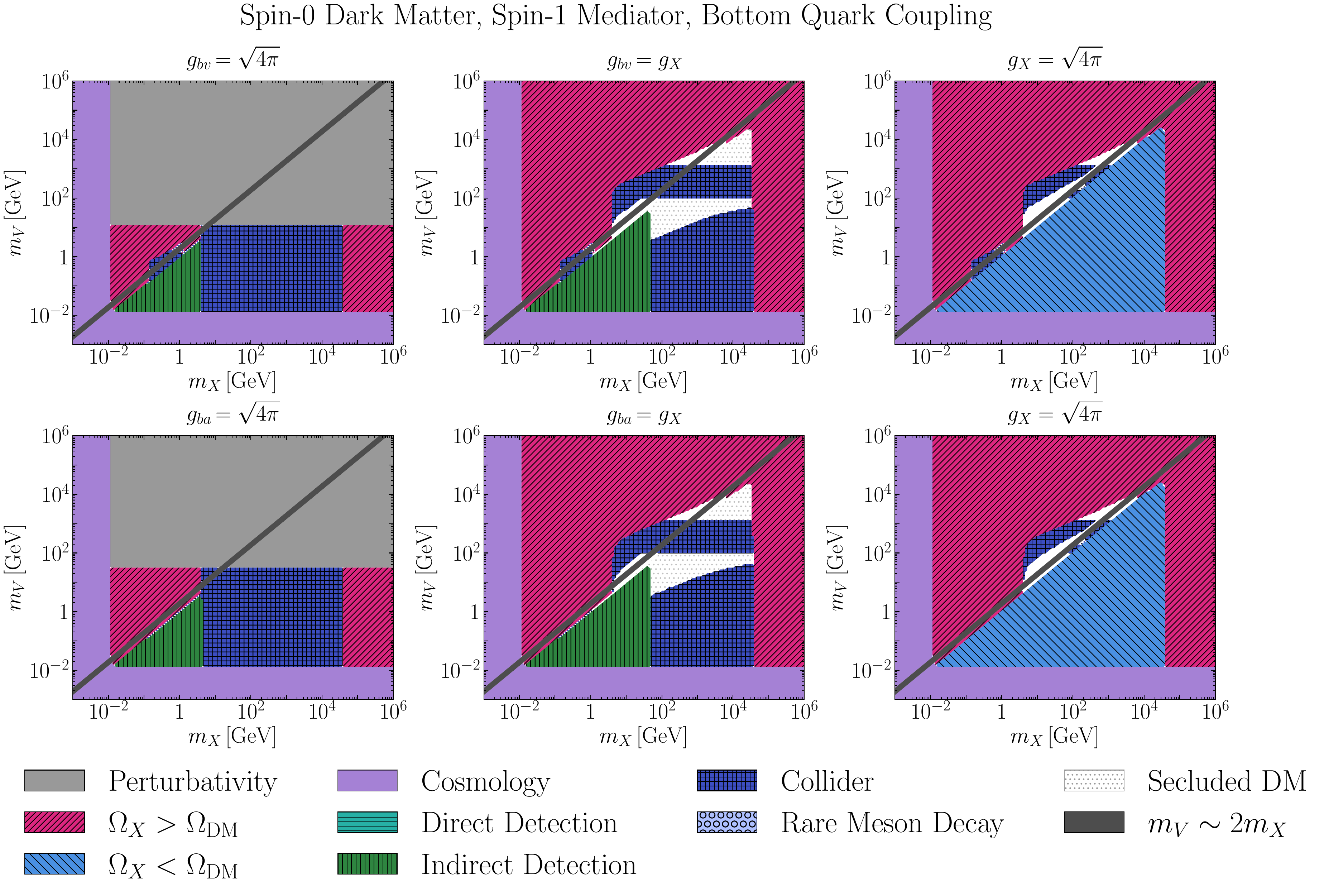}
    \caption{As in Fig.~\ref{fig:u_6_panel}, but for spin-0 dark matter with a spin-1 mediator that couples to bottom quarks, for vector or axial couplings. The meanings of the various colored regions are summarized in Sec.~\ref{sec:result} and the model is described in Sec.~\ref{sec:Dm0Med1}. The viable regions of parameter space are shown in white (and in white with black dots). See Table~\ref{tab:summary} for links to other figures.}
    \label{fig:b_scalar_vector_6_panel}
\end{figure}

\begin{figure}[!htbp]
    \centering
    \includegraphics[width=\textwidth]{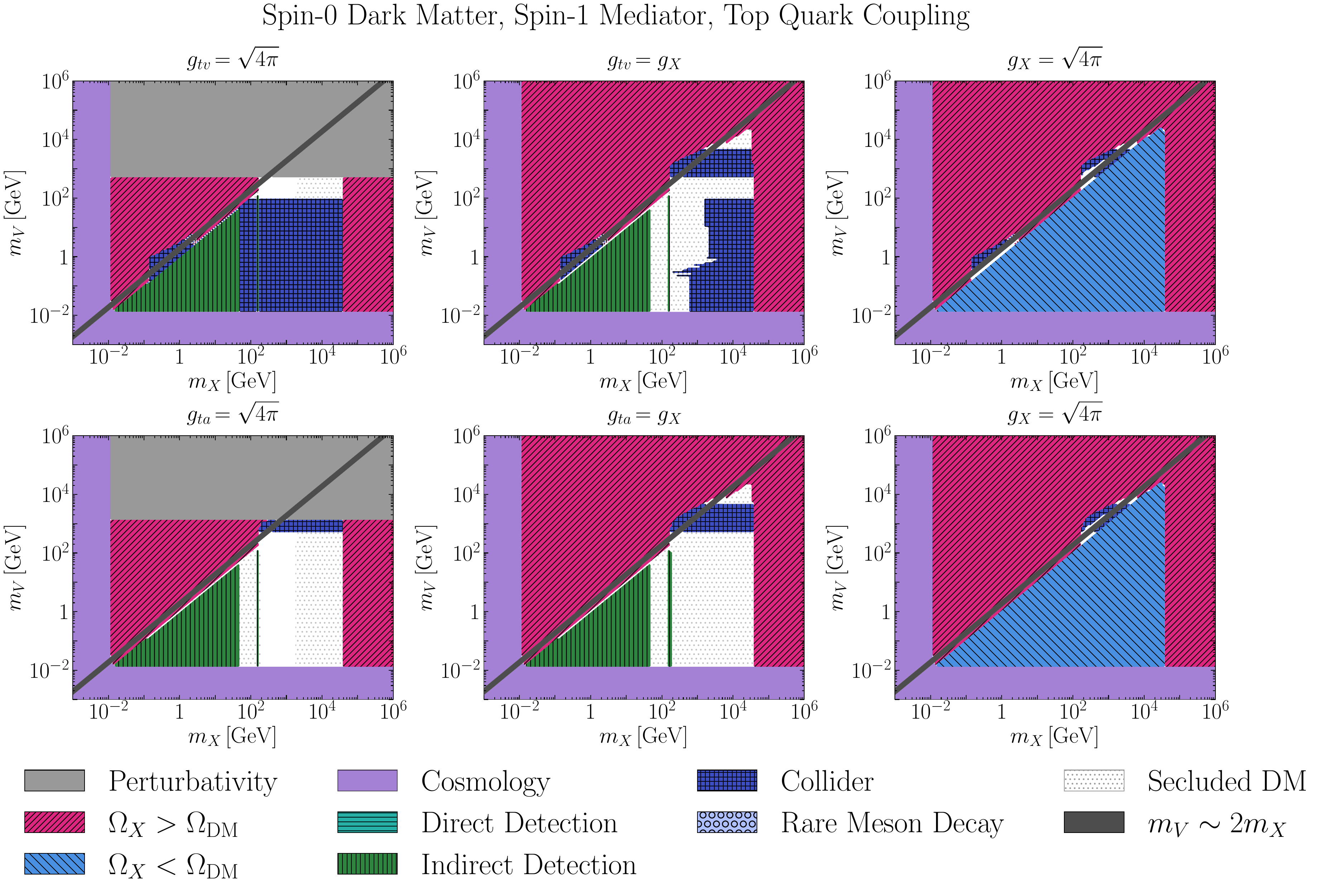}
    \caption{As in Fig.~\ref{fig:u_6_panel}, but for spin-0 dark matter with a spin-1 mediator that couples to top quarks, for vector or axial couplings. The meanings of the various colored regions are summarized in Sec.~\ref{sec:result} and the model is described in Sec.~\ref{sec:Dm0Med1}. The viable regions of parameter space are shown in white (and in white with black dots). See Table~\ref{tab:summary} for links to other figures.}
    \label{fig:t_scalar_vector_6_panel}
\end{figure}

\begin{figure}[!htbp]
    \centering
    \includegraphics[width=\textwidth]{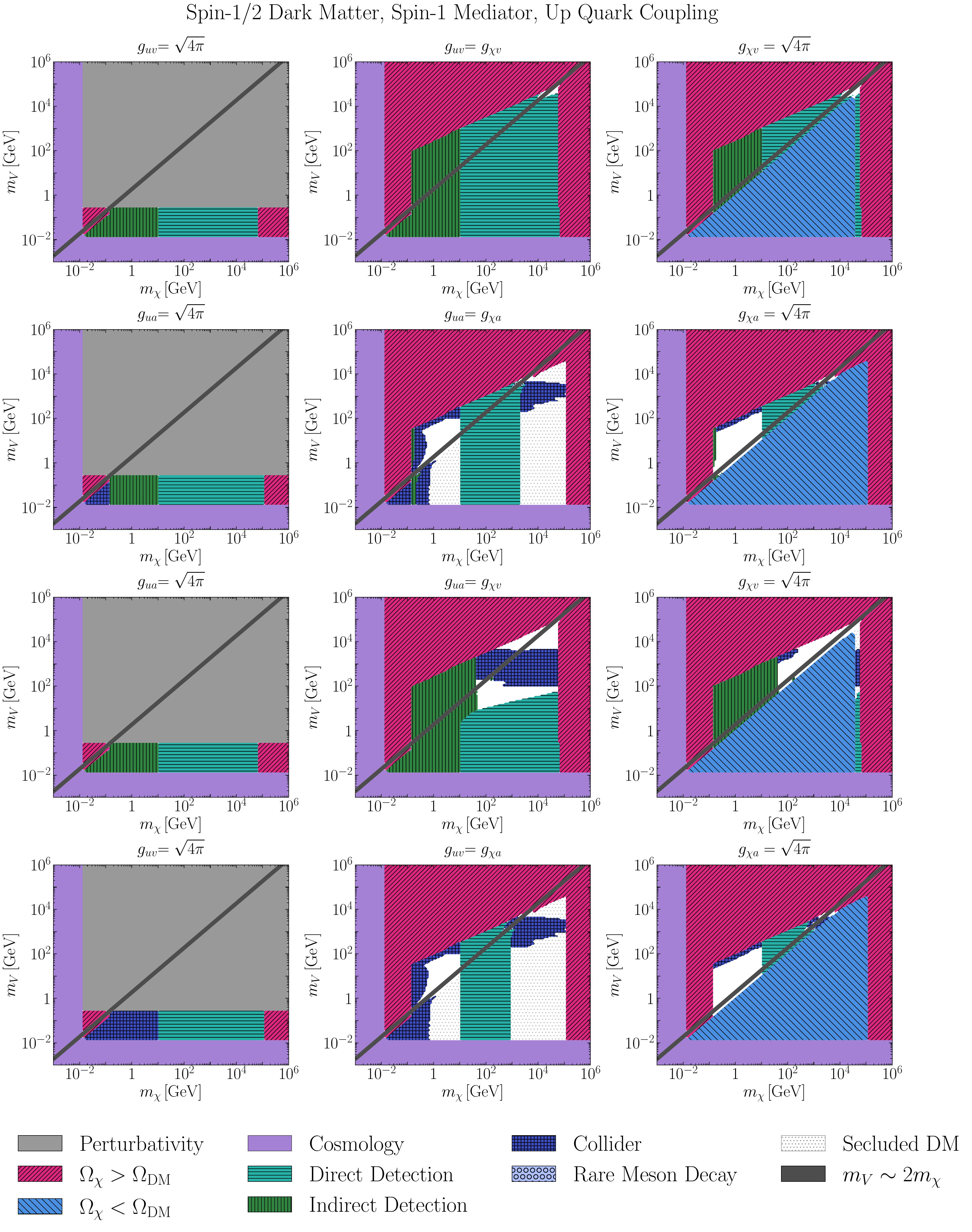}
    \caption{As in Fig.~\ref{fig:u_6_panel}, but for spin-1/2 dark matter with a spin-1 mediator that couples to up quarks, for various combinations of vector and axial couplings. The meanings of the various colored regions are summarized in Sec.~\ref{sec:result} and the model is described in Sec.~\ref{sec:Dm1/2Med1}. The viable regions of parameter space are shown in white (and in white with black dots). See Table~\ref{tab:summary} for links to other figures.}
    \label{fig:u_12_panel_2}
\end{figure}

\begin{figure}[!htbp]
    \centering
    \includegraphics[width=\textwidth]{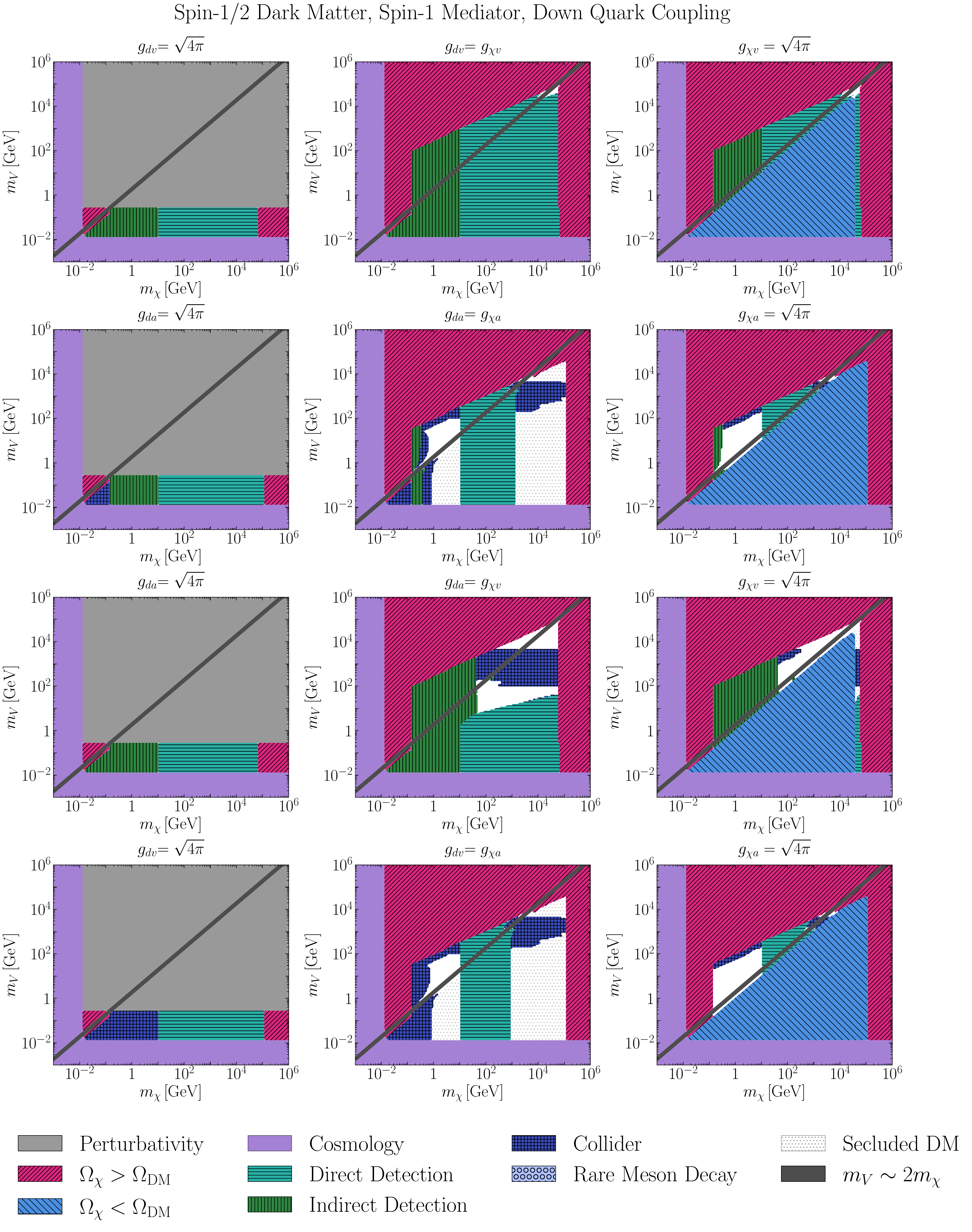}
    \caption{As in Fig.~\ref{fig:u_6_panel}, but for spin-1/2 dark matter with a spin-1 mediator that couples to down quarks, for various combinations of vector and axial couplings. The meanings of the various colored regions are summarized in Sec.~\ref{sec:result} and the model is described in Sec.~\ref{sec:Dm1/2Med1}. The viable regions of parameter space are shown in white (and in white with black dots). See Table~\ref{tab:summary} for links to other figures.}
    \label{fig:d_12_panel_2}
\end{figure}

\begin{figure}[!htbp]
    \centering
    \includegraphics[width=\textwidth]{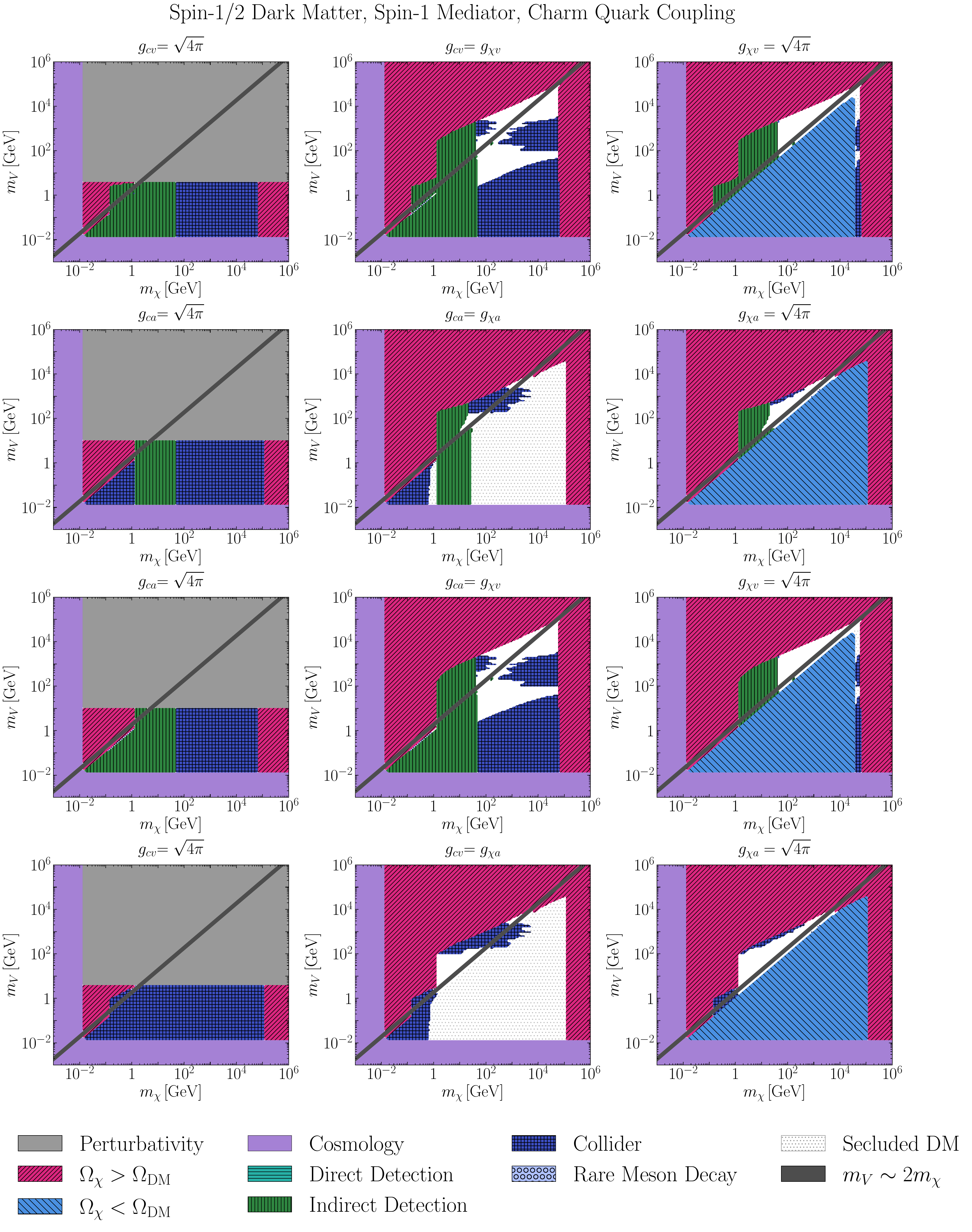}
    \caption{As in Fig.~\ref{fig:u_6_panel}, but for spin-1/2 dark matter with a spin-1 mediator that couples to charm quarks, for various combinations of vector and axial couplings. The meanings of the various colored regions are summarized in Sec.~\ref{sec:result} and the model is described in Sec.~\ref{sec:Dm1/2Med1}. The viable regions of parameter space are shown in white (and in white with black dots). See Table~\ref{tab:summary} for links to other figures.}
    \label{fig:c_12_panel_2}
\end{figure}

\begin{figure}[!htbp]
    \centering
    \includegraphics[width=\textwidth]{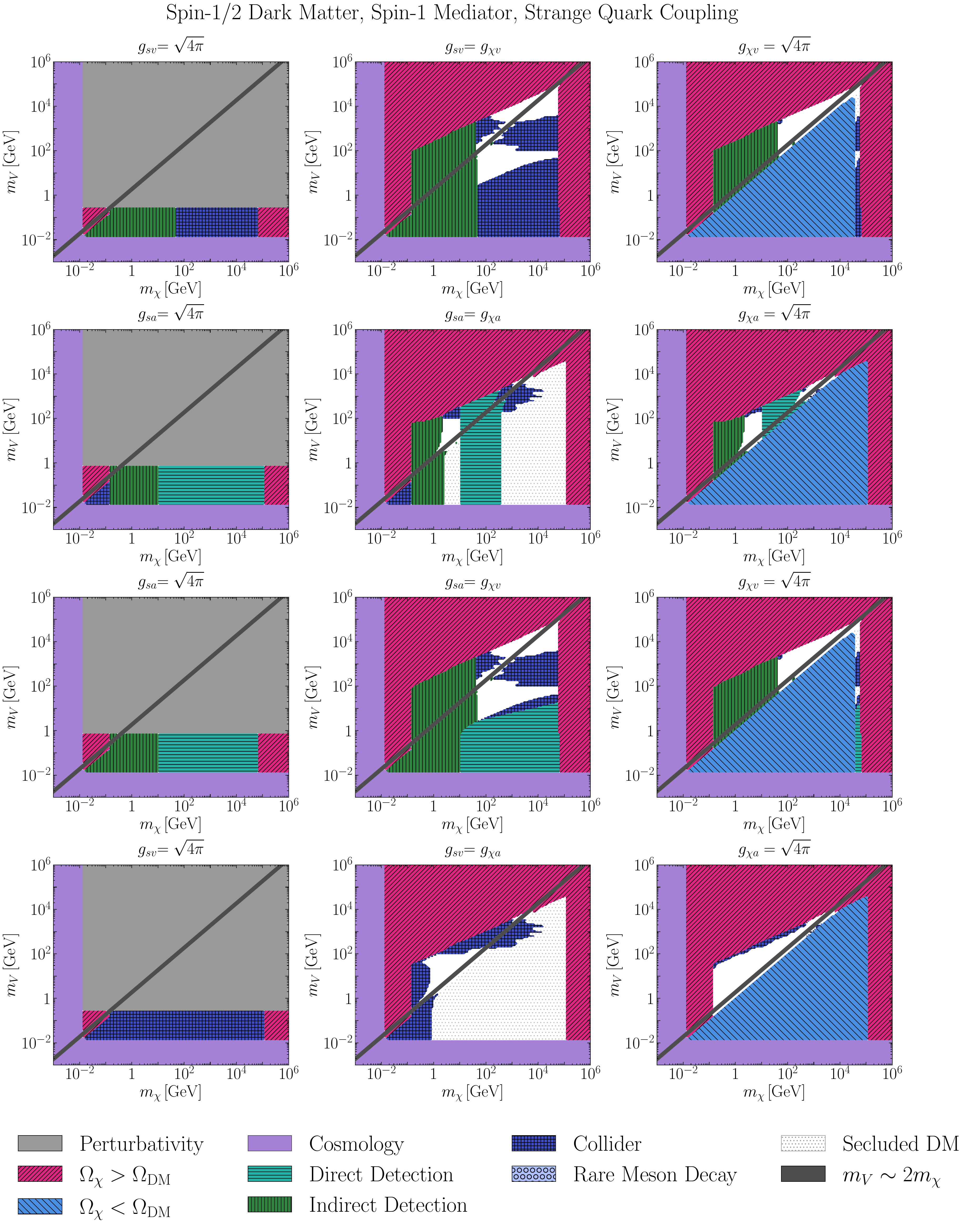}
    \caption{As in Fig.~\ref{fig:u_6_panel}, but for spin-1/2 dark matter with a spin-1 mediator that couples to strange quarks, for various combinations of vector and axial couplings. The meanings of the various colored regions are summarized in Sec.~\ref{sec:result} and the model is described in Sec.~\ref{sec:Dm1/2Med1}. The viable regions of parameter space are shown in white (and in white with black dots). See Table~\ref{tab:summary} for links to other figures.}
    \label{fig:s_12_panel_2}
\end{figure}

\begin{figure}[!htbp]
    \centering
    \includegraphics[width=\textwidth]{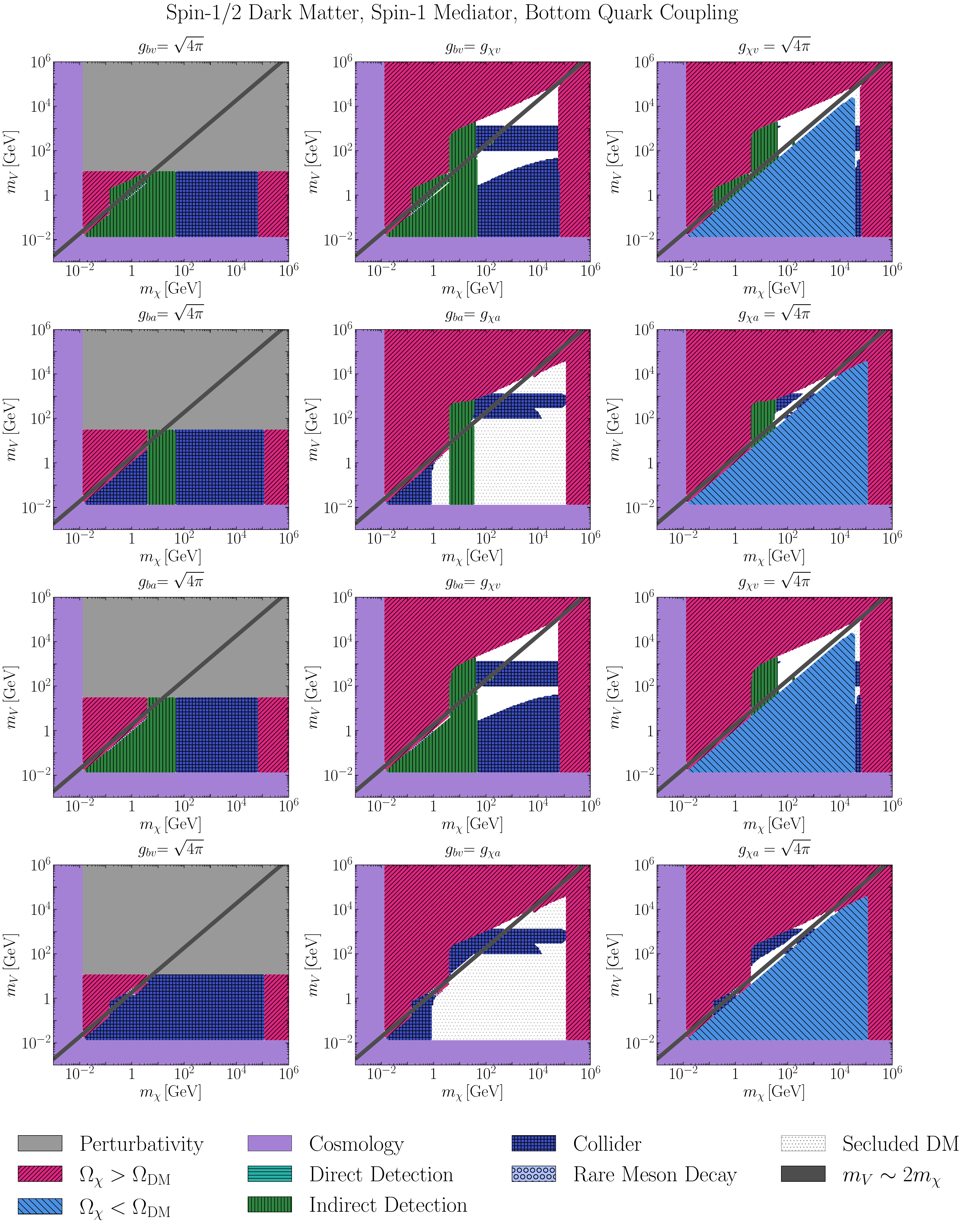}
    \caption{As in Fig.~\ref{fig:u_6_panel}, but for spin-1/2 dark matter with a spin-1 mediator that couples to bottom quarks, for various combinations of vector and axial couplings. The meanings of the various colored regions are summarized in Sec.~\ref{sec:result} and the model is described in Sec.~\ref{sec:Dm1/2Med1}. The viable regions of parameter space are shown in white (and in white with black dots). See Table~\ref{tab:summary} for links to other figures.}
    \label{fig:b_12_panel_2}
\end{figure}

\begin{figure}[!htbp]
    \centering
    \includegraphics[width=\textwidth]{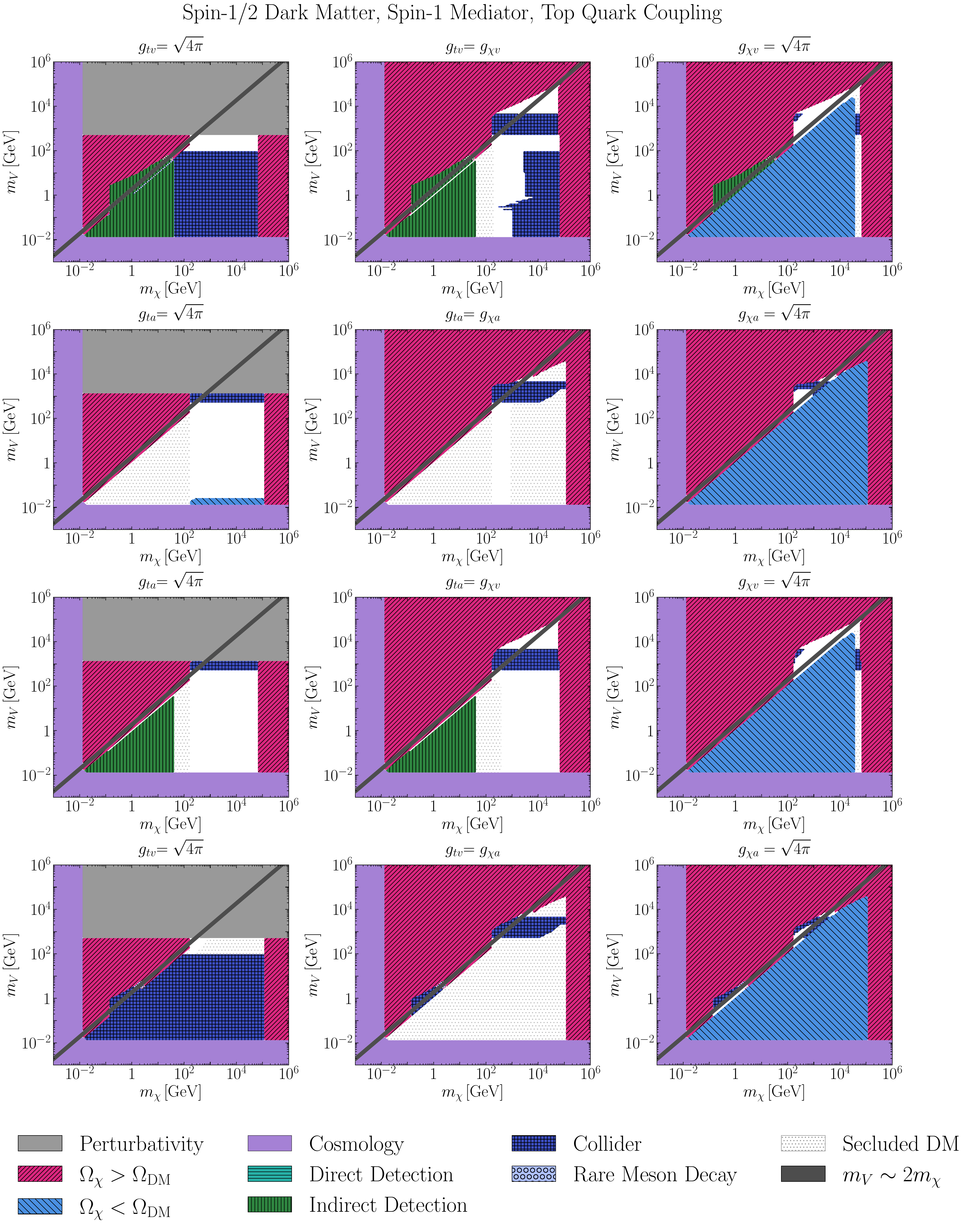}
    \caption{As in Fig.~\ref{fig:u_6_panel}, but for spin-1/2 dark matter with a spin-1 mediator that couples to top quarks, for various combinations of vector and axial couplings. The meanings of the various colored regions are summarized in Sec.~\ref{sec:result} and the model is described in Sec.~\ref{sec:Dm1/2Med1}. The viable regions of parameter space are shown in white (and in white with black dots). See Table~\ref{tab:summary} for links to other figures.}
    \label{fig:t_12_panel_2}
\end{figure}

We have systematically explored a broad range of simplified models in which dark matter is produced through thermal freeze-out. We have considered scenarios featuring spin-0 or spin-1/2 dark matter candidates, which couple to mediators of spin-0 or spin-1.
In Figs.~\ref{fig:u_6_panel}-\ref{fig:t_12_panel_2}, we take the mediator to couple to dark matter and to a single SM quark flavor, through renormalizable interactions which govern chemical decoupling between the dark and visible sectors in the early universe.
In each case, we scan over the model parameters and identify regions of parameter space which remain viable after applying constraints from early universe cosmology (Sec.~\ref{sec:Neff}), direct (Sec.~\ref{sec:direct}) and indirect detection (Sec.~\ref{sec:indirect}), collider searches (Sec.~\ref{sec:collider}), and rare meson decays (Sec.~\ref{sec:meson}).
By combining these probes, we find that most simplified models of quark-coupled thermal-relic dark matter are significantly constrained, although significant regions of parameter space remain viable.

In order for a thermal relic to be produced with the measured dark matter abundance without ruining the successful predictions of BBN, it must have a mass $10^{-2} \, \text{GeV}\lesssim m_\chi \lesssim 10^{5} \, {\rm GeV}$. Much of the parameter space of such models has been constrained in recent years by experiments directly searching for dark matter scattering with nuclei. In particular, we find that models with spin-0 mediators with scalar couplings, or spin-one mediators with vector couplings, are now severely constrained by such experiments. To a lesser degree, mediators with axial or pseudoscalar couplings are also constrained.

Searches for dark matter annihilation products (or the effects of the those products on the cosmic microwave background) have also restricted this class of models. These constraints most strongly impact dark matter candidates that are relatively light ($m_{\chi } 
< 40 \, {\rm GeV}$) and that annihilate efficiently at low velocities (i.e., are not $p$-wave suppressed). 

Of the dark matter parameter space that has not yet been constrained by direct or indirect detection efforts, substantial regions have been explored by collider searches, including those involving rare meson decays. This is particularly true of models with mediators that are heavier than $\sim 100 \, {\rm GeV}$ or lighter than $\sim 1-10 \, {\rm GeV}$, depending on the value of the dark matter's mass.

We remind the reader that we have limited ourselves in this study to dark matter candidates which froze-out of equilibrium in the early universe through self-annihilations mediated by particles that couple to Standard Model quarks. If the dark matter instead annihilates to gauge and/or Higgs bosons, their elastic scattering cross sections with nuclei will be suppressed by loop factors, allowing for more parameter space to remain viable~\cite{Hisano:2011cs,Hill:2014yka,Hill:2014yxa,Berlin:2015ymu}. Models in which the dark matter abundance is depleted through coannihilations~\cite{Griest:1990kh,Edsjo:1997bg} also remains largely unconstrained. Alternatively, the dark matter could plausibly be a thermal relic that is part of a highly sequestered hidden sector, with couplings to the Standard Model that are well beyond the reach of direct detection experiments or collider searches~\cite{Pospelov:2007mp,Hooper:2012cw,Berlin:2016vnh,Berlin:2016gtr}.

Although we have only considered models here in which the dark matter couples to individual Standard Model quark flavors, the union of these results can be straightforwardly adapted to consider more complete realizations with particular flavor structure (e.g. Higgs-like couplings). It is important to note that, for UV-complete models, there may be additional constraints and theoretical considerations that complement the analysis presented here. For example, in order to realize flavor-specific couplings in a gauge-invariant UV model, there may be additional flavor bounds or naturalness issues that we have not included. For further discussions of such issues, see Refs.~\cite{Kahn:2016vjr,Batell:2017kty,Batell:2018fqo,Batell:2021xsi}.

\begingroup
\renewcommand{\addcontentsline}[3]{}
\acknowledgments
\endgroup

We would like to thank Asher Berlin for collaboration during the early stage of this work, and for useful discussion throughout. We are grateful to Keisuke Harigaya, Phil Harris, Sung-Hak Lim, Olivier Mattelaer, Kevin Pedro, Matthew Reece, Ke-Pan Xie, and Xun-Jie Xu for useful discussions. Fermilab is operated by Fermi Forward Discovery Group, LLC under Contract No. 89243024CSC000002 with the U.S. Department of Energy, Office of Science, Office of High Energy Physics. IRW is also supported by DOE distinguished scientist fellowship grant FNAL 22-33. DH is supported by the Office of the Vice Chancellor for Research at the University of Wisconsin-Madison, with funding from the Wisconsin Alumni Research Foundation.

\clearpage
\thispagestyle{empty}
\vspace*{\fill}
\vspace*{-1cm}
\vspace*{\fill}

\clearpage
\thispagestyle{empty}
\vspace*{\fill}
\vspace*{-1cm}
\vspace*{\fill}

\clearpage
\thispagestyle{empty}
\vspace*{\fill}
\vspace*{-1cm}
\vspace*{\fill}

\clearpage
\thispagestyle{empty}
\vspace*{\fill}
\vspace*{-1cm}

\vspace*{\fill}

\clearpage
\thispagestyle{empty}
\vspace*{\fill}
\vspace*{-1cm}

\vspace*{\fill}

\clearpage
\thispagestyle{empty}
\vspace*{\fill}
\vspace*{-1cm}

\vspace*{\fill}

\clearpage

\clearpage

\thispagestyle{empty}
\vspace*{\fill}
\vspace*{-1cm}
\vspace*{\fill}

\clearpage
\thispagestyle{empty}
\vspace*{\fill}
\vspace*{-1cm}

\vspace*{\fill}

\clearpage
\thispagestyle{empty}
\vspace*{\fill}
\vspace*{-1cm}

\vspace*{\fill}

\clearpage
\thispagestyle{empty}
\vspace*{\fill}
\vspace*{-1cm}

\vspace*{\fill}

\clearpage
\thispagestyle{empty}
\vspace*{\fill}
\vspace*{-1cm}

\vspace*{\fill}

\clearpage
\thispagestyle{empty}
\vspace*{\fill}
\vspace*{-1cm}

\vspace*{\fill}

\clearpage

\appendix

\setcounter{equation}{0}
\setcounter{figure}{0}
\setcounter{table}{0}
\setcounter{section}{0}

\makeatletter
\renewcommand{\thesection}{A\arabic{section}}
\renewcommand{\theequation}{A\arabic{equation}}
\renewcommand{\thefigure}{A\arabic{figure}}
\renewcommand{\thetable}{A\arabic{table}}

% Reset hyperref anchors
\renewcommand{\theHfigure}{A\arabic{figure}}
\renewcommand{\theHtable}{A\arabic{table}}
\renewcommand{\theHequation}{A\arabic{equation}}
\makeatother

\clearpage

\section*{Appendix: Three-Flavor Meson Chiral Perturbation Theory}
\label{sec:chpt app}

In this appendix, we provide a brief review of the formulation and results of meson chiral perturbation theory (ChPT).
These formulations are derived from symmetry arguments, and we refer the readers to Ref.~\cite{Pich:1993uq,Scherer:2002tk,Scherer:2012xha} for a comprehensive and pedagogical review.
For the purpose of generality, we work in three-flavor ChPT at leading order, in which $s$ is treated as a light flavor.

The mesons appear as pseudo Nambu-Goldstone bosons (pNGB) arising from the spontaneous breaking of the approximated $SU(3)_L \times SU(3)_R \times U(1)_V$ symmetry, known as the chiral symmetry of QCD.
The meson fields are written in an octet under the basis of $(u, d, s)$ as
\begin{align}
    \label{eq:field}
    U    = \exp(i \frac{\Pi(x)}{F_0})\,,~~
    \Pi  \equiv \sum_a \lambda^a \Pi^a (x) =
    \begin{pmatrix}
        \pi^0 + \frac{1}{\sqrt{3}} \eta & \sqrt{2} \pi^+                    & \sqrt{2}K^+               \\
        \sqrt{2} \pi^-                  & - \pi^0 + \frac{1}{\sqrt{3}} \eta & \sqrt{2} K^0              \\
        \sqrt{2} K^-                    & \sqrt{2} K^0                      & - \frac{2}{\sqrt{3}} \eta
    \end{pmatrix}\,,
\end{align}
where $F_0 = 93.2$ MeV and $\lambda^a$ are Gellman matrices.
Such a field is embedded into the general Lagrangian of ChPT.
To the lowest order, the general effective Lagrangian is
\begin{align}
    \label{eq:L2}
    \mathcal{L} = \frac{F_0^2}{4} \tr \left( D_\mu U (D_\mu U)^\dagger \right) + \frac{F_0^2}{4} \left(  \chi U^\dagger + U \chi^\dagger \right)\,.
\end{align}
Here, the gauge interactions are included in the covariant derivative, while the interaction with the spin-0 mediator is included in $\chi$.
The definition is
\begin{align}
    D_\mu U & \equiv \partial_\mu U - i r_\mu U + i U l_\mu\,~~~,~~\chi \equiv    2 B_0 (- \chi_s + i \chi_p)\,,
\end{align}
where $B_0 = 2853.3$ MeV and $\chi_{s,p}$ stands for scalar and pseudoscalar coupling, respectively.
The notations are defined as follows.
$r_\mu$ and $l_\mu$ are gauge interactions with right-handed and left-handed quarks, respectively.
$s$ and $p$ are scalar and pseudoscalar interactions with quarks.
All of these elements should be written in a 3-by-3 matrix form in the $(u,d,s)$ basis.
For example, QED coupling is written as $l_\mu = r_\mu = e A_\mu Q$, with $Q = \rm diag (2/3, -1/3, -1/3)$.
The Dirac masses are regarded as a scalar interaction.

Here, we provide the effective Lagrangian derived from three-flavor ChPT for mediators coupling to up, down, or strange quarks.

{\bf Spin-0 mediator}:
The coupling between quarks and a spin-0 mediator is embedded in $\chi$ through
\begin{align}
    \chi_s = - M + g_{fs} \phi E_i\,, \quad \chi_p = g_{fp} \phi E_i\,,
\end{align}
where $M = \mathrm{diag}(m_u, m_d, m_s)$ is the quark mass matrix, $E_i$ denotes a matrix with 1 on the $i$-th diagonal element and 0 elsewhere, and $i$ labels the quark flavor that couples to the mediator.

If the mediator couples to up quarks, the effective Lagrangian is
\begin{align}
    \mathcal{L}_u = & g_{fs} B_0 \phi \left(\frac{\eta^2}{6}+\frac{\pi^0 \eta}{\sqrt{3}}+K^- K^++\frac{\left(\pi ^0\right)^2}{2}+\pi ^- \pi ^+ \right)  + g_{fp} F_0 B_0 \phi \left( \frac{\eta}{\sqrt{3}}+\pi^0 \right)\,.
\end{align}
If the mediator couples to down quarks, the effective Lagrangian is
\begin{align}
    \mathcal{L}_d = & g_{fs} B_0 \phi \left(\frac{\eta ^2}{6}-\frac{\pi ^0 \eta }{\sqrt{3}}+\left(K^0\right)^2+\frac{\left(\pi ^0\right)^2}{2}+\pi ^- \pi ^+ \right) + g_{fp} B_0 F_0 \phi \left( \frac{\eta }{\sqrt{3}}-\pi ^0 \right)\,.
\end{align}
If the mediator couples to strange quarks, the effective Lagrangian is
\begin{align}
    \mathcal{L}_s = & g_{fs} B_0 \phi \left( \frac{2 \eta ^2}{3}+\left(K^0\right)^2+K^- K^+ \right)  - g_{fp} B_0 F_0 \phi \frac{2 \eta }{\sqrt{3}}\,.
\end{align}

The scalar mediator-photon operator induced by the meson loop is then given by
\begin{align}
    g_{fs} B_0 \phi K^- K^+ \to g_{fs} B_0 \frac{\alpha}{48 \pi m_K^2} \phi F_{\mu \nu}F^{\mu \nu}\,,~~~~g_{fs} B_0 \phi \pi^- \pi^+ \to g_{fs} B_0 \frac{\alpha}{48 \pi m_\pi^2} \phi F_{\mu \nu}F^{\mu \nu}\,.
\end{align}

On the other hand, a pseudoscalar can mix both with the $\pi^0$ and $\eta$.
Specifically, the $\eta$ meson appearing above is a quantum state defined as $\eta^8 \equiv (\bar{u}u + \bar{d}d - 2 \bar{s}s)/\sqrt{6}$.
This meson mixes with $\eta^1 \equiv (\bar{u}u + \bar{d}d + \bar{s}s)/\sqrt{3}$ to give two physical mass eigenstates, $\eta$ and $\eta'$, with masses of $m_\eta = 547.86~\rm MeV$ and $m_{\eta'} = 957.78~\rm MeV$~\cite{ParticleDataGroup:2024cfk}.
One needs to carefully examine the $\eta$ when determining the fate of the pseudoscalar mediator in the relevant mass scale.

    {\bf Spin-1 mediator}:
The coupling between quarks and a spin-1 mediator is embedded into the covariant derivative, with
\begin{align}
    l_\mu = g_{fs} V_\mu + g_{fa} V_\mu\,~~~,~~~ r_\mu = g_{fs} V_\mu - g_{fa} V_\mu\,.
\end{align}
Both appear in the $i$-th diagonal element, where $i$ labels the quark flavor that couples to the mediator.

If the mediator couples to up quarks, the effective Lagrangian is
\begin{align}
    \mathcal{L} = & i g_{fv} V_\mu \left(-K^+ \partial^\mu K^- + K^- \partial^\mu K^+ - \pi^+ \partial^\mu \pi^- + \pi^- \partial^\mu \pi^+\right) + g^2_{fv} V_\mu V^\mu \left(K^- K^+ + \pi^- \pi^+\right) \nonumber                                                                          \\
                  & + F_0 g_{fa} V_\mu \left(\partial^\mu \pi^0 + \frac{1}{\sqrt{3}} \partial^\mu \eta\right)  + \frac{1}{3}g_{fa}^2 V_\mu V^\mu \left( 3 \left(K^- K^++\left(\pi ^0\right)^2+\pi ^- \pi ^+\right)+\eta ^2+2 \sqrt{3} \pi^0 \eta \right)\,.
\end{align}
If the mediator couples to down quarks, the effective Lagrangian is
\begin{align}
    \mathcal{L} = & i g_{fv} V_\mu \left(\pi ^+ \partial^\mu \pi^- -\pi^- \partial^\mu \pi^+\right)  + g_{fv}^2 V_\mu V^\mu \left( (K^0)^2 + \pi^- \pi^+ \right) \nonumber                                                                                    \\
                  & + F_0 g_{fa} V_\mu \left(\frac{1}{\sqrt{3}} \partial^\mu \eta- \partial^\mu \pi^0\right)  + \frac{1}{3} g_{fa}^2 V_\mu V^\mu \left(3 \left(\left(K^0\right)^2+\left(\pi ^0\right)^2+\pi ^- \pi ^+\right)+\eta ^2-2 \sqrt{3} \pi ^0 \eta \right)\,.
\end{align}
If the mediator couples to strange quarks, the effective Lagrangian is
\begin{align}
    \mathcal{L} = & i g_{fv} V_\mu \left( K^+ \partial^\mu K^- - K^- \partial^\mu K^+  \right) + g_{fv}^2 V_\mu V^\mu (K^- K^+ + (K^0)^2) \nonumber                                               \\
                  & - \frac{2}{\sqrt{3}}g_{fa} F_0 V_\mu \partial^\mu \eta + \frac{1}{3} g_{fa}^2 V_\mu V^\mu \left( 3 \left( K^- K^+ + (K^0)^2 \right)  + 4 \eta^2\right)\,.
\end{align}

As stated in the main text, spin-1 mediator cannot decay into a pair of photons, and thus no effective meson-loop operator is discussed here.
Again, a careful assessment of the role of the $\eta$ can be necessary in this case in the relevant mass scale.

\bibliographystyle{utphys3}
\bibliography{main}

\end{document}